\documentstyle[11pt,amssymb]{article}
\def\tabaddress#1{{\small\it\begin{tabular}[t]{c}#1
\\[1.2ex]\end{tabular}}}

\font\fr=eufm10 scaled \magstep 1 
\newtheorem{teor}{Theorem}
\newtheorem{prop}{Proposition}
\newtheorem{corol}{Corollary}
\newtheorem{definition}{Definition}
\newtheorem{lem}{Lemma}
\newtheorem{state}{Statement}
\newtheorem{assum}{Assumption}
\newtheorem{remark}{Remark}
\def\beq{\begin{equation}}
\def\eeq{\end{equation}}
\def\bea{\begin{eqnarray}}
\def\eea{\end{eqnarray}}
\def\beann{\begin{eqnarray*}}
\def\eeann{\end{eqnarray*}}
\def\ben{\begin{enumerate}}
\def\een{\end{enumerate}}
\def\bit{\begin{itemize}}
\def\eit{\end{itemize}}
\def\dst{\(\displaystyle}
\def\derpar#1#2{\frac{\partial{#1}}{\partial{#2}}}

\def\moment#1#2#3{{#1}_{#2}, \ldots, {#1}_{#3}}

\def\qed{\ifvmode\removelastskip\fi
{\unskip\nobreak\hfil\penalty50\hbox{}\nobreak\hfil \hbox{\vrule
height1.2ex width1.2ex}\parfillskip=0pt \finalhyphendemerits=0
\par\smallskip}}
\def\vf{\mbox{\fr X}}
\def\df{{\mit\Omega}}
\def\Lag{{\cal L}}
\def\lag{\pounds}

\def\d{{\rm d}}
\def\Real{\hbox{$\Bbb R$}} 

\def\Img{\mathop{\rm Im}\nolimits}
\def\inn{\mathop{i}\nolimits}
\def\Tan{{\rm T}}

\def\Lie{\mathop{\rm L}\nolimits}
\def\ls{(J^1E,\Omega_\Lag )}

\def\hso{(J^{1*}E,{\cal P},\Omega_{{\rm h}_0})}

\def\Cinfty{{\rm C}^\infty}
\def\proof{( {\sl Proof} )\quad}

\begin{document}

\title{SINGULAR LAGRANGIAN SYSTEMS ON JET BUNDLES}
\author{\sc Manuel de Le\'on\thanks{{\bf e}-{\it mail}:
 MDELEONL@IMAFF.CFMACC.CSIC.ES}
 \\
 \tabaddress{Instituto de Matem\'aticas y F\'\i sica Fundamental, CSIC\\
   C/ Serrano 123. E-28006 Madrid. Spain}
   \\
{\sc Jes\'us Mar\'\i n-Solano \thanks{{\bf e}-{\it mail}:
 JMARIN@ECO.UB.ES}},
 \\
 \tabaddress{Departamento de Matem\'atica Econ\'omica, Financiera
  y Actuarial, UB\\
  Av. Diagonal 690. E-08034 Barcelona. Spain}
   \\
{\sc Juan Carlos Marrero \thanks{{\bf e}-{\it mail}:
 JCMARRER@ULL.ES}},
 \\
 \tabaddress{Departamento de Matem\'atica Fundamental, Fac. Matem\'aticas,
  U. La Laguna\\
  La Laguna, Tenerife. Spain}
   \\
{\sc Miguel C. Mu\~noz-Lecanda\thanks{{\bf e}-{\it mail}:
 MATMCML@MAT.UPC.ES}},
{\sc Narciso Rom\'an-Roy\thanks{{\bf e}-{\it mail}:
 MATNRR@MAT.UPC.ES}},
 \\
 \tabaddress{Departamento de Matem\'atica Aplicada IV\\
  Edificio C-3, Campus Norte UPC\\
  C/ Jordi Girona 1. E-08034 Barcelona. Spain}}

\pagestyle{myheadings}
\markright{\sc M. de Le\'on {\it et al\/},
   \sl Singular Lagrangian systems on jet bundles\ldots }
\maketitle
\thispagestyle{empty}

\begin{abstract}
The jet bundle description of time-dependent mechanics is revisited.
The constraint algorithm for singular Lagrangians is discussed and
an exhaustive description of the constraint functions is given.
By means of auxiliary connections we give a basis of constraint
functions in the Lagrangian and Hamiltonian sides.
An additional description of constraints is also given 
considering at the same time
compatibility, stability and second order condition problems. 
Finally, a classification of the constraints in first and second class is 
obtained using a cosymplectic geometry setting.
Using the second class constraints, a Dirac bracket is introduced, extending
the well-known construction by Dirac.
\end{abstract}


\bigskip

\vbox{\raggedleft AMS s.\,c.\,(2000):
37C10, 37J05, 53C05, 53D05, 55R10, 58A20, 70G45, 70H03, 70H05, 70H45.
 \\
PACS (1999): 02.40.Vh, 45.20.Jj, 45.90.+t.}\null

 \clearpage

 \tableofcontents

 \section{Introduction}

For many years the problem of quantizing singular Lagrangians has
been the object of much attention. Indeed, if we have a regular Lagrangian
$\lag\colon\Tan Q \longrightarrow \Real$
defined on the velocity space $\Tan Q$ of a configuration manifold $Q$,
we get a nice Hamiltonian description on the phase space $\Tan^*Q$:
the Lagrangian energy is transported
from $\Tan Q$ to $\Tan^*Q$ by the Legendre transformation
and, using the canonical symplectic structure on $\Tan^*Q$
we obtain the corresponding Hamilton equations.
Next, we can use the standard quantization rules for the
canonical Poisson bracket on the cotangent bundle \cite{AM-78}.

When the Lagrangian is singular, however, (i.e. the Hessian matrix
of $\lag$ with respect to the velocities is nonregular)
we do not have a nice Hamiltonian description.
In fact, not all the momenta are available,
so we have some primary constraints
defining a submanifold of $\Tan^*Q$. Moreover,
given an initial data and its possible evolution, we have to ensure that
it would be again admissible (the tangency condition).
Dirac \cite{Dir-64} solved the problem by developing a constraint
algorithm which gives (in the favourable cases) a final constraint submanifold
where a solution of the dynamics exists. In addition,
Dirac classified the constraints
into two categories: first and second class constraints.
The second class constraints were
then used to modify the canonical Poisson bracket to obtain a new one
(now called the Dirac bracket) which allows us to use the quantization rules.
In this approach, an accurate description of the constraint functions
plays a crucial role. Dirac's main aim was to apply
this procedure to Field theories;
indeed, many field theories (for instance, electromagnetism) are singular.

After Dirac, a lot of work was done in order to geometrize his algorithm.
The first important step was the work by Gotay {\it et al}
\cite{GNH-78}, and its application to the Lagrangian formalism
\cite{GN-79,GN-80}. Extension to field theories
was always the main aim. Other algorithms were given later, in order to finding
consistent solutions of the dynamical equations
in the Lagrangian formalism of singular systems 
(including the second-order problem) \cite{BGPR-86,Ka-82}, and
afterwards, new geometric algorithms
were developed to be applied both in the Hamiltonian and
the Lagrangian formalisms \cite{CLR-87,GP-92,MMT-97,Mu-89,MR-92,SR-83}.

However, the ``Hamiltonian'' description of field theories,
termed the multisymplectic approach, is the natural extension of
time-dependent mechanics. Therefore, if we wish to understand
the constraint algorithm for field theories
in a covariant formalism, the first step would be
to extend the Dirac procedure to time-dependent Lagrangian systems.
Some work was provided in
\cite{CF-93,CLM-94,GMS-97,IM-92,ILDM-99,
LMM-96,LMM-96b,MS-98,MassVig-00,Vig-00}.
In this paper, we present a complete covariant description for this kind
of system, which generalizes the results of some of the
above-mentioned references. Let us explain this formalism:

We consider a configuration fibred manifold
$\pi : E \longrightarrow B$, where $B$ is an oriented 1-dimensional manifold
($\Real$ or $S^1$) with volume form $\eta$. 
The Lagrangian density is $\Lag = \lag \eta$,
where $\lag$ is a function $\lag : J^1E \longrightarrow \Real$. Here
$J^1E$ denotes the 1-jet prolongation of $E$ which is called the 
evolution phase space of the system. 
If the Lagrangian is regular, then the dynamics is provided by 
the Euler-Lagrange vector field
$X_{\cal L}$, which is the Reeb vector 
field of the cosymplectic structure $(\Omega_{\cal L}, 
\eta)$ on $J^1E$, where $\Omega_{\cal L}$ is the Poincar\'e-Cartan two-form
and $\eta$ is the pull-back to $J^1E$ of the volume form on $B$:
\begin{equation}\label{eq01}
i(X_{\cal L}) \Omega_{\cal L} = 0 , i(X_{\cal L}) \eta = 1.
\end{equation}
If ${\cal L}$ is not regular, then $(\Omega_{\cal L}, 
\eta)$ is no longer cosymplectic so that Eq. (\ref{eq01})
has no solution in general
(in addition, if such a solution exists it is not necessarily
a second order differential equation).
Assuming some weak regularity (constancy of the ranks of some distributions),
a constraint algorithm can be developed in such a way that,
in the better cases, we obtain a
sequence of submanifolds which ends in some final constraint
submanifold $M_f$ of $J^1E$. This is done in Section 3.

The constraint functions defining these submanifolds
were carefully determined in \cite{CLM-94}
when $E$ is trivially fibered, say
$\pi=pr_1: E = \Real \times Q \longrightarrow \Real$,
in which case $J^1E = \Real \times \Tan Q$.
Using a convenient decomposition of the Poincar\'e-Cartan 2-form 
$\Omega_{\cal L}$ based in the canonical connection in that trivial fibration,
the authors have characterized the constraint functions.
In this paper, we use a decomposition of $\Omega_{\cal L}$
based on an arbitrary auxiliar connection in $J^1E \longrightarrow B$.
The connection allows us to choose a basis of constraints,
and could be understood as a choice of reference.

We also examine the Hamiltonian setting (Section 4), and develop the
corresponding constraint algorithm. We give a characterization of
the constraints which can be connected to the Lagrangian ones by means
of the Legendre transformation ${\cal FL}$. This can be done in the case
of almost-regular Lagrangians.
The characterization obtained here is based again on the choice of
a connection in the bundle ${\cal P} \longrightarrow B$,
where ${\cal P}={\cal FL}(J^1E)$ is
a submanifold (the primary constraint submanifold)
of the restricted momentum dual bundle $J^{1*}E$. As in the Lagrangian 
setting, the connection enables us to choose a basis of constraints.
In addition, the 
Hamiltonian dynamics is obtained and related with the Lagrangian dynamics.
We also consider the second order differential equation problem, since
the Euler-Lagrange equations are of second order.

Let us say that both algorithms can be considered in abstract
as particular cases of a general algorithm developed in Section 2
for the case of a precosymplectic fiber manifold $F\to B$.

In Section 5 we consider the problem of finding the dynamics,
but imposing from the very beginning that this dynamics should simultaneously
satisfy the second-order differential equation condition.
We characterize the new constraints and relate them to the above approach.
The results of this Section are a generalization to the time-dependent case
of those given in \cite{CLR-87} and \cite{MR-92} for autonomous systems.

In Section 6, we classify the Hamiltonian constraints
into first and second class, and
define some geometrical projectors which allows us
 to introduce a Poisson bracket (the Dirac bracket) 
that has properties similar to the Dirac bracket in the autonomous case:
second class constraints become Casimir functions, and the evolution of
an observable is given by the bracket with a suitable Hamiltonian function.
Note that now we need, in addition, a suitable vector field
in order to describe the evolution
(the bracket is not enough to do this task). 

Finally, we include two examples in Section 7 to illustrate how 
the procedure for chosing constraints works.

The paper contains also three auxiliary Sections with the purpose of making
it selfcontained.

It is very important to point out that, throughout the paper,
an auxiliar connection is used for constructing
different geometrical structures.
In fact, this technique (the use of a connection) was used for the first
time in \cite{CCI-91}, in order to obtain (global) Hamiltonian functions,
and afterwards applied both in the Lagrangian 
and Hamiltonian formalisms for this
and other purposes (see \cite{EMR-sdtc,EMR-96,JMP,GMS-97,MS-98,Sd-95,Sd-98}).

We trust that the results contained in the present paper
will provide new insights into
the problem of characterizing the constraints in classical field theories,
and, consequently, in their quantization procedure.

 Manifolds are real, paracompact,
 connected and $C^\infty$. Maps are $C^\infty$. Sum over crossed repeated
 indices is understood.

 \section{The general case}
 \protect\label{gencase}

 \subsection{Statement of the problem and solution}
 \protect\label{algstate}

Let $\kappa\colon F\to B$ be a fiber manifold, where
$\dim\,F=2N+1$ and  the fibers have even dimension $2N$ ($B$ can
be $\Real$ or $S^1$). Let $\varpi\in\df^1(B)$ be a volume form,
and denote $\eta=\kappa^*\varpi$.

Let $\nabla$ be a connection in $\kappa\colon F\to B$; that is,
$\nabla$ is a $\kappa$-semibasic $1$-form on $F$ with values in
$\Tan F$ such that $\nabla^*\beta =\beta$, for every
$\kappa$-semibasic form $\beta\in\df^1(F)$. As is known,
a connection $\nabla$ always exists, and it defines a horizontal subbundle
${\rm H}(\nabla)\subset\Tan F$,
 such that $\Tan F={\rm H}(\nabla)\oplus {\rm V}(\kappa)$,
 where ${\rm V}(\kappa)$ is the $\kappa$-vertical subbundle. If
 $x\in F$, then ${\rm H}_x(\nabla)=\Img\,\nabla_x$.
If ${\cal Y}\in\vf(F)$ is the vector field spanning the horizontal
subbundle ${\rm H}(\nabla)$ such that $\inn({\cal Y})\eta=1$.
Then we can write $\nabla=\eta\otimes{\cal Y}$.

 For every $X\in\vf(F)$,
 $$
 \inn(X)\nabla=\inn(X)(\eta\otimes{\cal Y}) =
 (\inn(X)\eta){\cal Y}\equiv X^H\in\vf(F)
 $$
 is an {\sl horizontal vector field}; that is,
 a section of ${\rm H}(\nabla)\to F$.
 $X^H$ is the {\sl horizontal component} of $X$, and we
 write $X=X^H+X^V$, where $X^V=X-X^H\in\vf^{{\rm V}(\kappa)}(F)$
 (it is a $\kappa$-vertical vector field).

 Furthermore, if $\alpha\in\df^r(F)$, then
 $$
 \inn(\nabla)\alpha =
 \inn(\eta\otimes{\cal Y})\alpha =
 \eta\wedge\inn({\cal Y})\alpha\equiv\alpha^H.
 $$
 We have that
 $\alpha=\alpha^H+\alpha^V$, where $\alpha^V=\alpha-\alpha^H$
 is a $\kappa$-vertical $r$-form with respect to the connection $\nabla$,
 that is, it vanishes under the action of the horizontal vector
 fields associated with the connection $\nabla$, and in particular
 $\inn({\cal Y})\alpha^V=0$. Moreover, if $\moment{X}{1}{r}\in\vf(F)$
 are $\kappa$-vertical vector fields, then
 $\alpha^H(\moment{X}{1}{r})=0$.

 The problem we wish to solve can be posed in the following way:

 \begin{state}
 Given $\kappa\colon F\to B$, and $\eta$ as above, let $\Omega\in\df^2(F)$.
 We wish to find a submanifold $C\hookrightarrow F$
 and a section $X\colon C\to\Tan_CF$, verifying that
 $X_x\in\Tan_x C$, for every $x\in C$, and
 such that the following equations hold:
 \beq
 [\inn(X)\Omega]_x=0 \quad , \quad [\inn(X)\eta]_x=1
 \quad ,\quad \mbox{\rm  for every $x\in C$} \; .
 \label{fundeqs}
 \eeq
 \label{stat1}
 \end{state}

 A section $X\colon C\to\Tan_CF$ is a
 ``vector field on $F$ with support on $C$''
 (we will denote the set of these vector fields by
 $\vf (F,C)$).

 This is the problem given by $(F,\Omega,\eta)$. Now we are going to
 state the above system of equations in an equivalent way.

\begin{teor}
 Consider $(F,\Omega,\eta)$. Let $\nabla$ be a connection
 in $\kappa\colon F\to B$, and consider the induced splitting of $\Omega$.
 Then, the couple $(C,X)$ (where $C\hookrightarrow F$ is a submanifold,
 and $X\in\vf (F,C)$ is such that $X_x\in \Tan_xC,$ for every $x\in C$) is a
 solution for the problem stated by equations
 (\ref{fundeqs}) if, and only if, the following system holds:
 \beq
 [\inn(X)\Omega^V]_x=-[\inn({\cal Y})\Omega]_x
 \quad , \quad [\inn(X)\eta]_x=1 \quad ; \quad \mbox{for every
 $x\in C$}\; .
 \label{fundeqs1}
 \eeq
 \label{nsc1}
 \end{teor}
 \proof
 We have that
 $$
 \Omega=\Omega^H+\Omega^V=
 \eta\wedge[\inn({\cal Y})\Omega]+\Omega^V
 $$
 and then, if $X\in\vf(F)$,
 \beq
 \inn(X)\Omega=(\inn(X)\eta)(\inn({\cal Y})\Omega)-
 \eta[\inn(X)\inn({\cal Y})\Omega]+\inn(X)\Omega^V \; .
 \label{f0}
 \eeq

 \quad ($\Longrightarrow$)\quad
 If $(C,X)$ is a solution of equations
 (\ref{fundeqs}), from (\ref{f0}) we have that
 $$
 0=[\inn({\cal Y})\Omega]_x-\eta_x[\inn(X)\inn({\cal Y})\Omega ]_x+
[\inn(X)\Omega^V]_x
\quad ,\quad \mbox{\rm  for every $x\in C$}
 $$
 but $[\inn(X)\inn({\cal Y})\Omega]_x=0$, because
 $[\inn(X)\Omega]_x=0$, and the equations (\ref{fundeqs1}) hold.

 \quad ($\Longleftarrow$)\quad
 If $(C,X)$ is a solution of equations
 (\ref{fundeqs1}) then, from (\ref{f0}) we have that
 $$
 [\inn(X)\Omega]_x=
 -\eta_x[\inn(X)\inn({\cal Y})\Omega]_x=
 \eta_x[\inn(X)\inn(X)\Omega^V]_x=0
 \quad ,\quad \mbox{\rm  for every $x\in C$} \; .
 $$
\qed

  From the equivalence of equations (\ref{fundeqs}) (which are
independent of the choice of the connection $\nabla$) and
(\ref{fundeqs1}), we have:

\begin{corol}
The couple $(C,X)$, solution of equations (\ref{fundeqs1}), is
independent of the connection $\nabla$.
\end{corol}

 The case we will focus on is:

\begin{assum}
 $\Omega\in\df^2(F)$ and  $\eta\in\df^1(F)$ are closed forms,
(and $\nabla=\eta\otimes{\cal Y}$ is a fixed connection in the
fiber manifold $\kappa\colon F\to B$).
 \label{aclosed}
\end{assum}

 Then, we will denote
 $$
 \gamma\equiv\inn({\cal Y})\Omega \quad ,\quad
 \omega\equiv\Omega^V=\Omega-\eta\wedge\gamma \; .
 $$
 (Notice that
 $\inn( {\cal Y} )\gamma=0$, and $\inn( {\cal Y} )\omega=0$).
 Finally, with this notation, equations (\ref{fundeqs1}) become
  \beq
 [\inn(X)\omega]_x=-\gamma_x
 \quad , \quad
 [\inn(X)\eta]_x=1 \quad ; \quad \mbox{for every
 $x\in C$} \; .
 \label{fundeqs2}
 \eeq

Now, introducing the map (see Appendix \ref{A}) $$
\begin{array}{ccccc}
\flat_{(\eta,\omega)}&\colon&\Tan F&\to&\Tan^*F
\\
& & (x,v)& \mapsto &\inn(v)\omega_x+(\inn(v)\eta_x)\eta_x
\end{array}
$$
we can state the following fundamental result:

\begin{teor}
The necessary and sufficient condition for the existence of
a submanifold $C\hookrightarrow F$
and $X\in\vf (F,C)$, with
$X_x\in\Tan_x C$, for every $x\in C$,
such that the equations (\ref{fundeqs}) (or equivalently
 equations (\ref{fundeqs2})) hold, is that
 \beq
 \eta_x-\gamma_x\in\flat_{(\eta_x,\omega_x)}(\Tan_xC)
\quad ; \quad
\mbox{for every $x\in C$}
\label{bemolbis}
 \eeq
or, what is equivalent
 $$
 \langle \eta_x-\gamma_x,\Tan^\perp_xC\rangle = 0
\quad ; \quad
\mbox{for every $x\in C$}
 $$
where $\flat_{(\eta_x,\omega_x)}=\flat_{(\eta,\omega)|\Tan_xF}$ and
$\Tan^\perp_xC\equiv(\Tan_xC)^\perp:=[\flat_{(\eta_x,\omega_x)}(\Tan_xC)]^0$.
\label{nsc2}
\end{teor}
\proof
 ($\Longrightarrow$)\quad
 It is an immediate
consequence of Theorem \ref{nsc1} and the definition of
$\flat_{(\eta_x,\omega_x)}$.

\quad ($\Longleftarrow$)\quad
Suppose that a submanifold $C\hookrightarrow F$
and a vector field $X\in\vf(F,C)$ tangent to $C$ exist, such that
(\ref{bemolbis}) holds. This means that, for every $x\in C$,
 \beq
 \eta_x-\gamma_x=\flat_{(\eta_x,\omega_x)}(X_x)=
[\inn(X)\omega]_x+[\inn(X)\eta]_x\eta_x\; .
 \label{interm}
 \eeq
Contracting both members with ${\cal Y}$ we obtain
 $$
 [\inn( {\cal Y} )\eta]_x-[\inn( {\cal Y} )\gamma]_x=
[\inn( {\cal Y} )\inn(X)\omega]_x+ [\inn( {\cal Y})\eta\inn(X)\eta]_x
 $$
and, as $\inn( {\cal Y} )\eta=1$, $\inn( {\cal Y} )\gamma=0$, and
$\inn( {\cal Y} )\omega=0$, we conclude that
 $[\inn(X)\eta]_x=1$. Therefore, from (\ref{interm}) we have that
$[\inn(X)\omega]_x=-\gamma_x$, and the result follows from Theorem
\ref{nsc1}.
 \qed

 \begin{remark}
{\rm  Every $X\in\vf (F,C)$ can be extended to define
 a vector field $X\in\vf (F)$. Therefore, although we are really interested
 in finding vector fields on $F$ at support on $C$,
 from now on we will suppose that the vector fields
 are defined everywhere in $F$.}
 \end{remark}

\subsection{The constraint algorithm}
 \protect\label{ca}

 Now we will apply the last result in order to solve the problem
 stated above; that is, to find a submanifold $C\hookrightarrow F$,
 transverse to the projection $\kappa\colon F\to B$, and
 a vector field $X\in\vf (F)$, such that
\ben
\item
$[\inn(X)\Omega]_x=0$,  and
$[\inn(X)\eta]_x=1$,
for every $x\in C$.
\item
$X\vert_{C}$ is tangent to $C$. \een
 The procedure is algorithmic, and produces a sequence of
 subsets $\{ C_i\}$ of $F$. Then, we will assume that:

\begin{assum}
 Every subset $C_i$ of this sequence is a regular submanifold of $F$,
 and its natural injection is an embedding.
 \label{asub}
\end{assum}

Thus, we consider the submanifold $C_1\hookrightarrow F$ where a
solution exists; that is,
 $$
 C_1 = \{ x\in F \mid \exists X\in \Tan_xF\ : \
 \inn(X_x)\Omega_x=0, \hbox{ } \inn(X_x)\eta_x=1 \} \; .
 $$
Then there is a vector field $X\in\vf(F)$ such that
$[\inn(X)\Omega]|_{C_1}=0$, $[\inn(X)\eta]|_{C_1}=1$. But in
general $X$ is not tangent to $C_1$. We can consider therefore the
submanifold
 $$
 C_2 = \{ x\in C_1 \mid \exists X\in \Tan_xC_1 :
 \inn(X_x)\Omega_x=0, \hbox{ } \inn(X_x)\eta_x=1\}
 \; .
 $$
 Then there will be a vector field $X\in\vf(F)$ tangent to $C_1$ such that
 $[\inn(X)\Omega]|_{C_2}=0$, $[\inn(X)\eta]|_{C_2}=1$.
 Again, $X$ may not be tangent to $C_2$. Following this process,
 we obtain a sequence of constrained submanifolds
 \beq
 \cdots \stackrel{j^i_{i+1}}{\hookrightarrow} C_i
 \stackrel{j^{i-1}_i}{\hookrightarrow} \cdots
 \stackrel{j^1_2}{\hookrightarrow} C_1
 \stackrel{j_1}{\hookrightarrow} C_0\equiv F\; .
 \label{seqsubman0}
 \eeq
 For every $i\geq 1$, $C_i$ is called the {\sl $i$th constraint submanifold}.

 This procedure will be called the {\sl constraint algorithm}, and
 we have three possibilities:
 \bit
 \item
 There exists an integer $k>0$ such that $C_k = \emptyset$. This
 means that the equations are not consistent; that is, they have no solution.
 \item
 There exists an integer $k>0$ such that $C_k \neq \emptyset$,
 but $\dim C_k=0$. In this case, there is no dynamics. $C_k$
 consists of isolated points, and the solution of the equations is $X = 0$.
 \item
 There exists an integer $k>0$ such that $C_{k+1}=C_k\equiv C_f$, and
 $\dim C_f>0$. In such a case, there exists a vector field
 $X\in\vf(F)$, tangent to $C_f$, such that
 $$
 [\inn(X)\Omega]|_{C_f}=0 \; , \; [\inn(X)\eta]|_{C_f}=1
 $$
 In this case, the manifold $C_f$ is called the
 {\sl final constraint submanifold}.
 This is the situation which is of concern to us.
 \eit

 Next we wish to give an intrinsic characterization of the constraints
 which define the constraint submanifolds $C_i$. In order to do this,
 following the same pattern as in Theorem \ref{nsc2}, we conclude that:

\begin{prop}
 Every submanifold $C_i$ ($i\geq 1$) in the sequence (\ref{seqsubman0})
 can be defined as
 $$
 C_i=\{ x\in C_{i-1}\ \mid \
 \langle \eta_x-\gamma_x,\Tan_x^\perp C_{i-1}\rangle=0 \} \; .
 $$
 Therefore, if the distribution $\Tan^\perp C_{i-1}$
 has constant rank, and
 $\{Z^{(i-1)}_{1},\dots ,Z^{(i-1)}_{r}\}$ is a set of vector fields
 spanning locally this distribution,
 then $C_i$, as a submanifold of $C_{i-1},$ is defined locally as the zero set of the
 functions $\chi_j^{(i)}\in\Cinfty (F)$ given by
 $$
 \chi_j^{(i)}=\inn(Z^{(i-1)}_j)(\eta-\gamma)
 \quad , \quad (j=1,\ldots ,r) \; .
 $$
These functions are called {\rm $i$th-generation constraints}.
\label{cons}
\end{prop}

\subsection{Stability of vector fields solution.
  New characterization of constraints}
 \protect\label{svfs}

 The solutions of the equations (\ref{fundeqs}) are now analyzed.
 Thus, consider the problem posed in Statement \ref{stat1}.
 Once we have found the submanifold $C_1$ where a solution exists,
 we impose the tangency condition for the vector field solutions.
 This enables us to obtain a new characterization of the
 constraints defining the submanifolds of the sequence (\ref{seqsubman0}).
 To do this we need the following additional hypothesis
 (and the results stated in Appendices  \ref{A} and \ref{C}):

\begin{assum}
The distributions $\Tan^\perp C_0=\Tan^\perp
F=\ker\omega\cap\ker\eta$, and $\Tan_{C_1}^\perp C_0\cap\Tan C_1$
have constant rank. \label{arank}
\end{assum}

 From the first part of Lemma \ref{10'''} we deduce that these
 conditions are independent of the connection $\nabla$.

 If $X\in\vf(F)$ verifies that
\beq
 [\inn(X)\omega]\vert_{C_1}=-\gamma\vert_{C_1}
\quad , \quad
 [\inn(X)\eta]\vert_{C_1}=1
\label{C1}
\eeq
 then, from now on, we can assume that
$\inn(X)\eta=1$ everywhere in $F$. In fact, since if
$\inn(X)\eta\not=1$, we take $X'=X+[1-\inn(X)\eta]{\cal Y}$, and
it is clear that \dst X'\vert_{C_1}= X\vert_{C_1}\) , and
$\inn(X')\eta=\inn(X)\eta+1-\inn(X)\eta=1$. Therefore:

\begin{prop}
Let $X$ be a vector field on $F$  such that equations (\ref{C1})
hold, and verifying the above condition. \ben
\item
If $Z,Y\in\ker\,\omega\cap\ker\,\eta$, such that $Z$ is tangent to
$C_1$, then $$ \inn([X,Z]-[\Lie({\cal Y})\inn(Z)\gamma]{\cal
Y})\omega\vert_{C_1}- \eta[\inn([X,Z]-[\Lie({\cal
Y})\inn(Z)\gamma]{\cal Y})\eta]\vert_{C_1}=
\d[\inn(Z)\gamma]\vert_{C_1} $$ and, as a consequence, $$ \{
[X,Z]-[\Lie({\cal Y})\inn(Z)\gamma)]{\cal Y}+Y\}_x\in\Tan_x^\perp
C_1\;, \quad  \quad \mbox{\rm for every $x\in C_1$} \; . $$
\item
If $Z'\in\vf(C_0)$ such that $Z'\vert_x\in\Tan_x^\perp C_1$, for
every $x\in C_1$, then there exist locally
$Z,Y\in\ker\,\omega\cap\ker\,\eta$, such that $Z$ is tangent to
$C_1$, and $Z'\vert_{C_1}=\{ [X,Z]-[\Lie({\cal
Y})\inn(Z)\gamma]{\cal Y}+Y\}\vert_{C_1}$. \een \label{d}
\end{prop}
\proof
\ben
\item
It is clear that $Y_x\in\Tan_x^\perp C_1$, since
$Y\in\ker\,\omega\cap\ker\,\eta$, hence
$\flat_{(\eta_x,\omega_x)}(X_1)(Y_x)=0,$ for $X_1\in \Tan_{x_1}C_1.$

For $[X,Z]-[\Lie({\cal Y})\inn(Z)\gamma)]{\cal Y},$ we have
\begin{equation} \label{formula1}
 \begin{array}{lcl}
 \inn([X,Z]-[\Lie({\cal Y})\inn(Z)\gamma)]{\cal Y})\eta &\kern-5pt=&
\kern-5pt \Lie(X)\inn(Z)\eta-\inn(Z)\Lie(X)\eta- [\Lie({\cal
 Y})\inn(Z)\gamma]\inn({\cal Y})\eta\\&\kern-5pt=&\kern-5pt -\Lie({\cal Y})\inn(Z)\gamma
 \end{array}
 \end{equation}
 and
 \beq
 \inn([X,Z]-[\Lie({\cal Y})\inn(Z)\gamma]{\cal Y})\omega =
 \inn([X,Z])\omega-[\Lie({\cal Y})\inn(Z)\gamma])\inn({\cal Y})\omega=
 \inn([X,Z])\omega\; .
 \label{formula2}
 \eeq
However, from the second equality, for every $X'\in\vf(C_0)$, and
bearing in mind that $Z\in\ker\omega\cap \ker\eta$ and that
$\omega=\Omega-\eta\wedge \gamma,$  we have \beann
\inn(X')\inn([X,Z])\omega&=&
\inn(X')[\Lie(X)\inn(Z)-\inn(Z)\Lie(X)]\omega=
\inn(X')[-\inn(Z)\inn(X)\d\omega-\inn(Z)\d\inn(X)\omega]
\\ &=&
\inn(X')[-\inn(Z)\inn(X)\d\Omega-\inn(Z)\inn(X)(\eta\wedge\d\gamma)
-\inn(Z)\d\inn(X)\omega]
\\ &=&
-\inn(X')\inn(Z)\{[\inn(X)\eta]\d\gamma-\eta\wedge\inn(X)\d\gamma
+\d\inn(X)\omega\}
\\ &=&
-[\inn(X)\eta]\inn(X')\inn(Z)\d\gamma-
[\inn(X')\eta][\inn(Z)\inn(X)\d\gamma]-\inn(X')\inn(Z)\d\inn(X)\omega
\\ &=&
-[\inn(X)\eta]\inn(X')\inn(Z)\d\gamma-
[\inn(X')\eta][\inn(Z)\inn(X)\d\gamma]-
 \\ & &
\inn(X')\Lie(Z)\inn(X)\omega+
\inn(X')\d\inn(Z)\inn(X)\omega
\\ &=&
-[\inn(X)\eta]\inn(X')\inn(Z)\d\gamma-
[\inn(X')\eta][\inn(Z)\inn(X)\d\gamma]-
 \\ & &
\Lie(Z)\inn(X')\inn(X)\omega+ \inn([Z,X'])\inn(X)\omega \; .
\eeann On $C_1$, using (\ref{C1}), and taking into account that
$-\Lie(Z)[\inn(X')\inn(X)\omega]\vert_{C_1}=
[\Lie(Z)\inn(X')\gamma]\vert_{C_1}$ (since $Z$ is tangent to
$C_1$), we obtain \bea \inn(X')\inn([X,Z])\omega\vert_{C_1}&=&
-\inn(X')\inn(Z)\d\gamma\vert_{C_1}-
[\inn(X')\eta][\inn(Z)\inn(X)\d\gamma]\vert_{C_1}-
 \nonumber
\\ & &
\Lie(Z)\inn(X')\inn(X)\omega\vert_{C_1}+
\inn([Z,X'])\inn(X)\omega\vert_{C_1}
 \nonumber
\\ &=&
-\inn(X')\Lie(Z)\gamma\vert_{C_1}+\inn(X')\d\inn(Z)\gamma\vert_{C_1}-
[\inn(X')\eta][\inn(Z)\inn(X)\d\gamma]\vert_{C_1}+
 \nonumber
\\ & &
\Lie(Z)\inn(X')\gamma\vert_{C_1}-\inn([Z,X'])\gamma\vert_{C_1}
 \nonumber
\\ &=&
\inn([Z,X'])\gamma\vert_{C_1}+\Lie(X')\inn(Z)\gamma\vert_{C_1}-
 \nonumber
\\ & &
[\inn(X')\eta][\inn(Z)\inn(X)\d\gamma]\vert_{C_1}
-\inn([Z,X'])\gamma\vert_{C_1}
 \nonumber
\\ &=&
\Lie(X')\inn(Z)\gamma\vert_{C_1}-
[\inn(X')\eta][\inn(Z)\inn(X)\d\gamma]\vert_{C_1} \; .
\label{formula3} \eea Furthermore, using that
$\gamma=i({\cal Y}) \Omega$  and that $\Omega$ is closed, \bea
\inn(Z)\inn(X)\d\gamma\vert_{C_1}&=& \inn(Z)\inn(X)\Lie({\cal
Y})\Omega\vert_{C_1}
 \nonumber
\\ &=&
\Lie({\cal Y})\inn(Z)\inn(X)\Omega\vert_{C_1}+
\inn([{\cal Y},X])\inn(Z)\Omega\vert_{C_1}-
\inn([{\cal Y},Z])\inn(X)\Omega\vert_{C_1}
 \nonumber
\\ &=&
\Lie({\cal Y})\inn(Z)\gamma\vert_{C_1} \; ,
 \label{formula4}
\eea since $\inn(X)\Omega\vert_{C_1}=0$ (because $X$ is a
solution), $\inn(Z)\Omega\vert_{C_1}=0$ (by Lemma \ref{b},
Appendix \ref{C}), and
$\inn(Z)\inn(X)\Omega=\inn(Z)\inn(X)\omega+\inn(Z)\gamma=\inn(Z)\gamma$.
Therefore, from (\ref{formula2}), (\ref{formula3}) and
(\ref{formula4}), $$ \inn(X')\{\inn([X,Z]-[\Lie({\cal
Y})\inn(Z)\gamma)]{\cal Y})\omega\}\vert_{C_1}
=
\{\Lie(X')\inn(Z)\gamma-[\inn(X')\eta]\Lie({\cal
Y})\inn(Z)\gamma\}\vert_{C_1} \; . $$ Moreover, from
(\ref{formula1}), $$ \inn(X')\{\eta[\inn([X,Z]-[\Lie({\cal
Y})\inn(Z)\gamma] {\cal Y})\eta]\}\vert_{C_1}=
-[\inn(X')\eta]\Lie({\cal Y})\inn(Z)\gamma)\vert_{C_1} \; . $$
Hence we conclude

$$\inn(X')\left[\inn([X,Z]-[\Lie({\cal Y})\inn(Z)\gamma]{\cal
Y})\omega - \eta[\inn([X,Z]-[\Lie({\cal Y})\inn(Z)\gamma]{\cal
Y})\eta]\right]\vert_{C_1}= $$
 $$ =[\Lie(X')\inn(Z)\gamma]\vert_{C_1} =
\left[\inn(X')\d[\inn(Z)\gamma]\right]\vert_{C_1}\; , $$
 for every $X'\in\vf(C_0)$, so
$$\inn([X,Z]-[\Lie({\cal Y})\inn(Z)\gamma]{\cal
Y})\omega\vert_{C_1}- \eta[\inn([X,Z]-[\Lie({\cal
Y})\inn(Z)\gamma]{\cal Y})\eta]\vert_{C_1}=
\d[\inn(Z)\gamma]\vert_{C_1}\; . $$

Finally, notice that $\Tan_x^\perp C_1=\{ v_x\in\Tan_x
C_0\mid\langle
 \flat_{(\eta_x,\omega_x)}(u_x),v_x\rangle =0\;
 ,~\forall u_x\in\Tan_x C_1\}$. That is,
 $\langle\inn(v_x)\omega_x-[\inn(v_x)\eta_x]\eta_x,
u_x\rangle=0$, for every $u_x\in\Tan_x C_1$. Taking
 $v_x=\{ [X,Z]-[\Lie({\cal Y})\inn(Z)\gamma]{\cal Y}+Y\}_x$,
 we obtain that $\langle\inn(v_x)\omega_x-[\inn(v_x)\eta_x]\eta_x,
u_x\rangle= [\Lie(W)\inn(Z)\gamma]\vert_x =0$, if $U$ is a vector
 field tangent to $C_1$ such that $U(x)=u_x$ (notice that
 $\inn(Z)(\eta-\gamma)=-\inn(Z)\gamma$ is a constraint function
 defining $C_1$ in $C_0$). Therefore, $v_x\in\Tan_x^\perp C_1$.

\item
Consider a local basis $\{ Z^{(0)}_1,\ldots ,Z^{(0)}_{r_1}\}$
 of the distribution $\ker\,\omega\cap\ker\,\eta$, in an open set
$U\subset F$. Then the $1$-forms
$\d\inn(Z^{(0)}_i)\gamma\vert_{C_1},$ $i=1,\dots ,r_1$  generate
the annihilator of $\Tan_{U\cap C_1}C_1$. Therefore $$
\{\inn(Z')\omega-[\inn(Z')\eta)]\eta\}\vert_{C_1}= \tilde
f^i[\d\inn(Z^{(0)}_i)\gamma]\vert_{C_1} $$ where $\tilde
f^i\in\Cinfty (U\cap C_1)$ are arbitrary functions, which can be
extended to functions $ f^i\in\Cinfty (U)$. Thus, we consider
$Z= f^iZ^{(0)}_i$, then $\inn(Z)\gamma= f^i\inn(Z^{(0)}_i)\gamma$,
and $$ [\d\inn(Z)\gamma]\vert_{C_1}=
 f^i\vert_{C_1}[\d\inn(Z^{(0)}_i)\gamma]\vert_{C_1}+
 (\d f^i)\vert_{C_1}\inn(Z^{(0)}_i)\gamma\vert_{C_1}=
 f^i\vert_{C_1}[\d\inn(Z^{(0)}_i)\gamma]\vert_{C_1}.
$$ Hence, comparing these last equations, we conclude that
 \beq
\{\inn(Z')\omega-[\inn(Z')\eta)]\eta\}\vert_{C_1}=
[\d\inn(Z)\gamma]\vert_{C_1} \label{fm1} \eeq
 and in this way,
for every $Z^{(0)}_i$,
 \beq\label{fm2}
 0 = \{\inn(Z^{(0)}_i)(\inn(Z')\omega-[\inn(Z')\eta)]\eta)\}\vert_{C_1}=
 \Lie(Z^{(0)}_i)\inn(Z)\gamma\vert_{C_1}\; .
 \eeq
 Now,
 \[
 \inn(Z)\Lie(Z_i^{(0)})\gamma\vert_{C_1} =
 \{\inn(Z)\inn(Z^{(0)}_i)\d\gamma +
 \inn(Z)\d\inn(Z^{0)}_i)\gamma\}\vert_{C_1} =
 \{\inn(Z)\inn(Z^{(0)}_i)\d\gamma +
 \Lie(Z)\inn(Z^{0)}_i)\gamma\}\vert_{C_1}\; .
 \]
 Also,
 \[
 \inn([Z^{(0)}_i,Z])\gamma\vert_{C_1} = \{\Lie
 (Z^{(0)}_i)\inn(Z)\gamma -
 \inn(Z)\Lie(Z^{(0)}_i)\gamma\}\vert_{C_1}\; .
 \]
 Therefore,
 \[
 \inn(Z)\inn(Z^{(0)}_i)\d\gamma\vert_{C_1} =
 \{\inn(Z)\Lie(Z^{(0)}_i)\gamma - \Lie(Z)\inn(Z^{(0)}_i)\gamma
 \}\vert_{C_1} =
 \]
 \[
 = \{\Lie(Z^{(0)}_i)\inn(Z)\gamma -
 \Lie(Z)\inn(Z^{(0)}_i)\gamma-
 \inn([Z^{(0)}_i,Z])\gamma\}\vert_{C_1}\; .
 \]
 Taking into account the first item of Lemma \ref{c} (Appendix \ref{C}),
 we have
 \[
 0=\inn(Z)\inn(Z^{(0)}_i)\d\gamma\vert_{C_1}=
\Lie(Z^{(0)}_i)\inn(Z)\gamma\vert_{C_1}-
\Lie(Z)\inn(Z^{(0)}_i)\gamma\vert_{C_1}-\inn([Z^{(0)}_i,Z])\gamma\vert_{C_1}\;
.
 \]
 Hence, $\Lie(Z)\inn(Z^{(0)}_i)\gamma\vert_{C_1} =
 \Lie(Z^{(0)}_i)\inn(Z)\gamma\vert_{C_1} -
 \inn([Z^{(0)}_i,Z])\gamma\vert_{C_1}$. But
 $\Lie(Z^{(0)}_i)\inn(Z)\gamma\vert_{C_1}=0 $ (see (\ref{fm2})) and, using the
 second item of the Lemma \ref{c} and the characterization of $C_1$,
 $\inn([Z^{(0)}_i,Z])\gamma\vert_{C_1}=0$. Therefore, $Z$ is tangent to
$C_1$. Finally, from (\ref{stat1}), (\ref{fm1}) and the first part
of this Proposition, we conclude that
 \[
 \{\inn(Z'-[X,Z]+[\Lie({\cal Y})\inn(Z)\gamma]{\cal Y})\omega-
[\inn(Z'-[X,Z]+[\Lie({\cal Y})\inn(Z)\gamma]{\cal
Y})\eta]\eta\}\vert_{C_1} =
 \]
 \[
 = \{\inn(Z')\omega - [\inn(Z')\eta)]\eta\}\vert_{C_1} -
 \{\inn([X,Z]-[\Lie({\cal Y})\inn(Z)\gamma]{\cal Y})\omega-
[\inn([X,Z]-[\Lie({\cal Y})\inn(Z)\gamma]{\cal
Y})\eta)]\eta\}\vert_{C_1} = \]
 \[
 = \d\inn(Z)\gamma\vert_{C_1} - \d\inn(Z)\gamma\vert_{C_1} = 0
 \]
 and this means that $$
Z'-[X,Z]+[\Lie({\cal Y})\inn(Z)\gamma]{\cal Y}\in\Tan_x^\perp C_0=
\ker\,\omega_x\cap\ker\,\eta_x \quad , \quad \forall x\in C_1\;
,$$ i.e., $Z'\vert_{C_1} = \{ [X,Z]-[\Lie({\cal
Y})\inn(Z)\gamma]{\cal Y} + Y\}\vert_{C_1}$, with $Y\in \ker
\,\omega\cap \ker\,\eta.$
 \qed \een

Using this Proposition we can construct a local system of generators
for $\Tan^\perp C_1$. In fact:

\begin{corol}
\item
Let $X\in\vf(F)$ be such that equations (\ref{C1}) hold, $r_0$ the
rank of the distribution $\Tan_{C_1}^\perp C_0\cap\Tan C_1$, and
\dst\{ Z^{(0)}_1,\ldots ,Z^{(0)}_{r_0},Z^{(0)}_{r_0+1},\ldots
,Z^{(0)}_{r_1}\}\)
 a local basis for $\ker\,\omega\cap\ker\,\eta$
(in a neighbourhood of a point $x\in C_1$),
where \dst\{Z^{(0)}_i\}_{i=1,...,r_0}\) are tangent to $C_1$.
Then $\Tan^\perp C_1 $ is locally generated by the vector fields
 $$
 \{ Z^{(0)}_1,\ldots ,Z^{(0)}_{r_0},Z^{(0)}_{r_0+1},\ldots ,Z^{(0)}_{r_1},
[X,Z^{(0)}_i]-[\Lie({\cal Y})\inn(Z^{(0)}_i)\gamma]{\cal
Y}\}_{i=1, \ldots ,r_0}\; .
 $$
\end{corol}

\begin{remark}
{\rm  From now on, we will denote by $\vf^\perp(C_i)$
 the set of vector fields of $\vf (F,C_i)$
 which span locally the distribution $\Tan^\perp C_i$.
 This set can be characterized as
 $$
 \vf^\perp(C_i)=\{ Z\in\vf (F)\ \mid\
 j_i^*[\inn(Z)\omega-(\inn(Z)\eta)\eta]=0\}
 $$
  where $j_i\colon C_i\hookrightarrow F$ denotes the corresponding
  embedding.}
\end{remark}

Since $C_2=\{ x\in C_1\ \mid \ \langle\eta_x-\gamma_x,\Tan_x^\perp
C_1\rangle=0\}$ and
\dst\inn(Z^{(0)}_j)(\eta-\gamma)\vert_{C_1}=0\) , if we introduce
the functions $$ \chi^{(2)}_i:= \inn([X,Z^{(0)}_i]-[\Lie({\cal
Y})\inn(Z^{(0)}_i)\gamma]{\cal Y})(\eta-\gamma) $$ we conclude
that a basis of constraints for $C_2$ is made by those functions
which do not vanish on $C_1$. Bearing in mind that
$\inn([X,Z^{(0)}_k])\eta\vert_{C_1}=0$,
 $\inn({\cal Y})\gamma\vert_{C_1}=0$ and $\inn({\cal Y})\eta=1,$
they can be characterized as \beq \chi^{(2)}_i=
-\inn([X,Z^{(0)}_i])\gamma-({\cal Y}[\inn(Z^{(0)}_i)\gamma])\; .
\label{chi2} \eeq

 Next we will analyze the solutions of the dynamical equations.
 First we describe the set of solutions on $C_1$.

 \begin{prop}
 The general solution of the equations (\ref{fundeqs}) on the
 submanifold $C_1\hookrightarrow F$ where they are compatible is
 $X^{(0)}+Y^{(0)}$, where $X^{(0)}\in\vf(F)$ is a particular solution
 and $Y^{(0)}\in\vf(F)$ is an arbitrary element of
 $\ker\,\Omega\cap\ker\,\eta$.
 \end{prop}
 \proof
 It is a consequence of the linearity of $\inn(X)\Omega$ and
 $\inn(X)\eta$ on the vector field $X$.
 \qed

Taking into account the above results, we can write these
solutions in the form $$
X=X^{(0)}+Y^{(0)}=X^{(0)}+f^iZ^{(0)}_i+f^kZ^{(0)}_k \quad
(i=1,\ldots ,r_0\ ;\ k=r_0+1,\ldots ,r_1)\; . $$
 Now we must identify the points of $C_1$ where there are
 solutions tangent to $C_1$.
 Consider the set of independent constraints $\{\chi_i^{(1)}\}$
 defining locally $C_1$ as
 a closed submanifold of $C_0=F$, which by Proposition \ref{cons} are
 described  in the following way
 $$
 \{\inn(Z^{(0)}_i)(\eta-\gamma)\}_{i=1,...,r_0}
 \quad , \quad
 \{\inn(Z^{(0)}_k)(\eta-\gamma)\}_{k=r_0+1,...,r_1} \; .
 $$
 Then the tangency condition imposes that
 \bea
 0&=& \Lie(X^{(0)}+f^iZ^{(0)}_i+f^kZ^{(0)}_k)\inn(Z^{(0)}_{i'})(\eta-\gamma)
 \vert_{C_1} \nonumber
 \\ &=&
 -\Lie(X^{(0)})\inn(Z^{(0)}_{i'})\gamma\vert_{C_1}-
 f^i\Lie(Z^{(0)}_i)\inn(Z^{(0)}_{i'})\gamma\vert_{C_1}-
 f^k\Lie(Z^{(0)}_k)\inn(Z^{(0)}_{i'})\gamma\vert_{C_1}
 \label{first}
\\
0&=& \Lie(X^{(0)}+f^iZ^{(0)}_i+f^kZ^{(0)}_k)\inn(Z^{(0)}_{k'})(\eta-\gamma)
 \vert_{C_1} \nonumber
 \\ &=&
 -\Lie(X^{(0)})\inn(Z^{(0)}_{k'})\gamma\vert_{C_1}-
 f^i\Lie(Z^{(0)}_i)\inn(Z^{(0)}_{k'})\gamma\vert_{C_1}-
 f^k\Lie(Z^{(0)}_k)\inn(Z^{(0)}_{k'})\gamma\vert_{C_1}
 \label{second}
 \eea
 where we have taken into account that, as a consequence of Lemma \ref{a},
 $\Lie(Z^{(0)})\eta\vert_{C_1}=0$, for every $Z^{(0)}\in\vf^\perp(C_0)$.
 We will need the following result:

\begin{lem}
The matrix
 $(A_{kk'})\equiv\left(\Lie(Z^{(0)}_k)\inn(Z^{(0)}_{k'})\gamma\right)$
 ($k,k'=r_0+1,\ldots ,r_1$)
 is regular, for every $x\in C_1$ (in the corresponding neighbourhood).
\end{lem}
\proof Let us suppose that the rows of this matrix are dependent
for a point $x\in C_1.$ Then,  $$ \mu^k
Z_k^{(0)}(x_1)(\inn(Z^{(0)}_{k'})\gamma)=0 \quad, \quad \mbox{ for
all $k'$, with }  \mu^k\in \Real\; . $$
 Then, by Lemma \ref{c}, if $Z^{(0)}_i$, $i=1,\ldots,r_0$,
 are tangent to $C_1$, we have that
 \[
 \Lie(Z^{(0)}_k)\inn(Z^{(0)}_i)\gamma\vert_{C_1}=
 \inn([Z^{(0)}_k,Z^{(0)}_i])\gamma\vert_{C_1} +
 \inn(Z^{(0)}_i)\Lie(Z^{(0)}_k)\gamma\vert_{C_1}
 \]
 \beq
 =
 \inn(Z^{(0)}_i)\inn(Z^{(0)}_k)\d\gamma\vert_{C_1} +
 \inn(Z^{(0)}_i)\d(\inn(Z^{(0)}_k)\gamma)\vert_{C_1} =
 \Lie(Z^{(0)}_i)\inn(Z^{(0)}_k)\gamma\vert_{C_1}=0, \label{raro}
 \eeq
 and hence $$
\mu^k Z^{(0)}_k(x)(\inn(Z^{(0)}_i)(\gamma))=0, \quad \mbox{\rm for
every $i=1,\ldots ,r_1$}. $$
 Therefore $\mu^k(Z^{(0)}_k)_x\in\Tan_xC_1$ and hence
$\mu^k(Z^{(0)}_k)_x\in\Tan_x^\perp C_0\cap\Tan_xC_1$, for every
$x\in C_1$, but as $\{(Z^{(0)}_1)_x,\ldots ,(Z^{(0)}_{r_0})_x\}$
is a basis of $\Tan_x^\perp C_0\cap\Tan_xC_1$, we must conclude
that $\mu^k=0$, for every $k=r_0+1,\ldots ,r_1$. \qed

Since $\Lie(Z_k^{(0)})
\inn(Z_i^{(0)})\gamma\vert_{C_1}=\Lie(Z_i^{(0)})\inn
(Z_k^{(0)})\gamma\vert_{C_1}=0,$ system (\ref{second}) becomes
 $$
0=\Lie(X^{(0)})\inn(Z^{(0)}_{k'})\gamma\vert_{C_1}+
 f^k\Lie(Z^{(0)}_k)\inn(Z^{(0)}_{k'})\gamma\vert_{C_1}
$$
 which is a linear system on the coefficients $f^k.$ As a
consequence of the last Lemma, this system has a unique solution
that gives us the coefficients $f^k\vert_{C_1}$. Then the general
solution on $C_1$ can be written as
 $$
X=X^{(0)}+f^kZ^{(0)}_k+f^iZ^{(0)}_i\equiv X^{(1)}+Y^{(1)} \quad
(i=1,\ldots ,r_0\ ;\ k=r_0+1,\ldots ,r_1)
 $$
 where $X^{(1)}\equiv
X^{(0)}+f^kZ^{(0)}_k$ and $Y^{(1)}$ denote the determined and the
undetermined parts of the solution, respectively.

Furthermore, since
 \[
 \Lie(Z^{(0)}_i)\inn(Z^{(0)}_{i'})\gamma\vert_{C_1} =
 \inn([Z^{(0)}_i,Z^{(0)}_{i'}])\gamma\vert_{C_1} +
 \inn(Z^{(0)}_{i'})\Lie(Z^{(0)}_i)\gamma\vert_{C_1}
 \]
 \[
 =-\inn({\cal Y})\inn([Z^{(0)}_i,Z^{(0)}_{i'}])\Omega\vert_{C_1} +
 \inn(Z^{(0)}_{i'})\inn(Z^{(0)}_i)\d\gamma\vert_{C_1} -
 \inn(Z^{(0)}_{i'})\d[\inn({\cal Y})\inn(Z^{(0)}_i)\Omega]\vert_{C_1}
 = 0\]
 (where we have used that $\gamma=\inn({\cal Y})\Omega$
 and Lemmas \ref{b} and \ref{c}),
 system (\ref{first}) reduces to $$
0=\Lie(X^{(0)})\inn(Z^{(0)}_{i'})\gamma\vert_{C_1} $$ and, if
these equalities do not hold, they give the constraint which
defines locally the submanifold $C_2$ as a submanifold of $C_1$. In
fact, we have that
 \beann
\Lie(X^{(0)})[\inn(Z^{(0)}_{i'})\gamma]\vert_{C_1}&=&
\inn(X^{(0)})\d[\inn(Z^{(0)}_{i'})\gamma]\vert_{C_1}
\\ &=&
\inn(X^{(0)})\{\inn([X,Z^{(0)}_{i'}]- [\Lie({\cal
Y})\inn(Z^{(0)}_{i'})\gamma]{\cal Y})\omega \\
 & - &
\eta[\inn([X,Z^{(0)}_{i'}]- [\Lie({\cal
Y})\inn(Z^{(0)}_{i'})\gamma]{\cal Y})\eta]\}\vert_{C_1}
\\ &=&
\inn(X^{(0)})\inn([X,Z^{(0)}_{i'}])\omega+
[\Lie({\cal Y})\inn(Z^{(0)}_{i'})\gamma]\inn(X^{(0)})\eta\vert_{C_1}
\\ &=&
-\inn([X,Z^{(0)}_{i'}])\inn(X^{(0)})\omega\vert_{C_1}+
\Lie({\cal Y})\inn(Z^{(0)}_{i'})\gamma\vert_{C_1}
\\ &=&
\inn([X,Z^{(0)}_{i'}])\gamma\vert_{C_1}+ \Lie({\cal
Y})\inn(Z^{(0)}_{i'})\gamma\vert_{C_1} \eeann
 where we employ the first item of Proposition \ref{d} and
$\inn(X^{(0)})\eta\vert_{C}=1$, $\inn(X)\eta=1$ (since
$X$ is a particular solution). Observe that this result is just
the characterization of the $2nd$-generation constraints given in
(\ref{chi2}).

 This study has been done for the submanifold $C_1$,
 but it can be extended to every level of the constraint algorithm.
 As in the case of the submanifold $C_1$, we will assume that for every
 submanifold $C_i$ $(i\geq 0)$ of the sequence (\ref{seqsubman0}),
 the distributions
 $\Tan^\perp C_i$ and $\Tan_{C_{i+1}}^\perp C_i\cap \Tan C_{i+1}$ have
 constant rank (note that from the second part of Lemma
 \ref{10'''}, we deduce that the first of these conditions is
 independent of the connection $\nabla$). Under the above
 conditions, we have

 \begin{teor}
 Consider the sequence $\{ C_i\}$
 of submanifolds (\ref{seqsubman0}).
 For every $i\geq 1$, let $r_{(i-1)},r_i$ be the
 ranks of the distributions $\Tan^\perp C_{i-1}$ and
 $\Tan^\perp_{C_i} C_{i-1}\cap\Tan C_i$ respectively,
 and consider a local basis for $\vf^\perp(C_{i-1})$
 (in a neighbourhood of a point $x\in C_i$)
 $$
 \{ Z^{(i-1)}_1,\ldots ,Z^{(i-1)}_{r_{(i-1)}},
 Z^{(i-1)}_{r_{(i-1)}+1},\ldots ,Z^{(i-1)}_{r_i}\}
 $$
 where \dst\{Z^{(i-1)}_j\}_{j=1,...,r_{(i-1)}}\) are tangent to $C_i$.
 Finally, let $X=X^{(i-1)}+Y^{(i-1)}$ be the
 general solution on the submanifold $C_i$,
 where $X^{(i-1)}$ and $Y^{(i-1)}$ denote
 the determined and the undetermined parts of the solutions,
 respectively.
 \ben
 \item
 Every submanifold $C_{i+1}$  in this sequence
 can be defined (in $C_i$) as the zero set of the so-called
 {\rm $(i+1)$th-generation constraints},
 $\{\chi^{(i+1)}\}\subset\Cinfty (F)$, which are characterized
 in the following equivalent ways:
 \ben
 \item
 For every $Z^{(i)}\in\vf^\perp(C_i)$,
 $$
 \chi^{(i+1)}=\inn(Z^{(i)})(\eta-\gamma)\; .
 $$
 \item
 For every $j\in \{1,\dots ,r_{(i-1)}\},$
  $$
 \chi^{(i+1)}=
 \Lie(X^{(i-1)})[\inn(Z_j^{(i-1)})(\eta-\gamma)]\; .
 $$
 \item
 For every $j\in \{1,\dots ,r_{(i-1)}\}$,
 $$
 \chi^{(i+1)}=
 \inn([X^{(i-1)},Z_j^{(i-1)}])\gamma+\Lie({\cal Y})[\inn(Z_j^{(i-1)})\gamma]
 \; .
 $$
 \een
 \item
 For every $k\in \{r_{(i-1)}+1,\dots ,r_i\}$ the stability condition
 $$
 0=
\Lie(X^{(i-1)}+Y^{(i-1)})[\inn(Z_k^{(i-1)})(\eta-\gamma)]\vert_{C_i}
  $$
 determines (partially or totally) $Y^{(i-1)}$ (on $C_i$).
 \een
 \label{totalcons}
 \end{teor}

 \section{Lagrangian formalism for mechanical systems}
 \protect\label{lf}

 In this section we wish to apply the above results in order to study
 the evolution equations of non-autonomous singular Lagrangian
 systems. By using the constraint algorithm we will obtain a final
 constraint submanifold defined by the
 so-called dynamical Lagrangian constraints. However, this
 problem shows special features since, in addition
 to solving the dynamical equations, we must find solutions
 satisfying the so-called {\sl Second-order condition}. This
 condition will be studied in Sections \ref{seccion4.5} and \ref{sodeprob}.

 \subsection{General description of Lagrangian
 systems}\label{seccion3.1}

 (For more details on the jet bundle
 description of non-autonomous Lagrangian systems
 see, for instance, \cite{EMR-91}, \cite{EMR-sdtc},
 \cite{EMR-96}, \cite{GMS-97}, \cite{GMS-97b},
 \cite{LMM-96b} and \cite{Sa-89}).

 Let $\pi\colon E\to B$ be the {\sl configuration fibered manifold}
 of a non-autonomous mechamical system,
 with volume form $\varpi\in\df^1(B)$,
 and $E$ is an $(n+1)$-dimensional manifold.
 $\pi^1\colon J^1E\to E$ is the
 jet bundle of local sections of $\pi$,
 which is called the {\sl evolution phase space} of the system.
 The map $\bar\pi^1 = \pi \circ \pi^1\colon J^1E \longrightarrow B$
 defines another structure of differentiable
 fibered manifold. Finally, $(t,q^\rho,v^\rho)$
 (with $\rho = 1,\ldots,n$) will denote
 natural local systems of coordinates in $J^1E$
 adapted to the projection $\pi\colon E\to B$, and such that
 the form $\eta:=\bar\pi^{1*}\varpi$ can be locally written
 as $\eta=\d t$.

Note that $J^1E$ is an embedded submanifold of the tangent bundle
$\Tan E$ of $E$. In fact, the map $\iota:J^1E\rightarrow \Tan E$ defined
by
\[
\iota(j_{t_0}^1\phi)=\phi_{*t_0}(V_{\varpi}(t_0))
\]
for $t_0\in B$  and $\phi$ a local section of $\pi:E\rightarrow
B$, is an embedding. Here, $V_\varpi$ denotes the vector field on
$B$ characterized by the condition $\varpi(V_\varpi)=1.$ If we
consider fibered coordinates $(z,q^\rho,v^\rho)$ on $J^1E$ and
$(t,q^\rho, \dot{t},v^\rho)$ on $\Tan E$ then the local expression of
$\iota$ is
\[
\iota(t,q^\rho,v^\rho)=(t,q^\rho,1,v^\rho).
\]

 The dynamical information is given by introducing a
 {\sl Lagrangian density} which is a $\bar\pi^1$-semibasic $1$-form
 on $J^1E$. Then there is a function $\lag\in\Cinfty (J^1E)$
 such that $\Lag =\lag\eta$, which is called the
 {\sl Lagrangian function} associated with $\Lag$ and $\eta$.
 The {\sl Poincar\'e-Cartan $1$ and $2$-forms} associated with the
 Lagrangian density $\Lag$ are defined using the {\sl vertical
 endomorphism} ${\cal V}$ of the bundle $J^1E$:
 $$
 \Theta_{\Lag}:=\inn({\cal V})\Lag+\Lag\in\df^1(J^1E)
 \quad ;\quad
 \Omega_{\Lag}:= -\d\Theta_{\Lag}\in\df^2(J^1E)\; .
 $$
 Then a {\sl Lagrangian system} is a couple $\ls$.

 In a natural chart in $J^1E$ we have that
 $$
 {\cal V}= (\d q^\rho-v^\rho\d t)\otimes\derpar{}{v^\rho}\otimes\derpar{}{t}
 $$
 and therefore
\[ \begin{array}{lcl}
  \Theta_{\Lag}&=&\derpar{\lag}{v^\rho}\d q^\rho -
 \left(\derpar{\lag}{v^\rho}v^\rho-\lag\right)\d t\; ,
\end{array}\]
\begin{equation}
\begin{array}{lcl}
\Omega_{\Lag}&=& -\displaystyle\frac{\partial^2\lag}{\partial
v^\rho\partial v^\nu} \d v^\rho\wedge\d v^\nu
-\displaystyle\frac{\partial^2\lag}{\partial q^\rho\partial v^\nu}
\d q^\rho\wedge\d q^\nu+\displaystyle
\frac{\partial^2\lag}{\partial v^\rho\partial v^\nu} v^\nu \d
v^\rho\wedge\d t+
\\ & &
 \left(\displaystyle\frac{\partial^2\lag}{\partial q^\rho\partial v^\nu}v^\nu -
 \derpar{\lag}{q^\rho}+
 \displaystyle\frac{\partial^2\lag}{\partial t\partial v^\rho}
 \right)\d q^\rho\wedge\d t\; .
 \label{lagforms}
 \end{array}
 \end{equation}
Notice also that
\[
\eta \wedge \Omega_\Lag^{n} = (-1)^{\frac{n(n+1)}{2}}n!\det
\left(\frac{\partial^2\lag}{\partial v^\rho\partial v^\nu}
\right)_{\rho,\nu=1,\dots,n} \d t \wedge\d q^1\wedge\dots\wedge
\d q^n \wedge\d v^1 \wedge \dots \wedge\d v^n  \; .
\]
 As usual, we say that a Lagrangian function $\lag$ (and hence
 the corresponding Lagrangian system)
 is {\sl regular \/} if its associated form
 $\Omega_{\Lag}$ has maximal rank, or what is equivalent,
 the couple $(\Omega_\Lag,\eta)$ is a {\sl cosymplectic structure}.
 This is also equivalent to demanding that
 \dst{\rm det}
 \left(\frac{\partial^2{\lag}}{\partial{v^\rho}\partial{v^\nu}}\right)\)
 is different from zero at every point.
 There exists a more general notion of
 regularity in the recent geometrical approach based on Lepagean
 forms \cite{Olga1,Olga2,Olga4} (see also
 \cite{Krupka}). For Lagrangian systems of order 1, both
 definitions coincide.

 A variational problem can be posed from the Lagrangian density
 $\Lag$, which is called the {\sl Hamilton principle} of the
 Lagrangian formalism: the dynamical trajectories are
 canonical liftings of the sections of
 $\pi$ which are critical for the functional
 ${\bf L}\colon\Gamma_c(B,E)\to\Real$ defined by
 ${\bf L}(\phi):=\int_{B}(j^1\phi)^*\Lag$,
 for every $\phi\in\Gamma_c(B,E)$
 ($\Gamma_c(B,E)$ denotes the set of compact supported sections of $\pi$,
 and $j^1\phi$ is the canonical lifting of $\phi$ to $J^1E$).
 Therefore, it can be proved that
 the critical sections of the Hamilton variational principle
 are the integral curves of an {\sl holonomic} vector field
 $X_{\Lag}\in\vf(J^1E)$,
 (also called a {\sl Second Order Differential Equation} or SODE),
 satisfying that:
 $$
 \inn(X_{\Lag})\Omega_{\Lag} = 0
 \quad , \quad
 \inn(X_{\Lag})\eta \not= 0
 $$
 The second equation means that the vector field $X_\Lag$ is
 $\bar\pi^1$-transverse. It is usual to write this condition in the form
 $$
\inn(X_{\Lag})\eta=1
 $$
 and, locally, this is equivalent to fixing the parametrization of the
 integral curves, taking as parameter the coordinate $t$.
  From now on we will follow this convention.

\begin{remark}
{\rm  Let $\phi\in\Gamma_c(M,E)$ be a section such that $j^1\phi$
 is an integral curve of a vector field $X_\Lag$ which is
 a solution to the above equations. Then $\phi$
 verifies the {\sl Euler-Lagrange equations}:
 $$
 \derpar{\lag}{q^\rho}\Big\vert_{j^1\phi}-
 \frac{\d}{\d t}\derpar{\lag}{v^\rho}\Big\vert_{j^1\phi} = 0
 \quad , \quad
 \mbox{(for $\rho=1,\ldots ,n$)}
 $$
 and conversely.}
 \end{remark}

We can characterize SODE vector fields in $J^1E$ in a more
suitable way, by using the vertical endomorphism.
 In fact, as
 ${\cal V}\in\df^1(J^1E)\otimes\Gamma(J^1E,{\rm V}(\pi^1))\otimes
 \Gamma(J^1E,\bar\pi^{1*}\Tan B)$,
 we can contract the last factor with elements of
 $\Gamma(J^1E,\bar\pi^{1*}\Lambda^1\Tan^*B)$, thus obtaining
 an element of
 $\df^1(J^1E)\otimes\Gamma(J^1E,{\rm V}(\pi^1))$,
 in a natural way. Therefore, we define the {\sl canonical endomorphism}
 ${\cal J}\in\df^1(J^1E)\otimes\Gamma(J^1E,{\rm V}(\pi^1))$
 by making
 $$
 {\cal J}=\inn({\cal V})\eta
 $$
 whose local expression in a natural chart of $J^1E$ is
 \[
 {\cal J}=\inn({\cal V})\eta=(\d q^\rho-v^\rho\d t)\otimes
 \frac{\partial}{\partial v^\rho}\; .
 \]
 Now, it can be proved that a vector field $D\in\vf (J^1E)$
 is a SODE if, and only if, the following conditions hold
 $$
 {\cal J}(D)=0 \quad \hbox{and} \quad
 \inn(D)\eta=1 \; .
 $$

So, in Lagrangian mechanics,
we search for vector fields $X_{\Lag}\in\vf(J^1E)$ such that:
\ben
\item
They are solutions of the so-called {\sl dynamical Lagrangian equations}
(\ref{eq01})
$$
 \inn(X_{\Lag})\Omega_{\Lag} = 0
 \quad , \quad
 \inn(X_{\Lag})\eta=1\; .
$$
 \vspace{-15pt}
\item
$X_{\Lag}$ are holonomic (SODE); that is, ${\cal J}(X_\Lag)=0.$
\een

 As is well known, if $\ls$ is a regular Lagrangian system,
 then there exists a unique Euler-Lagrange vector field
 $X_\Lag\in\vf(J^1E)$ for this system (in fact, $X_\Lag$
 is the {\sl Reeb's vector field}
 of the cosymplectic structure $(\Omega_\Lag,\eta)$).
 This result does not hold for non-regular Lagrangian systems,
 and if $\ls$ is singular, $(\Omega_\Lag,\eta)$ is no longer cosymplectic,
 and the equations (\ref{eq01}) have no  solution in general.
 Even if it exists, it will be neither unique nor a SODE.
 In the best of cases, these vector fields can exist only in a subset
 of points of $J^1E$, and the most interesting situation is when
 this subset is a submanifold of $J^1E$. Thus, the problem we want to solve
 is the following: to look for a submanifold $S\hookrightarrow J^1E$,
 transverse to the projection $\bar\pi^1\colon J^1E\to B$,
 and a vector field $X_\Lag\in\vf (J^1E,S)$
 such that
\ben
\item
$[\inn(X_\Lag)\Omega_\Lag]_{\bar y}=0$, and
$[\inn(X_{\Lag})\eta]_{\bar y}=1$,
for every ${\bar y}\in S$.
\item
$X_\Lag$ is a SODE on the points of $S$; that is, ${\cal
J}(X_\Lag)\vert_S=0.$
\item
$X_\Lag$ is tangent to $S$.
\een
 Conditions 1 and 3 are called {\sl compatibility} and
 {\sl stability} or {\sl consistency conditions} for the dynamical
 equations, respectively. The
 problem can then be solved by first considering only these conditions
 and afterwards adding the second one, or taking all of them simultaneously.

\subsection{Solving the dynamical Lagrangian equations.
Lagrangian constraint algorithm and dynamical Lagrangian constraints}
 \protect\label{sdelf}

 First we consider the problem of finding
 a submanifold $M\hookrightarrow J^1E$, transverse to the projection
 $\bar\pi^1\colon J^1E\to B$, and a vector field $X\in\vf (J^1E,M)$
 such that
\ben
\item
$[\inn(X)\Omega_\Lag]_{\bar y}=0$,  and
$[\inn(X)\eta]_{\bar y}=1$,
for every ${\bar y}\in M$.
\item
$X$ is tangent to $M$.
\een

 To solve this problem, we apply the algorithmic procedure developed
 in Section \ref{ca}, with the hypothesis there assumed.
 Following this process, we obtain a sequence of constrained submanifolds
 \beq
 \cdots \stackrel{\imath_{r+1}^r}{\hookrightarrow} M_i
 \stackrel{\imath_r^{r-1}}{\hookrightarrow} \cdots
 \stackrel{\imath_2^1}{\hookrightarrow} M_1
 \stackrel{\imath_1}{\hookrightarrow} M_0\equiv J^1E\; .
 \label{seqsubman}
 \eeq
 For every $i$, $M_i$ is called the
 {\sl $i$th Lagrangian dynamical constraint submanifold}, and
 this procedure will be called the
 {\sl Lagrangian dynamical constraint algorithm}.

 The only case that concerns us is when
 the algorithm ends by giving a submanifold $M_f$ (with
 $\dim M_f>0$), which is called the
 {\sl final dynamical constraint submanifold} for the Lagrangian problem.
 In such a case, there exists a vector field
 $X_\Lag\in\vf(J^1E)$, tangent to $M_f$, such that
 \beq
 [\inn(X_\Lag)\Omega_\Lag]|_{M_f}=0 \; , \;
 [\inn(X_\Lag)\eta]|_{M_f}=1\; .
 \label{uuu}
 \eeq

 Next we wish to give an intrinsic characterization of the constraints
 which define the constraint submanifolds $M_i$. In order to do this,
 we apply the results stated in Section \ref{gencase}.
 First we take a connection $\bar\nabla$ in the fibered manifold
 $\bar\pi^1\colon J^1E\to B$; that is, an horizontal subbundle of
 $\Tan (J^1E)$ or, what is equivalent in this case, a
 $\bar\pi^1$-transverse vector field ${\cal Y}\in\vf (J^1E)$,
 which can be selected such that $\inn({\cal Y})\eta=1$ holds.

 \begin{remark}
 {\rm
 It is interesting to point out that this construction can be made
 starting from a connection $\nabla$ in
 $\pi\colon E\to B$ or, what is equivalent in this case, a
 $\pi$-transverse vector field ${\cal Y}_E\in\vf (E)$.
 Then this connection induces another one in the fibered manifold
 $\bar\pi^1\colon J^1E\to B$, which is associated with the
 $\bar\pi^1$-transverse vector field
 ${\cal Y}=j^1{\cal Y}_E\in\vf (J^1E)$
(the canonical lifting of ${\cal Y}_E$ to $J^1E$).}
\end{remark}

 Now, let $\gamma_\Lag\in\df^1(J^1E)$ and $\omega_\Lag\in\df^2(J^1E)$
 be the forms defined as
 $$
\gamma_\Lag:=\inn({\cal Y})\Omega_\Lag
 \quad , \quad
\omega_\Lag:=\Omega_\Lag-\eta\wedge\gamma_\Lag\; .
 $$

 Thus, for every submanifold $M_i$ and every ${\bar y}\in M_i$,
 $\Tan_{\bar y}M_i$ is a vector subspace of $\Tan_{\bar y}J^1E$,
 and we can consider the orthogonal complement $\Tan_{\bar y}^\perp M_i$
 with respect to the couple $(\eta_{\bar y},
 (\omega_\Lag)_{\bar y})$; that is
 $$
\Tan_{\bar y}^\perp M_i= [\flat_{(\eta_{\bar
y},(\omega_\Lag)_{\bar y})}(\Tan_{\bar y}M_i)]^0= \{
Y\in\Tan_{\bar y}J^1E\ \mid \ [\inn(Y)(\omega_\Lag)_{\bar y}-
\eta_{\bar y}(\inn(Y)\eta_{\bar y})]\vert_{\Tan_{\bar y}M_i}=0
\}\; .
 $$
 Note that
 $\Tan^\perp_{\bar y}M_0=\ker\,(\omega_\Lag)_{\bar y}\cap\ker\,\eta_{\bar y}$.
 Denote by $\vf^\perp(M_i)$ the set of sections of the vector bundle $\Tan^\perp M_i\to M_i$.
 Therefore, as a direct application of Proposition \ref{cons}
 and Theorem \ref{totalcons} to the present case, we have:

 \begin{teor}
 Let $\ls$ be a singular Lagrangian system, and
 consider the sequence $\{ M_i\}$ of submanifolds (\ref{seqsubman}).
 For every $i\geq 1$, let $r_{(i-1)},r_i$ be the
 ranks of the distributions $\Tan^\perp M_{i-1}$ and
 $\Tan^\perp_{M_i} M_{i-1}\cap\Tan M_i$ respectively,
 and consider a local basis for $\vf^\perp(M_{i-1})$
 (in a neighbourhood of a point $x\in M_i$)
 $$
 \{ {\cal Z}^{(i-1)}_1,\ldots ,{\cal Z}^{(i-1)}_{r_{(i-1)}},
 {\cal Z}^{(i-1)}_{r_{(i-1)}+1},\ldots ,{\cal Z}^{(i-1)}_{r_i}\}
 $$
 where \dst\{{\cal Z}^{(i-1)}_j\}_{j=1,...,r_{(i-1)}}\) are tangent to $M_i$.
 Finally, let $X_\Lag=X_\Lag^{(i-1)}+Y_\Lag^{(i-1)}$ be the
 general solution on the submanifold $M_i$,
 where $X_\Lag^{(i-1)}$ and $Y_\Lag^{(i-1)}$ denote
 the determined and the undetermined parts of the solutions, respectively.
  \ben
 \item
 The submanifold $M_1$ where the dynamical Lagrangian equations
 are compatible can be defined as the zero set of the so-called
 {\rm $1st$-generation dynamical Lagrangian constraints},
 $\{\zeta^{(1)}\}\subset\Cinfty (J^1E)$, which are characterized as
 $$
 \zeta^{(1)}=\inn({\cal Z}^{(0)})(\eta-\gamma_\Lag) \ ,\
 \hbox{for every}\ {\cal Z}^{(0)}\in\vf^\perp(J^1E)\; .
 $$
 \vspace{-20pt}

 \item
 Every submanifold $M_{i+1}$ in this sequence
 can be defined (in $M_i$) as the zero set of the so-called
 {\rm $(i+1)$th-generation dynamical Lagrangian constraints},
 $\{\zeta^{(i+1)}\}\subset\Cinfty (J^1E)$, which are characterized
 in the following equivalent ways:
 \ben
 \item
 For every ${\cal Z}^{(i)}\in\vf^\perp(M_i)$,
 $$
 \zeta^{(i+1)}=\inn({\cal Z}^{(i)})(\eta-\gamma_\Lag)\; .
 $$
 \item
 For every $j\in \{1,\dots ,r_{(i-1)}\},$
 $$
 \zeta^{(i+1)}=
 \Lie(X_\Lag^{(i-1)})[\inn({\cal Z}_j^{(i-1)})(\eta-\gamma_\Lag)]\;
 .
 $$
 \item
 For every $j\in \{1,\dots ,r_{(i-1)}\}$,
 $$
 \zeta^{(i+1)}=
 \inn([X_\Lag^{(i-1)},{\cal Z}_j^{(i-1)}])\gamma_\Lag+
 \Lie({\cal Y})[\inn({\cal Z}_j^{(i-1)})\gamma_\Lag]\; .
 $$
 \een
 \item
 For every $k\in \{r_{(i-1)}+1,\dots, r_i\} $, the stability condition
 $$
 0= \Lie(X_\Lag^{(i-1)}+Y_\Lag^{(i-1)})
 [\inn({\cal Z}_k^{(i-1)})(\eta-\gamma_\Lag)]\vert_{M_i}
 $$
 determines (partially or totally) $Y_\Lag^{(i-1)}$ (on $M_i$).
 \een
 \label{lagdyncons}
 \end{teor}

\section{Hamiltonian formalism for mechanical systems}

 Now we consider the Hamiltonian formalism
 of a non-autonomous mechanical system associated with a singular Lagrangian
 and, in particular, the problem of finding solutions of the
 {\sl Hamilton equations}.

 (For more details on the jet bundle
 description of the Hamiltonian formalism of non-autonomous mechanical systems
 see, for instance,  \cite{EMR-91}, \cite{EMR-sdtc}, \cite{GMS-97},
 \cite{GMS-97b}, \cite{LMM-96b}, \cite{Ra1} and \cite{Sd-98}).

\subsection{The momentum dual bundle and the Legendre map.
 Hyper-regular and almost regular systems}\label{seccion4.1}

In order to associate a Hamiltonian system with a
non-autonomous Lagrangian system, we first need to introduce
two new elements: a suitable {\sl dual bundle} of $J^1E$, and
a {\sl Legendre map}.

 In the jet bundle description of non-autonomous
 dynamical systems, there are several choices for the
 momentum bundle where the covariant Hamiltonian formalism
 takes place (see \cite{EMR-00} for a general review).
 In this work, following \cite{CCI-91}, first we take the bundle
 $\Tan^*E$,
 which is called the {\sl extended momentum dual bundle} associated with
 $\pi\colon E\to B$. The natural projections are denoted
 by $\sigma^1\colon\Tan^*E\to E$ and
 $\bar\sigma^1\colon\Tan^*E\to B$.
 Then, if $\Lambda_0^1\Tan^*E$ denotes the
 bundle of $\pi$-semibasic $1$-forms in $E$, we define the bundle
 $$
 J^{1*}E:=\Tan^*E/\Lambda_0\Tan^*E\simeq{\rm V}^*(\pi)
 $$
 which is called the {\sl restricted momentum dual bundle}
 associated with $\pi\colon E\to B$.
 We denote the natural projections by $\tau^1\colon J^{1*}E\to E$,
 and $\bar\tau^1:=\pi\circ\tau^1\colon J^{1*}E\to B$.
 In addition, we have the canonical projection
 $\mu\colon\Tan^*E\to J^{1*}E$.

 Natural coordinates in $\Tan^*E$ and
 $J^{1*}E$ (adapted to the bundle structures) will be denoted by
 $(t,q^\rho,p,p_\rho)$ and $(t,q^\rho,p_\rho)$, respectively.
 Then we also write $\tilde\eta:=\bar\tau^{1*}\varpi=\d t$.

 The construction of the Legendre map can be made following different
 but equivalent ways.
 For instance, we can follow a procedure similar to
 the case of autonomous mechanics, and define this map as
 a ``fiber derivative'' of the Lagrangian density
 (see, for instance, \cite{EMR-00}).
 Alternatively, we can define it using the theory of affine dual bundles
 (as in \cite{CCI-91} and \cite{SC-90}).
 Another more simple way consists in constructing this map
 by means of the Poincar\'e-Cartan form as follows:
 taking into account that $\Theta_{\Lag}$ can be seen as
 a $1$-form on $J^1E$ along the
 projection $\pi^1\colon J^1E\to E$, the {\sl extended Legendre map}
 associated with a Lagrangian density $\Lag$ is the $\Cinfty$-map
 $\widetilde{{\cal F}\Lag} \colon J^1E\to\Tan^*E$
 defined by
\begin{equation}\label{23'}
[\widetilde{{\cal F}\Lag}(\bar y)](Z)= (\Theta_{\Lag})_{\bar
y}(\bar Z)
\end{equation}
 where $\bar y\in J^1E$,
 $Z\in\Tan_{\pi^1(\bar y)}E$, and
 $\bar Z\in\Tan_{\bar y}J^1E$ is such that
 $\pi^1_{*\bar{y}}(\bar Z_\nu)=Z_\nu$.
 Therefore, the {\rm restricted Legendre map}
 associated with a Lagrangian density $\Lag$
 is the $\Cinfty$-map
 ${\cal F}\Lag\colon J^1E\to J^{1*}E$ defined by
 $$
 {\cal F}\Lag:=\mu\circ\widetilde{{\cal F}\Lag}
 $$

 In natural coordinates, the local expressions of these Legendre maps are
 \bea
\begin{array}{ccccccc}
\widetilde{{\cal F}\Lag}^*t = t &\  ,\ &
\widetilde{{\cal F}\Lag}^*q^\rho = q^\rho &\  , \ &
\widetilde{{\cal F}\Lag}^*p_\rho = \derpar{\lag}{v^\rho}  &\  , \ &
\widetilde{{\cal F}\Lag}^*p = \lag -v^\rho\derpar{\lag}{v^\rho}
 \\
 {\cal F}\Lag^*t =t & \ , \ & {\cal F}\Lag^*q^\rho = q^\rho
 & \ , \ & {\cal F}\Lag^*p_\rho = \derpar{\lag}{v^\rho} & &
 \end{array}\; .
 \label{legmaps}
 \eea

 The cotangent bundle $\Tan^*E$ is endowed with canonical forms:
 the {\sl Liouville forms}
 $\Theta\in\df^1(\Tan^*E)$ and  $\Omega:=-\d{\bf \Theta}\in\df^2(\Tan^*E)$
(which is a symplectic form), whose local expressions are
 $$
 \Theta=p\d t+p_\nu\d q^\nu
 \ , \
 \Omega=-\d p\wedge\d t-\d p_\nu\wedge\d q^\nu\ .
 $$
Using (\ref{lagforms}) and (\ref{legmaps}), we obtain that
 \beq
{\cal F}\Lag^*\Theta=\Theta_{\Lag}
 \quad ,\quad
{\cal F}\Lag^*\Omega=\Omega_{\Lag}\; .
 \label{pback}
 \eeq

The matrix of the tangent map ${\cal F}\Lag_*$
in a natural coordinate system is
 \begin{equation}\label{25'}
 \left(\matrix{{\rm Id} & 0 & 0 \cr 0 & {\rm Id} & 0 \cr
 \frac{\partial^2\lag}{\partial t\partial v^\rho} &
 \frac{\partial^2\lag}{\partial q^\nu\partial v^\rho} &
 \frac{\partial^2\lag}{\partial v^\nu\partial v^\rho}\cr}\right)
 \end{equation}
 where the sub-matrix
 \dst\left(\frac{\partial^2\lag}{\partial v^\nu\partial v^\rho}\right)\)
 is the {\sl partial Hessian matrix} of $\lag$.
 Thus, one deduces that (see \cite{LMM-96b})
 \begin{equation}
 \ker {\cal F}{\cal L}_*=\ker \widetilde{{\cal FL}_*}=\ker\Omega_{\cal
 L}\cap\vf^{V(\pi^1)}(J^1E)
 \end{equation}
 Note that the Lagrangian system $(J^1E,\Omega_{\Lag})$ is {\sl regular},
 that is, $(\Omega_{\Lag},\eta)$ is a cosymplectic structure
 (or equivalently the partial Hessian matrix
 \dst\left(\frac{\partial^2\lag}{\partial v^\nu\partial v^\rho}\right)\)
 is regular everywhere in $J^1E$) if and only if ${\cal F}\Lag$
 is a local diffeomorphism (see also \cite{CCI-91}
 for a different definition of this concept).
 Then, following a well-known terminology in mechanics we define:

\begin{definition}
Let $\ls$ be a Lagrangian system.
 \ben
\item
$\ls$ is said to be a {\rm regular} or {\rm non-degenerate}
Lagrangian system if ${\cal F}\Lag$ is a local diffeomorphism.

As a particular case, $\ls$ is said to be a {\rm hyper-regular}
Lagrangian system if ${\cal F}\Lag$ is a global diffeomorphism.
\item
Elsewhere $\ls$ is said to be a {\rm singular} or {\rm degenerate}
Lagrangian system.
 \een
\end{definition}

 \begin{prop}
 {\rm (See \cite{EMR-00,LMM-96b})}.
 Let $\ls$ be a hyper-regular Lagrangian system. Then:
 \ben
 \item
 $\widetilde{{\cal F}\Lag}(J^1E)$
 is a 1-codimensional embedded submanifold of $\Tan^*E$
 which is transverse to the projection $\mu$.
 \item
 The map $\widetilde{{\cal F}\Lag}$ is a diffeomorphism on its image,
 and the map $\mu$ restricted to $\widetilde{{\cal F}\Lag}(J^1E)$
 is also a diffeomorphism.
 \een
 \label{hrprop}
 \end{prop}

 For dealing with singular Lagrangians we must assume minimal
 ``regularity'' conditions. Hence we introduce the following
 terminology:

 \begin{definition}\cite{CLM-94,LMM-96b}
 A singular Lagrangian system $\ls$ is said to be
 {\rm almost regular} if:
 \ben
  \item
 $\tilde{\cal P}:=\widetilde{{\cal F}\Lag}(J^1E)$
 is a regular submanifold of $T^*E$.

 (We will denote by
 $\tilde\jmath_0\colon \tilde{\cal P}\hookrightarrow\Tan^*E$
 the corresponding embedding).
 \item
 $\widetilde{\cal F}\Lag$ is a submersion onto its image
 (with connected fibers).
 \item
 For every $\bar y\in J^1E$, the fibers
 ${{\cal F}\Lag}^{-1}({\cal F}\Lag (\bar y))$
 are connected submanifolds of $J^{1*}E$.
  \een
 \end{definition}

 (This definition is slightly different from that in references \cite{GMS-97},
 and \cite{Sd-95}).

 Let $\ls$ be an almost regular Lagrangian system. Denote
 $$
{\cal P}:={\cal F}\Lag(J^1E)\; .
 $$
 Then let  $\jmath_0\colon{\cal P}\hookrightarrow J^{1*}E$
 be the canonical inclusion, and
 $\mu_0\colon\tilde{\cal P}\to{\cal P}$
 the restriction of the map $\mu$. Finally, define the restriction
 mappings
 $$
 {\cal F}\Lag_0\colon J^1E\to{\cal P} \ ,\
 \widetilde{{\cal F}\Lag}_0\colon J^1E\to\tilde{\cal P}\; .
 $$
With these definitions, it follows that

\begin{prop}
{\rm (See \cite{EMR-00, LMM-96b})}.
 Let $\ls$ be an almost regular Lagrangian system. Then:
 \ben
 \item
 For every $\bar y\in J^1E$,
 $$
 \widetilde{{\cal F}\Lag_0}^{-1}(\widetilde{{\cal F}\Lag_0}(\bar y))=
 {\cal F}\Lag_0^{-1}({\cal F}\Lag_0(\bar y))\; .
 $$
 \item
 There exists a unique differentiable structure on ${\cal
 P}$ such that $\mu_0:\tilde{\cal P}\rightarrow {\cal P}$ is a
 diffeomorphism.
 \item
 ${\cal P}$ is a submanifold of $J^{1*}E$, and
 $\jmath_0\colon{\cal P}\hookrightarrow J^{1*}E$
 is an embedding.
 \item
 The restriction mapping ${\cal F}\Lag_0:J^1E\rightarrow {\cal P}$ is a
 submersion with connected fibers.
 \een
 \label{arprop}
 \end{prop}

 \subsection{Hamiltonian system associated with an almost regular
 Lagrangian system}\label{seccion4.2}

 If $\ls$ is an almost regular Lagrangian system,
 the submanifold $\jmath_0\colon {\cal P}\hookrightarrow J^{1*}E$,
 is a fiber manifold over $E$ (and $B$). The
 corresponding projections will be denoted
 $\tau^1_0\colon {\cal P}\to E$ and $\bar\tau^1_0\colon{\cal P}\to B$,
 satisfying that $\tau^1\circ\jmath_0=\tau^1_0$ and
 $\bar\tau^1\circ\jmath_0=\bar\tau^1_0$.
 We will denote $\eta^0:=\bar\tau^{1*}_0\varpi\equiv\d t$.

 Now, given the diffeomorphism ${\rm h}_0=\mu_0^{-1}$,
 we define the Hamilton-Cartan forms
  $$
 \Theta_{{\rm h}_0}=(\tilde\jmath_0\circ{\rm h}_0)^*\Theta
 \quad ; \quad
 \Omega_{{\rm h}_0}=-\d\Theta_{{\rm h}_0}=
 (\tilde\jmath_0\circ{\rm h}_0)^*\Omega
 $$
 and, taking into account the commutativity of the diagram
 $$
\begin{array}{ccccc}
\begin{picture}(15,52)(0,0)
\put(0,0){\mbox{$J^1E$}}
\end{picture}
&
\begin{picture}(65,52)(0,0)
 \put(17,28){\mbox{${\cal F}\Lag_0$}}
 \put(24,7){\mbox{$\widetilde{{\cal F}\Lag_0}$}}
 \put(0,7){\vector(2,1){65}}
 \put(0,4){\vector(1,0){65}}
\end{picture}
&
\begin{picture}(15,52)(0,0)
 \put(5,0){\mbox{$\tilde{\cal P}$}}
 \put(5,42){\mbox{${\cal P}$}}
 \put(5,13){\vector(0,1){25}}
 \put(10,38){\vector(0,-1){25}}
 \put(-7,22){\mbox{$\mu_0$}}
 \put(12,22){\mbox{${\rm h}_0$}}
 \end{picture}
&
 \begin{picture}(65,52)(0,0)
 \put(0,44){\vector(1,0){65}}
 \put(30,49){\mbox{$\jmath_0$}}
 \put(0,5){\vector(1,0){65}}
 \put(30,9){\mbox{$\tilde\jmath_0$}}
 \end{picture}
&
 \begin{picture}(10,52)(0,0)
 \put(0,41){\mbox{$J^{1*}E$}}
 \put(0,0){\mbox{$\Tan^*E$}}
 \put(10,13){\vector(0,1){25}}
 \put(0,22){\mbox{$\mu$}}
 \end{picture}
\end{array}
 $$
 from (\ref{pback}) we obtain that the following relations hold
 \beq
 {\cal F}\Lag_0^*\Theta_{{\rm h}_0}=\Theta_{\Lag}
 \quad ,\quad
 {\cal F}\Lag_0^*\Omega_{{\rm h}_0}=\Omega_{\Lag}\; .
 \label{pback0}
 \eeq
 Then $\hso$
 is the {\sl Hamiltonian system}
 associated with the almost regular Lagrangian system $\ls$.

 We can state a variational problem for $\hso$
 ({\sl Hamilton-Jacobi principle}\/):
 in a way analogous to the Lagrangian formalism, now
 using sections of $\bar\tau^1_0\colon{\cal P}\to B$, and the form
 $\Theta_{{\rm h}_0}$.
 So we look for sections $\psi_0\in\Gamma_c(B,{\cal P})$
 which are critical for the functional
 \dst{\bf H}^0(\psi):=\int_{B}\psi^*\Theta_{{\rm h}_0}\) ,
 for every $\psi\in\Gamma_c(B,{\cal P})$.
 Therefore, it can be proved that
 the critical sections of the Hamilton-Jacobi variational principle
 are the integral
 curves of a vector field $X_{{\rm h}_0}\in\vf({\cal P})$
 (which is called a {\sl Hamiltonian vector field} for the system $\hso$)
 satisfying that:
 $$
 \inn(X_{{\rm h}_0})\Omega_{{\rm h}_0} = 0
 \quad , \quad
 \inn(X_{{\rm h}_0})\eta^0 \not= 0\; .
 $$
 As above, the second equation means that the vector field
 $X_{{\rm h}_0}$ is
 $\bar\tau^1_0$-transverse, and  we will take this condition in the form
 $$
 \inn(X_{{\rm h}_0})\eta^0=1\; .
 $$

 Now, as $\ls$ is almost regular and $(\Omega_{{\rm h}_0},\eta^0)$
 is no longer cosymplectic, the above equations have no  solution in general.
 In the most favourable cases, Hamiltonian vector fields for $\hso$
 can exist only in a subset of points of ${\cal P}$,
 and the most interesting situation is when
 this subset is a submanifold of ${\cal P}$.

 \begin{remark}
{\rm
 The construction of a Hamiltonian system
 associated with a hyper-regular Lagrangian system $\ls$
 follows the same pattern, but now ${\cal P}\equiv J^{1*}E$,
 $\tilde{\cal P}\equiv\Tan^*E$, and ${\cal F}\Lag$ is a diffeomorphism.
 Then, the Hamiltonian section is now
 $$
 {\rm h}_0:= \widetilde{{\cal F}\Lag}\circ{\cal F}\Lag^{-1}
 $$
 which is a diffeomorphism connecting $J^{1*}E$ and
 $\widetilde{{\cal F}\Lag}(J^1E)$
 (observe that it is just the inverse of
 $\mu$ restricted to $\widetilde{{\cal F}\Lag}(J^1E)$).
 (See \cite{LMM-96b} for details).}
 \end{remark}

\subsection{Hamiltonian constraint algorithm and equivalence
 with the Lagrangian formalism}

 The problem we want to solve is the following: to look for
 a submanifold $P\hookrightarrow{\cal P}$, transverse to the projection
 $\bar\tau^1_0\colon{\cal P}\to B$, and a vector field
 $X_{{\rm h}_0}\in\vf ({\cal P},P)$
 such that \ben
\item
 $[\inn(X_{{\rm h}_0})\Omega_{{\rm h}_0}]_{\tilde y}=0$,
 and $[\inn(X_{{\rm h}_0})\eta^0]_{\tilde y}=1$,
 for every ${\tilde y}\in P$.
\item
 $X_{{\rm h}_0}$ is tangent to $P$.
\een
 Item 1 is the {\sl compatibility condition}, and item 2 is the
 {\sl stability} or {\sl consistency condition} for the Hamiltonian
 equations.

 Once again we apply the algorithmic procedure developed
 in Section \ref{ca}, with the hypothesis there assumed.
 Thus, following this process,
 we obtain a sequence of constrained submanifolds
 \beq
 \cdots \stackrel{\jmath_{i+1}^i}{\hookrightarrow} P_i
 \stackrel{\jmath_i^{i-1}}{\hookrightarrow} \cdots
 \stackrel{\jmath_2^1}{\hookrightarrow} P_1
 \stackrel{\jmath_1}{\hookrightarrow} P_0\equiv{\cal P}
 \stackrel{\jmath_0}{\hookrightarrow} J^{1*}E\; .
 \label{seqsubman1}
 \eeq
 For every $i$, $P_i$ is called the
 {\sl $(i+1)$th Hamiltonian constraint submanifold}, and this procedure
 is called the {\sl Hamiltonian constraint algorithm}.

 The only case interesting for us is when
 the algorithm ends giving a submanifold $P_f$ (with
 $\dim P_f>0$), which is called the
 {\sl final constraint submanifold} for the Hamiltonian problem.
 Then, there exists a vector field
 $X_{{\rm h}_0}\in\vf ({\cal P},P)$, tangent to $P_f$, such that
 \beq
 [\inn(X_{{\rm h}_0})\Omega_{{\rm h}_0}]|_{P_f}=0 \; , \;
 [\inn(X_{{\rm h}_0})\eta^0]|_{P_f}=1\; .
\label{uuuu}
 \eeq

Now, consider the sequences of dynamical Lagrangian constraint
submanifolds (\ref{seqsubman}) and of Hamiltonian constraint
submanifolds (\ref{seqsubman1}). Then, we have (see
\cite{LMM-96b}):

\begin{teor}
 At every level $i=1,\ldots ,f$
 of the Lagrangian dynamical constraint algorithm and
 the Hamiltonian constraint algorithm we have:
\ben
\item
 ${\cal F}\Lag_{i-1}^{-1}(P_i)=M_i$ (and hence
 ${\cal F}\Lag_{i-1}(M_i)=P_i$).
\item
 The induced mapping by ${\cal F}\Lag_{i-1}$ from $M_i$ to $P_i$,
 denoted by ${\cal F}\Lag_i\colon M_i \to P_i$,
 is a submersion.
\item
 For every $\bar y_i\in M_i$,
 ${\cal F}\Lag_i^{-1}({\cal F}\Lag_i(\bar y_i))=
 {\cal F}\Lag_0^{-1}({\cal F}\Lag_0(\bar y_i))$.
 Hence
 $$\ker\,({\cal F}\Lag_{i*})_{\bar y_i}=\ker\,({\cal F}\Lag_{0*})_{\bar y_i}=
 (\ker\,\Omega_\Lag\cap \vf^{V(\pi_1)}(J^1E))_{\bar{y}_i}.$$
\vspace{-15pt}

 \item
 There is a basis $\{\zeta_j^{(i)}\}\subset\Cinfty(J^1E)$ for the set of
 $i$th-generation dynamical Lagrangian constraints
 which is ${\cal F}\Lag_0$-related with a basis
 $\{\xi_j^{(i)}\}\subset\Cinfty({\cal P})$
 for the set of $i$th-generation Hamiltonian constraints; that is,
 $$
 {\cal F}\Lag^*_0\xi_j^{(i)}=\zeta_j^{(i)}\; .
 $$
\een

Thus, the codimension of $M_i$ (as a submanifold of $J^1E$)
coincides with the codimension of $P_i$ (as a submanifold of
${\cal P}$).
 \label{flprolagcons}
 \end{teor}
 \proof
 We make the proof for the first level ($i=1$). For the other
 cases it follows the same pattern.
 \ben
 \item
 First we prove that ${\cal F}\Lag_0(M_1)\subset P_1$.
 In fact, for every $\bar y\in M_1$, we have that there exists
 $X_{\bar y}\in\Tan_{\bar y}J^1E$ such that
 $$
 \inn(X_{\bar y})(\Omega_\Lag)_{\bar y}=0
 \quad ,\quad
 \inn(X_{\bar y})\eta_{\bar y}=1.
 $$
Thus, if ${\cal F}\Lag_0({\bar y})=\tilde{y}$ and
$\tilde{X}_{\tilde{y}}=({\cal F}\Lag_{0*})_{\bar y}(X_{\bar y})$,
then using ${\cal F}\Lag_0^*\Omega_{h_0}=\Omega_\Lag,$ and
${\cal F}\Lag_0^*\eta^0=\eta,$ we obtain that
 $$
 \inn(\tilde X_{\tilde y})(\Omega_{{\rm h}_0})_{\tilde y}=0
 \quad ,\quad
 \inn(\tilde X_{\tilde y})\eta^0_{\tilde y}=1
 $$
 which implies that $\tilde y\in P_1$.

 Conversely, for every $\tilde y\in P_1$, the fiber
 ${\cal F}\Lag_0^{-1}(\tilde y)$ is entirely contained in $M_1.$
 In fact, if $\bar{y}\in {\cal F}{\cal L}_0^{-1}(\tilde{y})$ and
 $\tilde{X}_{\tilde{y}}\in \Tan_{\tilde{y}} {\cal P}$ is such that
 $\inn_{\tilde{x}_{\tilde{y}}} \Omega_{{\rm h}_0}(\tilde{y})=0$ and
 $\inn_{\tilde{X}_{\tilde{y}}}\eta^0(\tilde{y})=1$, then since
 ${\cal F}\Lag_0$ is a submersion, there exists $X_{\bar{y}}\in
 \Tan_{\bar{y}}J^1E$ satisfying $({\cal F}\Lag_{0*})_{\bar
 y}(X_{\bar{y}})=\tilde{X}_{\tilde{y}}$ and, reasoning as above, we
 conclude that $\inn_{X_{\bar y}} \Omega_\Lag (\bar y)=0,$
 $i_{X_{\bar y}} \eta_{\bar y}=1.$ Thus, $\bar y\in M_1$ and
 therefore ${\cal F}\Lag_0^{-1}(P_1)\subseteq M_1.$

 As a consequence of both results we obtain that
 ${\cal F}\Lag_0^{-1}(P_1)=M_1$, and consequently
 ${\cal F}\Lag_0(M_1)=P_1$.
\item
 For every $\tilde y\in P_1$, as ${\cal F}\Lag_0$ is a submersion,
 then there exists an open neighborhood $U_0\subseteq {\cal P}$ with
 $\tilde y\in U_0$, and a differentiable mapping
 $s_0\colon U_0\to J^1E$, such that ${\cal F}\Lag_0\circ s_0={\rm Id}_{U_0}$.
 We take $U_1=U_0\cap P_1$, and $s_1=s_0\mid_{U_1}\colon U_1\to J^1E$
 is a differentiable mapping.

 If $\tilde y'\in U_1$, then there exists $\bar y'\in M_1$ such that
 ${\cal F}\Lag_0(\bar y')=\tilde y'$, so
 ${\cal F}\Lag_0(s_1(\tilde y'))={\cal F}\Lag_0[s_0(\tilde y')]=
 \tilde y'={\cal F}\Lag_0(\bar y')$.
 Therefore,
 $s_1(\tilde y')\in{\cal F}\Lag_0^{-1}({\cal F}\Lag_0(\bar y'))\subseteq M_1$.
 That is, $s_1\colon U_1\to J^1E$ takes
 values in $M_1$, and because $M_1$ is supposed to be an embedded
 submanifold of $J^1E$, then $s_2\colon U_1\to M_1$ is
 also a differentiable mapping. Moreover,
 ${\cal F}\Lag_1\circ s_1={\rm Id}_{U_1}$.
 Then, $s_1\colon U_1\to M_1$ is a local section of
 ${\cal F}\Lag_1\colon M_1 \to P_1$ in $\tilde y\in P_1$, and
 therefore
 ${\cal F}\Lag_1$ is a submersion.
 \item
Using the first item we deduce that ${\cal F}\Lag_i^{-1}({\cal
F}\Lag_i(\bar{y}_i))={\cal F}\Lag_0^{-1}({\cal F}{\cal L}_0({\bar
y}_i))$ and therefore
\[
\ker({\cal F}{\cal L}_{i*})_{\bar{y}_i}=\ker ({\cal
F}\Lag_{0*})_{\bar{y}_i}.
\]
Furthermore, in
 \cite{LMM-96b}, it was proved that
 \[
 \ker({\cal F}\Lag_{0*})=\ker \Omega_\Lag\cap \vf^{V(\pi_1)}(J^1E).
 \]
  \item
 We have the following diagram
 $$
 \begin{array}{ccc}
 J^1E & \stackrel{{\cal F}\Lag_0}{\longrightarrow} & {\cal P} \\
 \uparrow \imath_1 & & \uparrow \jmath_1 \\
 M_1 & \stackrel{{\cal F}\Lag_1}{\longrightarrow} & P_1
 \end{array}
 $$
 Consider the basis $\{ \xi_j^{(1)}\}$ of
 $1st$-generation Hamiltonian constraints, we must prove that
 the set $\{{\cal F}\Lag^*_0\xi_j^{(1)}\equiv\zeta_j^{(1)}\}$
 is a basis of $1st$-generation dynamical Lagrangian constraints.
 First observe that, as $\jmath_1^*\xi_j^{(1)}=0$, we obtain that
 $$
 0={\cal F}\Lag_1^* \jmath_1^*\xi_j^{(1)}=
 \imath_1^*{\cal F}\Lag_0^*\xi_j^{(1)}\equiv \imath_1^*\zeta_j^{(1)}
 $$
 and thus the functions $\zeta_j^{(1)}$ are dynamical Lagrangian constraints.
 Furthermore, the number of independent
 $1st$-generation Hamiltonian constraints
 is $\dim\,{\cal P}-\dim\, P_1\equiv r$. If $m$ is the
 dimension of the fibers of ${\cal F}\Lag_0$, then
 $\dim\, J^1E=\dim\, J^{1*}E=\dim\,{\cal P}+m$, and
 as a consequence of the items 1 and 3,
 $\dim\, M_1=\dim\, P_1+m$. Therefore, the number of
 independent $1st$-generation dynamical Lagrangian constraints is
 $$
 \dim\, J^1E-\dim\, M_1=\dim\,{\cal P}+m-\dim\, P_1-m=r\; .
 $$
 Notice that the constraints are independent because ${\cal F}\Lag$
 is a submersion and thus ${\cal F}\Lag_0^*$ is injective.
\qed \een
In this way we can draw the diagram
 \beq
 \begin{array}{lclcc}
 M_0=J^1E & \stackrel{{\cal F}\Lag_0}{\longrightarrow} &
 {\cal P}\equiv P_0 & \stackrel{i_0}{\hookrightarrow} & J^{1*}E \\
 \uparrow \imath_1 & & \uparrow \jmath_1 & & \\
 M_1 & \stackrel{{\cal F}\Lag_1}{\longrightarrow} & P_1 & & \\
 \uparrow \imath_2^1 & & \uparrow \jmath_2^1 & & \\
 \vdots & & \vdots & & \\
 \uparrow \imath_{f-1}^{f-2} & & \uparrow \jmath_{f-1}^{f-2} & & \\
 M_{f-1} & \stackrel{{\cal F}\Lag_{f-1}}{\longrightarrow} & P_{f-1} & & \\
 \uparrow \imath_f^{f-1} & & \uparrow \jmath_f^{f-1} & & \\
 M_f & \stackrel{{\cal F}\Lag_f}{\longrightarrow} & P_f & &
 \end{array}
 \label{diageq}
 \eeq

 Finally, for the vector fields solutions of the Lagrangian and Hamiltonian
 dynamical problems we have:

 \begin{teor}
 Let $M_f$ be the final dynamical constraint submanifold for
 the Lagrangian problem, and $P_f$ is the final constraint submanifold
 for the Hamiltonian counterpart.
 \ben
 \item
 Let $X_\Lag\in\vf (J^1E,M_f)$ such that:
 \ben
\item
 $X_\Lag$ is tangent to $M_f$.
\item
 $X_\Lag$ is a solution of the dynamical Lagrangian equations (\ref{uuu}).
\item
 $X_\Lag$ is ${\cal F}\Lag$-projectable.
 \een
 Then ${\cal F}\Lag_{f*}X_\Lag=X_{{\rm h}_0}\in\vf (J^{1*}E,P_f)$
 is tangent to $P_f$
 and it is a solution of the Hamiltonian equations (\ref{uuuu}).
 \item
 Conversely, if $X_{{\rm h}_0}$
 is a vector field tangent to $P_f$
 and it is a solution of the Hamiltonian equations (\ref{uuuu}),
 then there exists a vector field  $X_\Lag$ on $M_f$ such that
 ${\cal F}\Lag_{f*}X_\Lag=X_{{\rm h}_0}$, and it is a solution
 of the dynamical Lagrangian equations (\ref{uuu}).
 \een
 \label{eqsol}
 \end{teor}
 \proof
 The results follow using (\ref{pback0}), that ${\cal
 F}\Lag_f:M_f\to P_f$ is a surjective submersion and the fact that
 ${\cal F}\Lag_0^*\eta^0=\eta$ (see \cite{LMM-96b}).
 \qed

 In this case, the integral curves of $X_\Lag$ are not, in general,
 the solutions of the Euler-Lagrange equations,
 since they are not holonomic ($X_\Lag$ is not a SODE,
 necessarily). The problem of finding a solution satisfying the SODE condition,
 that is, an Euler-Lagrange vector field,
 is studied in Section \ref{seccion4.5}.

 \subsection{Intrinsic characterization of Hamiltonian constraints}

 We want to give an intrinsic characterization of the Hamiltonian constraints
 which define the constraint submanifolds $P_i$, and hence
 we will apply again the results given in Sections \ref{algstate} and \ref{ca}.
 First we take a connection $\tilde\nabla$ in
 $\bar\tau^1_0\colon {\cal P}\to B$ or, what is equivalent, a
 $\bar\tau^1_0$-transverse vector field $\tilde{\cal Y}\in\vf ({\cal P})$,
 which can be selected such that $\inn(\tilde{\cal Y})\eta^0=1$ holds.

 The connections $\tilde\nabla$ in
 $\bar\tau^1_0\colon {\cal P}\to B$, and $\bar\nabla$ in
 $\bar\pi^1\colon J^1E\to B$ (given in Section \ref{sdelf})
 can be chosen ${\cal F}\Lag_0$-related.
 This means that the vector field ${\cal Y}\in\vf(J^1E)$
 associated with $\bar\nabla$ can be selected in such a way that
 ${\cal F}\Lag_{0*}{\cal Y}=\tilde{\cal Y}$.
 This is always possible, since ${\cal F}\Lag_0$ is a
 surjective submersion.
  From now on this fact will be assumed.

 In this case, $\tilde\nabla$ can be constructed
 starting from a connection $\nabla$ in
 $\pi\colon E\to B$, as in the Lagrangian case.

 Observe also that since ${\cal F}\Lag_0^*\eta^0=\eta$,
 if $\inn({\cal Y})\eta=1$, then $\inn(\tilde{\cal Y})\eta^0=1$ too.

 Now, let $\gamma_{{\rm h}_0}\in\df^1(J^{1*}E)$ and
 $\omega_{{\rm h}_0}\in\df^2(J^{1*}E)$
 be the forms defined as
 $$
\gamma_{{\rm h}_0}:=\inn(\tilde{\cal Y})\Omega_{{\rm h}_0}
 \quad , \quad
\omega_{{\rm h}_0}:=\Omega_{{\rm h}_0}- \eta^0\wedge\gamma_{{\rm
h}_0}\; .
 $$
Using (\ref{pback0}) and after
 the comment in the first item above, it is obvious
that
 \beq
 {\cal F}\Lag_0^*\omega_{{\rm h}_0}=\omega_\Lag
 \quad , \quad
 {\cal F}\Lag_0^*\gamma_{{\rm h}_0}=\gamma_\Lag\; .
 \label{omegagamma}
 \eeq

 Suppose that $\bar{y}_i\in M_i$ (respectively,
 $\bar{y}_{i+1}\in M_{i+1}$) and that
 $\tilde{y}_i={\cal F}\Lag_0(\bar{y}_i)$ (respectively,
 $\tilde{y}_{i+1}={\cal F}\Lag_0(\bar{y}_{i+1})).$ Using
 (\ref{omegagamma}) and the fact that ${\cal
 F}\Lag_0^*(\eta^0)=\eta=(\bar{\pi}^1)^*(\varpi)$, we deduce that
$$ \begin{array}{rcl} ({\cal
F}\Lag_{0*})_{\bar{y}_i}(\Tan_{\bar{y}_i}^\perp
M_i)&=&\Tan_{\tilde{y}_i}^\perp P_i \\[8pt] ({\cal
F}\Lag_{0*})_{\bar{y}_{i+1}}(\Tan_{\bar{y}_{i+1}}^\perp M_i\cap
\Tan_{{\bar y}_{i+1}}M_{i+1})&=&\Tan_{\tilde{y}_{i+1}}^\perp P_i\cap
\Tan_{\tilde{y}_{i+1}} P_{i+1}.
\end{array}
$$

Moreover, from Lemmas \ref{10'} and \ref{a} (see Appendices
\ref{A} and \ref{C}) and Theorem \ref{flprolagcons}, we obtain
that $$
\begin{array}{rcl}
\ker({\cal F}\Lag_{0*})_{\bar{y}_i}&\subseteq&
(\ker\Omega_\Lag\cap \ker\eta)_{\bar{y}_i}\subseteq
\Tan^\perp_{\bar{y}_i}M_0\subseteq \Tan^\perp_{{\bar y}_i}M_i,\\
\ker({\cal F}\Lag_{0*})_{\bar{y}_i}&\subseteq&
\Tan^{\perp}_{\bar{y}_{i+1}}M_0\cap \Tan_{\bar{y}_{i+1}}M_{i+1}\subseteq
\Tan^\perp_{{\bar y}_{i+1}}M_i\cap \Tan_{\bar{y}_{i+1}}M_{i+1}.
\end{array}
$$ Thus, if $m$ is the dimension of the fibers of ${\cal
F}\Lag_0,$ it follows that $$\begin{array}{rcl} \dim \Tan^\perp_{\bar
y_i}M_i&=&\dim \Tan^{\perp}_{\tilde{y}_i} P_i + m\\ \dim
(\Tan^\perp_{\bar{y}_i+1}M_i\cap \Tan_{{\bar y}_{i+1}}M_{i+1})&=&\dim
(\Tan^\perp_{\tilde{y}_{i+1}}P_i \cap \Tan_{\tilde{y}_{i+1}}P_{i+1}) +
m.
\end{array}$$
Therefore, we conclude that the rank of the distributions $\Tan^\perp
M_i$ and $\Tan^\perp_{M_{i+1}} M_i\cap \Tan M_{i+1}$ is constant if and
only if the rank of the distributions $\Tan^\perp P_i$ and
$\Tan_{P_{i+1}}^\perp P_i \cap \Tan P_{i+1}$ is constant. We will assume
that the distributions $\Tan^\perp P_i$ and $\Tan^\perp_{P_{i+1}}P_i\cap
\Tan P_{i+1}$ have constant rank.

 Denote by $\vf^\perp(P_i)$
 the set of sections of the vector bundle $\Tan^\perp P_i\to P_i$.
 Then, as a direct application of Proposition \ref{cons}
 and Theorem \ref{totalcons} to the present case, we have:

\begin{teor}
 Let $(J^{1*}E,\Omega_{{\rm h}_0})$ be the Hamiltonian system associated with
 an almost regular Lagrangian system, and
 consider the sequence $\{ P_i\}$ of submanifolds (\ref{seqsubman1}).
 For every $i\geq 1$, let $r_{(i-1)},r_i$ be the
 ranks of the distributions $\Tan^\perp P_{i-1}$ and
 $\Tan_{P_i}^\perp P_{i-1}\cap\Tan P_i$ respectively,
 and consider a local basis for $\vf^\perp(P_{i-1})$
 (in a neighbourhood of a point $x\in P_i$)
 $$
 \{\tilde{\cal Z}^{(i-1)}_1,\ldots ,\tilde{\cal Z}^{(i-1)}_{r_{(i-1)}},
 \tilde{\cal Z}^{(i-1)}_{r_{(i-1)}+1},\ldots ,\tilde{\cal Z}^{(i-1)}_{r_i}\}
 $$
 where \dst\{\tilde{\cal Z}^{(i-1)}_j\}_{j=1,...,r_{(i-1)}}\)
 are tangent to $P_i$. Finally, let
 $X_{{\rm h}_0}=X_{{\rm h}_0}^{(i-1)}+Y_{{\rm h}_0}^{(i-1)}$ be the
 general solution on the submanifold $P_i$,
 where $X_{{\rm h}_0}^{(i-1)}$ and $Y_{{\rm h}_0}^{(i-1)}$ denote
 the determined and the undetermined parts of the solutions, respectively.
 \ben
 \item
 The submanifold $P_1$, where the Hamiltonian equations
 are compatible can be defined as the zero set of the so-called
 {\rm $1st$-generation Hamiltonian constraints},
 $\{\xi^{(1)}\}\subset\Cinfty ({\cal P})$, which are characterized as
 $$
 \xi^{(1)}=\inn(\tilde{\cal Z}^{(0)})(\eta^0-\gamma_{{\rm h}_0})
 \ ,\
 \hbox{for every}\ \tilde{\cal Z}^{(0)}\in\vf^\perp({\cal P})\; .
 $$
 \vspace{-15pt}

\item
 Every submanifold $P_{i+1}$ in this sequence
 can be defined (in $P_i$) as the zero set of the so-called
 {\rm $(i+1)$th-generation Hamiltonian constraints},
 $\{\xi^{(i+1)}\}\subset\Cinfty ({\cal P})$, which are characterized
 in the following equivalent ways:
 \ben
 \item
 For every $\tilde{\cal Z}^{(i)}\in\vf^\perp(P_i)$,
 $$
 \xi^{(i+1)}=\inn(\tilde{\cal Z}^{(i)})(\eta^0-\gamma_{{\rm h}_0})\;
 .
 $$
 \item
 For every $j\in \{1,\dots , r_{(i-1)}\}$,
 $$
 \xi^{(i+1)}=
 \Lie(X_{{\rm h}_0}^{(i-1)})
 [\inn(\tilde{\cal Z}_j^{(i-1)})(\eta^0-\gamma_{{\rm h}_0})]\; .
 $$
 \item
 For every $j\in \{1,\dots ,r_{(i-1)}\}$,
 $$
 \xi^{(i+1)}=
 \inn([X_{{\rm h}_0}^{(i-1)},\tilde{\cal Z}_j^{(i-1)}])\gamma_{{\rm h}_0}+
 \Lie(\tilde{\cal Y})[\inn(\tilde{\cal Z}_j^{(i-1)})\gamma_{{\rm h}_0}]\; .
 $$
 \een
 \item
 For every $k\in \{r_{(i-1)}+1,\dots ,r_{i}\}$ the stability
 condition
 $$
 0= \Lie(X_{{\rm h}_0}^{(i-1)}+Y_{{\rm h}_0}^{(i-1)})
 [\inn(\tilde{\cal Z}_k^{(i-1)})(\eta^0-\gamma_{{\rm h}_0})]\vert_{P}
 $$
 determines (partially or totally) $Y_{{\rm h}_0}^{(i-1)}$ (on $P_i$).
 \een
 \label{hamcons}
 \end{teor}

\subsection{The second order differential equation
problem}\label{seccion4.5}

Assume that $(J^1E,\Omega_\Lag)$ is an almost regular Lagrangian
system and that $M_f$ is the final constraint submanifold (in the
Lagrangian setting). Then there exists a vector field $X_{\cal L}^f$
 on $M_f$ such that
 $\inn(X_\Lag^f)\Omega_\Lag\vert_{M_f}=0$, and
$\inn(X_\Lag^f)\eta\vert_{M_f}=1$.
 But $X_\Lag^f$ will not in
general be a SODE on $M_f$, and thus the integral curves of
$X_\Lag^f$ will not satisfy, in general, the Euler-Lagrange
equations. In order to solve this problem, we will construct a
submanifold $S$ of $M_f$ where there exists a unique vector field
$X_\Lag^S$ such that $X_\Lag^S$ is a SODE on $S$ and
 $$
\inn(X_\Lag^S)\Omega_\Lag\vert_S=0\quad ,\quad
 \inn(X_\Lag^S)\eta\vert_S=1.
$$
  From Theorem \ref{eqsol}, we know that we can choose the vector field
$X_\Lag^f$ on $M_f$ such that it projects via ${\cal F}\Lag_f$
onto a vector field $X_{{\rm h}_0}^f$ on $P_f$. Then we consider
the subset $S$ of $M_f$ defined by
\begin{equation}\label{31'}
S=\{\bar x\in M_f/ ({\cal J}(X_\Lag^f))(\bar x)=0\}.
\end{equation}

In \cite{LMM-96b}, it was proved that
\begin{equation}\label{320}
 {\cal J}(X_\Lag^f)(\bar x)\in\ker({\cal FL}_f)_{*\bar x}
=\ker({\cal FL}_0)_{*\bar x},\makebox[1cm]{} \mbox{ for every
}\bar x\in M_f
\end{equation}
and that for every $\bar x\in M_f$,
 $S\cap {\cal FL}_f^{-1}({\cal FL}_f(\bar x))=
 S\cap {\cal FL}_0^{-1}({\cal FL}_0(\bar x))$ is a
point  $s_{X_{\cal L}^f}(\tilde{x}).$ This point is characterized
by the condition
\[
\iota(s_{X_\Lag^f}(\tilde{x}))=(\bar{\pi}^1)_{*x}(X_\Lag^f(\bar
x)),
\]
where $\iota:J^1E\rightarrow \Tan E$ is the canonical embedding (see
Section \ref{seccion3.1}).

The above result allows us to introduce a well-defined map
$s_{X_\Lag^f}:P_f\rightarrow M_f$ such that $S=s_{X_\Lag^f}(P_f)$
and ${\cal F}\Lag_f\circ s_{X_\Lag^f}=Id.$ In fact,
$s_{X_\Lag^f}:P_f\to M_f$ is a global section of the submersion
${\cal F}\Lag_f:M_f\rightarrow P_f$ and therefore $S$ is an
embedded submanifold of $M_f$ and the map
$s_{X_\Lag^f}:P_f\rightarrow S$ is a diffeomorphism (for more
details, see \cite{LMM-96b}).

Moreover, we have
\begin{teor}\label{7'}\cite{LMM-96b}
\begin{enumerate}
\item
There exists a unique vector field $X_{\cal L}^S
=(s_{X_\Lag^f})_*(X_{{\rm h}_0}^f)$ tangent to $S$ which satisfies
the following conditions
\[
\inn(X_\Lag^S)\Omega_\Lag\vert_S=0\; , \;
\inn(X_\Lag^S)\eta\vert_S=1\; , \;\;
{\cal J}(X_\Lag^S)\vert_S=0.
\]
\item The integral sections of $X_\Lag^S$ satisfy the Euler-Lagrange
equations.
\end{enumerate}
\end{teor}

Next, we will give a local description of the submanifold $S$ and
of the vector field $X_{\cal L}^S.$

If $m$ is the dimension of the fibers of the submersion ${\cal
FL}_0:J^1E\to {\cal P}$  then it is clear that the rank of the
partial Hessian matrix
$\left(\displaystyle\frac{\partial^2\lag}{\partial v^\nu\partial
v^\rho}\right)$ is $n-m$ (see (\ref{25'})). We can suppose without
loss of the generality, that the first $n-m$ rows of this matrix
are independent. With this hypothesis and using (\ref{25'}), we
deduce that for every $j\in\{1,\dots, m\}$ there exist local real
functions $\{W_j^i\}_{1\leq i\leq n-m}$ such that $\{W_j\}_{1\leq
j\leq m}$ is a local basis of $\ker({\cal FL}_0)_*,$ where
\begin{equation}\label{32'}
W_j=\frac{\partial}{\partial v^{n-m+j}} + 
W_j^i\frac{\partial}{\partial v^i}, \makebox[1cm]{} \mbox{ for }
j\in \{1,\dots ,m\}.
\end{equation}
Now, if
\[
X_{\cal L}^f=\frac{\partial}{\partial t} +
A^i\frac{\partial }{\partial q^i} + \bar A^i\frac{\partial }{\partial v^i}
\]
then, since $X_{\cal L}^f$ is ${\cal FL}_f$-projectable, it
follows that the functions $A^i$ are constant on the fibers of
${\cal FL}_f:M_f\to P_f.$ But, as ${\cal FL}_f^{-1}({\cal
FL}_f(\bar x))={\cal FL}_0^{-1}({\cal FL}_0(\bar x))$ for every
$\bar x\in M_f$ (see Theorem \ref{flprolagcons}), we obtain that
\begin{equation}\label{32''}
W_j\vert_{M_f}(A^i)=0,\makebox[1cm]{} \mbox{ for } j\in \{1,\dots ,m\}
\mbox{ and } i\in \{1,\dots ,n\}.
\end{equation}

Furthermore,
 \[ {\cal J}(X_{\Lag}^f)=(A^i-v^i)\frac{\partial }{\partial v^i}.\]
Thus, from (\ref{320}) and (\ref{32'}), we have that
\begin{equation}\label{32'''}
{\cal J}(X_\Lag^f)=(A^{n-m+j}-v^{n-m+j})W_j\vert_{M_f}.
\end{equation}
Note that the functions
\[
\xi_j^S=A^{n-m+j}-v^{n-m+j}, \;\;\; 1\leq j\leq m
\]
are independent on $M_f$. In fact (see (\ref{31'}) and
(\ref{32''})), \[ d\xi_j^S(W_i\vert_{P_f})=-\delta_{ij}.\]

 Moreover,
using (\ref{31'}) and (\ref{32'''}), we conclude that
$\{\xi_j^S\}_{1\leq j\leq m}$ is a set of local independent
constraint functions defining $S$ as a submanifold of $M_f$, that
is,
\[
S=\{\bar x\in M_f/ (A^{n-m+j}-v^{n-m+j})(\bar x)=0, \forall j\in
\{1,\dots ,m\}\}.
\]
Finally, a direct computation proves that the vector field
$X_\Lag^S$ on $S$ is given by
\[
X_{\Lag}^S=X^f_{\Lag}\vert_S-X^f_{\cal L}\vert_S(\xi_j^S)W_j\vert_S=
X^f_\Lag\vert_S-(X^f_\Lag\vert_S(A^{n-m+j})-\bar{A}^{n-m+j})W_j\vert_S.
\]

\section{Finding Euler-Lagrange vector fields.
            Dynamical and SODE Lagrangian constraints}
 \protect\label{sodeprob}

 In Section \ref{seccion4.5} it was shown that there exists an
 embedded submanifold $S$ of $M_f$ and a unique vector field
 tangent to $S$ which is a SODE and verifies the Euler-Lagrange
 equations. However, there is another different  approach,
 consisting in imposing the SODE condition on each step
 of the constraint algorithm. Next we will develop this procedure,
 which is a generalization of the method given in \cite{MR-92} for
 autonomous Lagrangian systems (see also \cite{CLR-87}).

\subsection{SODE condition: SODE first order generation
constraints}

 In Section \ref{lf} we have applied the results of Section \ref{gencase}
 to obtain
 a constraint submanifold $M_1$ such that there exist vector fields
 $\{X_\Lag^{(0)}+Y_\Lag^{(0)}\; ,~Y_\Lag^{(0)}\in\ker\Omega_\Lag\cap\ker\eta\}$
 satisfying the dynamical Lagrangian equations (\ref{eq01}) or,
 equivalently,
 \begin{equation}
 \label{oferta}
 \inn(X_\Lag^{(0)}+Y_\Lag^{(0)})\omega_\Lag|_{M_1} =
 -\gamma_\Lag|_{M_1}~~~,~~~ \inn(X_\Lag^{(0)}+Y_\Lag^{(0)})\eta|_{M_1}=1\; .
 \end{equation}
However, in general, these solutions on the constrained
submanifold $M_1$ do not satisfy the SODE condition on $M_1$.
Hence, we will look for the points in $M_1$ where such a solution
exists.

Thus, we consider the submanifold
\[
S_1=\{ \bar x \in
M_1\mid\exists\tilde{Y}_\Lag^{(0)}\in\ker\Omega_\Lag\cap\ker\eta~\hbox{such
that}~({\cal J}(X_\Lag^{(0)}+\tilde{Y}_\Lag^{(0)}))(\bar x)=0\}\; .
\]
That is, $S_1$ is the maximal set of points of $M_1$ where there
exist a vector field
$D=X_\Lag^{(0)}+\tilde{Y}_\Lag^{(0)}\in\vf(J^1E;S_1)$ such that
\begin{description}
\item[(a)]
 $\inn(D)\omega_\Lag|_{S_1}=-\gamma_\Lag|_{S_1}$,
$\inn(D)\eta|_{S_1}=1$, and
 \item[(b)]
 $D$ satisfies the SODE condition on $S_1$
 (${\cal J}(D)|_{S_1}=0$ and $\inn(D)\eta|_{S_1}=1$).
\end{description}

Notice that, if $D\in\vf(J^1E)$ is a vector field satisfying (a)
and (b), and $W\in\ker\omega_\Lag\cap\ker\pi^1_*$, then $D+W$
satisfies (a) and (b) too. In particular, since
$\ker\Omega_\Lag\cap\ker\eta \subset \ker\omega_\Lag\cap\ker\eta$,
then $\ker\Omega_\Lag\cap\ker\pi^1_* \subset
\ker\omega_\Lag\cap\ker\pi^1_*$. Hence, if
$W\in\ker\Omega_\Lag\cap\ker\pi^1_*$, $D+W$ also satisfies (a) and
(b). Conversely, if $D_1$ and $D_2$ satisfy (a) and (b), then
$D_1-D_2$ is an element of $\ker\Omega_\Lag\cap\ker\pi^1_* \subset
\ker\omega_\Lag\cap\ker\pi^1_*$.

Now, we are going to describe the submanifold $S_1$ inside $M_1$
as the vanishing of a family of functions. For this purpose, we will
use a particular kind of connections:

\begin{definition}
A connection $\bar{\nabla}=\eta\otimes{\cal Y}$ in $\bar{\pi}^1 \colon
J^1E\to B$ is said to be a {\rm second-order connection} if ${\cal
Y}$ is a SODE.
\end{definition}

For second-order connections, we will prove the following results:

 \begin{lem}
 \label{con2or}
Let $\bar{\nabla} = \eta\otimes{\cal Y}$ be a second order connection in
the fibered manifold $\bar{\pi}^1 \colon J^1E\to B$. Then the 2-form
$\omega_\Lag=\Omega_\Lag-\eta\wedge\gamma_\Lag$ satisfies
\[ \omega_\Lag ({\cal J}(X),Y) + \omega_\Lag (X,{\cal J}(Y))=0\;,
\] for every $X,Y\in\vf(J^1E)$.
\end{lem}
 \proof
 First, notice that for every $X\in\vf(J^1E)$,
$\inn({\cal J}(X))\eta=0$ (as ${\cal J}(X)\in\ker\pi^1_*$).
 Moreover, we know that for
every $X,Y\in\vf(J^1E)$ we have $\Omega_\Lag ({\cal J}(X),Y) +
\Omega_\Lag (X,{\cal J}(Y))=0$ (see \cite{IM-92}), hence
 $$
\inn({\cal J}(X))\gamma_\Lag = \inn({\cal J}(X))\inn({\cal
Y})\Omega_\Lag = - \inn(X)\inn({\cal J}({\cal Y}))\Omega_\Lag =0
 $$
 since ${\cal J}(Y)=0$ (${\cal Y}$ is a SODE). Then,
$(\eta\wedge\gamma_\Lag)({\cal J}(X),Y)=0$. Therefore,
\[
\omega_\Lag ({\cal J}(X),Y) = \Omega_\Lag ({\cal J}(X),Y)+
(\eta\wedge\gamma_\Lag)({\cal J}(X),Y) = -\Omega_\Lag(X,{\cal
J}(Y) = -\omega_\Lag(X,{\cal J}(Y)) \; .
\]
 \qed

 \begin{lem}
 \label{con2or2}
Let ${\cal M}$ be the subset of $\vf(J^1E)$ defined by
\[
{\cal M}=\{ Z\in\vf(J^1E)\mid {\cal J}(Z)\in\ker{\cal F}\Lag_*\}
\]
and assume that the connection $\bar{\nabla} = \eta\otimes{\cal Y}$ is
of second order (${\cal J}({\cal Y})=0$). Then, given
$X\in\vf(J^1E)$ verifying $\inn(X)\inn(Z)\omega_\Lag=0$, for every
$Z\in{\cal M}$, there exists a $\pi^1$-vertical vector field $V$
such that $X-V\in\ker\omega_\Lag$. Such a vector field $V$ is
unique up to an element of $\ker{\cal F}\Lag_*$.
 \end{lem}
 \proof
 Recall that $\ker{\cal F}\Lag_*=
\ker\Omega_\Lag\cap\ker\pi^1_*$. First, notice that if $\bar{\nabla}$
is a second order connection, then $\ker{\cal F}\Lag_* =
\ker\Omega_\Lag\cap\ker\pi^1_* = \ker\omega_\Lag\cap\ker\pi^1_*$.
We know that $\ker\Omega_\Lag\cap\ker\pi^1_* \subset
\ker\omega_\Lag\cap\ker\pi^1_*$. But if
$V\in\ker\omega_\Lag\cap\ker\pi^1_*$, then there exists a vector
field $U\in\vf(J^1E)$ such that ${\cal J}(U)=V$ ($V$ is
$\pi^1$-vertical). Then
 \beann
\inn(V)\Omega_\Lag &=& \inn(V)\omega_\Lag +
\inn(V)(\eta\wedge\gamma_\Lag) =
 - (\inn(V)\inn({\cal Y})\Omega_\Lag)\eta
 \\ &=&
 - (\inn({\cal J}(U))\inn({\cal Y})\Omega_\Lag)\eta
 = (\inn(U)\inn({\cal J}({\cal Y}))\Omega_\Lag)\eta = 0
 \eeann
 (since ${\cal Y}$ is of second order).

 Now, given an arbitrary vector field $X$,
 the necessary and sufficient condition for the equation
 $\inn(V)\omega_\Lag = \inn(X)\omega_\Lag$ to be solved for a
 $\pi^1$-vertical vector field $V\in\vf(J^1E)$ is that for every
 $U\in\vf(J^1E)$ verifying $\inn(U)\inn(V)\omega_\Lag=0$ (with $V$ a
 $\pi^1$-vertical vector field), then
 $\inn(U)\inn(X)\omega_\Lag=0$.
 Now, notice that, if $\inn(U)\inn(V)\omega_\Lag=0$, for every
 $V$, then
 $\inn(U)\inn({\cal J}(Y))\omega_\Lag=-\inn({\cal J}(U))\inn(Y)\omega_\Lag=0$,
 for every $Y\in\vf(J^1E)$; that is, $U\in{\cal M}$. Therefore, if $X$
 verifies $\inn(X)\inn(Z)\omega_\Lag=0$, for $Z\in{\cal M}$,
 then $\inn(U)\inn(X)\omega_\Lag =0$ and equation
 $\inn(V)\omega_\Lag = \inn(X)\omega_\Lag$ has solution. Finally,
 for every couple $V_1,V_2$ of such $\pi^1$-vertical fields, their
 difference is $\pi^1$-vertical and belongs to $\ker\omega_\Lag$.
 \qed

 \begin{lem}
 \label{con2or3}
Let ${\cal M}'$ be the subset of $\vf(J^1E)$ defined by
\[
{\cal M}'=\{ Z\in\vf(J^1E)\mid Z\in\ker\bar{\pi}^1~\hbox{ and }~
{\cal J}(Z)\in\ker{\cal F}\Lag_*\}
\]
and let $\bar{\nabla} = \eta\otimes{\cal Y}$ be a second order
connection. Then for every $X\in\ker\bar{\pi}^1_*$ verifying
$\inn(X)\inn(Z')\omega_\Lag=0$, for every $Z'\in{\cal M}$, there
exists a $\pi^1$-vertical vector field $V$ (unique up to an
element of $\ker{\cal F}\Lag_*$) such that
$X-V\in\ker\omega_\Lag$.
 \end{lem}
 \proof
 First, notice that if $V$ is a $\pi^1$-vertical vector field,
 then there exists a $\bar{\pi}^1$-vertical vector field $Y'$
 such that ${\cal J}(Y')=V$: if $Y\in\vf(J^1E)$ is such that
 ${\cal J}(Y)=V$ and $D$ is an arbitrary SODE, then the vector
 field $Y'=Y-(\inn(Y)\eta)D$ is $\bar{\pi}^1$-vertical and
 ${\cal J}(Y')=V$. Now, the proof is similar to that of the above lemma,
 simply observing that,
 if $\inn(U)\inn(V)\omega_\Lag=0$, for every
 $V$ $\pi^1$-vertical, then
 $\inn(U)\inn({\cal J}(Y'))\omega_\Lag=
-\inn({\cal J}(U))\inn(Y')\omega_\Lag=0$,
 for every $\bar{\pi}^1$-vertical vector field $Y'$,
 that is, $U\in{\cal M}'$.
 \qed

Now, we have the following.

 \begin{prop}
 Let $Y\in\vf(J^1E)$ be such that $X_\Lag^{(0)} +Y$
 satisfies the SODE condition; that is, ${\cal J}(X_\Lag^{(0)} +Y)
 = 0$ and $\inn(X_\Lag^{(0)} +Y)\eta = 1$, and let
 $\bar{\nabla}$ be a second order connection. Then
 \[
S_1=\{ \bar x \in M_1\mid (\inn(Z)\inn(Y)\omega_\Lag)(\bar x)=0\; ,~\forall
Z\in{\cal M}\}\; ,
 \]
\label{submS1}
\end{prop}
\proof
 Consider the set $C=\{ \bar x \in M_1\mid
(\inn(Z)\inn(Y)\omega_\Lag)(\bar x)=0\; ,~\forall Z\in{\cal M}\}$. Then
the proof can be done in three steps:
 \begin{enumerate}
 \item
 $C$ is independent of the chosen vector field $Y$.

Take $T=Y_1-Y_2$, where $Y_1$ and $Y_2$ verify that $X_\Lag^{(0)}
+Y_i$ are SODE for every $i=1,2$.
 Then $T$ is a $\pi_1$-vertical vector
field. So there exists a vector field $U\in\vf(J^1E)$ such that
$T={\cal J}(U)$. Now, using Lemma \ref{con2or} we have
\[
\omega_\Lag(T,Z)=\omega_\Lag({\cal J}(U),Z)=-\omega_\Lag(U,{\cal
J}(Z))=0\; ,
\]
since ${\cal J}(Z)\in\ker{\cal F}\Lag_* =
\ker\Omega_\Lag\cap\ker\pi^1_*=\ker\omega_\Lag \cap \ker\pi^1_*$.
\item
 $C\subset S_1$.

 Let $\bar x \in C$, then
$(\inn(Z)\inn(Y)\omega_\Lag)(\bar x)=0$ for every $Z\in{\cal M}$.  From
Lemma \ref{con2or2} we know that at each point $\bar x \in C$, there
exists a $\pi^1$-vertical vector $v$ such that
$Y_{\bar x}-v\in\ker\omega_\Lag|_{\bar x}$. Hence
$(X_\Lag^{(0)}+Y)_{\bar x}-v=(X_\Lag^{(0)})_{\bar x}+(Y_{\bar x}-v)$
 is of second order on $M_1$, and it is also a solution on $M_1$,
 so $\bar x \in S_1$.
\item
 $S_1\subset C$.

Let $\bar x \in S_1$. Then there exists
$\tilde{Y}_\Lag^{(0)}\in\ker\Omega_\Lag\cap\ker\eta\subset
\ker\omega_\Lag\cap\ker\eta$ such that
$(X_\Lag^{(0)}+\tilde{Y}_\Lag^{(0)})_{\bar x}$
 is a solution of the dynamical
equations and satisfies the SODE condition on $S_1$. In general ${\cal
J}(X_\Lag^{(0)}+\tilde{Y}_\Lag^{(0)})_{\bar x}\neq 0$
 for $\bar x \notin S_1$, but a new
vector field $Y$ can be constructed such that
$Y_{\bar x}=(\tilde{Y}_\Lag^{(0)})_{\bar x}$, if $\bar x \in S_1$, and
 $[{\cal J}(X_\Lag^{(0)}+Y)]_{\bar x}=0$,
for every $\bar x \in J^1E$. Then $(\inn(Z)\inn(Y)\omega_\Lag)(\bar x)=0$,
for every $\bar x \in S_1$ (since $[\inn(Y)\omega_\Lag](\bar x)=
[\inn(\tilde{Y}_\Lag^{(0)})\omega_\Lag](\bar x)=0$,
 for every $\bar x\in S_1$), so $\bar x \in C$.
\qed
\end{enumerate}

\begin{corol}
 With the hypothesis of the above proposition, for
 every SODE $D$ we have
\[
S_1=\{ \bar x \in M_1\mid
(\inn(Z)\inn(X_\Lag^{(0)}-D)\omega_\Lag)(\bar x)=0,~\forall Z\in{\cal
M}\}\; .
\]
\end{corol}
\proof
 As $Y=X_\Lag^{(0)}-D$, the proof is trivial.
 \qed

Given $Y\in\vf (J^1E)$, such that $X_\Lag^{(0)}+Y$ satisfies the
SODE condition, for every $Z\in{\cal M}$, the function $$
\phi^{(1)}:=\inn(Z)\inn(Y)\omega_\Lag\in\Cinfty(J^1E) $$ (or
equivalently $\phi^{(1)}=\inn(Z)\inn(X_\Lag^{(0)}-D)\omega_\Lag$,
for a SODE $D\in\vf (J^1E)$), is called a {\sl $1st$-generation
SODE Lagrangian constraint}.

\subsection{General expressions of first order generation constraints.
 ${\cal F}\Lag$-projectability}

 Using a second order connection $\bar{\nabla}$ to make the splitting of the
 form $\Omega_\Lag$, we can give a unified description
 of both the dynamical and SODE Lagrangian constraints.

 \begin{teor}
 For any SODE $D$ we have
\[
S_1=\{ \bar x \in J^1E\mid [\inn(Z)(\inn(D)\omega_\Lag +
\gamma_\Lag)](\bar x)=0,~\forall Z\in{\cal M}\}\; .
\]
 The function $\inn(Z)(\inn(D)\omega_\Lag+\gamma_\Lag))$ is called
 a {\rm $1st$-generation Euler-Lagrange constraints}, and
 $S_1$ is called the {\rm submanifold of
 $1st$-generation Euler-Lagrange constraints}.
 \label{s1enj1}
\end{teor}
\proof
 Let $\bar x \in M_1$. If $D=X_\Lag^{(0)}+Y$ for an arbitrary
 vector field $Y$, then
\[
[\inn(Z)(\inn(D)\omega_\Lag + \gamma_\Lag)](\bar x) =
[\inn(Z)(\inn(X_\Lag^{(0)})\omega_\Lag + \inn(Y)\omega_\Lag +
\gamma_\Lag)](\bar x) = [\inn(Z)\inn(Y)\omega_\Lag](\bar x)
\]
which are the SODE constraints defining $S_1$ as a submanifold of
$M_1$.

Next, we prove that, if $Z\in\vf^\perp(J^1E)$, then $Z\in{\cal
M}$; that is, ${\cal J}(Z)\in\ker{\cal
F}\Lag_*=\ker\omega_\Lag\cap \ker\pi^1_*$. In fact, we know that
$Z\in\vf^\perp(J^1E)$ if, and only if,
\[
\inn(Z)\inn(U)\omega_\Lag + (\inn(U)\eta)(\inn(Z)\eta)=0\; ,
\]
for every $U\in\vf(J^1E)$. Then $Z\in{\cal M}$ because
\[
\inn(U)\inn({\cal J}(Z))\omega_\Lag=- \inn({\cal
J}(U))\inn(Z)\omega_\Lag = (\inn(Z)\eta)(\inn({\cal
J}(U))\eta)=0\; ,
\]
Taking this into account, we obtain
\[
\inn(Z)(\inn(D)\omega_\Lag+\gamma_\Lag) =
\inn(Z)\inn(D)\omega_\Lag + \inn(Z)\gamma_\Lag =
-(\inn(Z)\eta)(\inn(D)\eta) + \inn(Z)\gamma_\Lag =
\inn(Z)(\gamma_\Lag - \eta)\; ,
\]
which are the dynamical constraints defining $M_1$ as a
submanifold of $J^1E$. \qed

Summarizing, we have arrived at a submanifold $S_1$ of $J^1E$ and
a family of vector fields
 \beq
\{ {\mit\Gamma}_\Lag:=
 X_\Lag^{(0)}+\tilde{Y}_\Lag^{(0)}+V^{(0)};~ V^{(0)}\in\ker{\cal F}\Lag_*\}
 \label{family}
 \eeq
such that
 $$
 [\inn({\mit\Gamma}_\Lag)\Omega_\Lag]|_{S_1}=0 \quad , \quad
 [\inn({\mit\Gamma}_\Lag)\eta]|_{S_1}=1 \quad , \quad
 [{\cal J}({\mit\Gamma}_\Lag)]|_{S_1}=0\; .
 $$
 Observe that $X_\Lag^{(0)}$ and $\tilde{Y}_\Lag^{(0)}$ are
 fixed vector fields, whereas $V^{(0)}\in\ker\,{\cal F}\Lag_*$
 is arbitrary.

 The submanifold $S_1$ is characterized as
 \[
S_1=\{ \bar x \in J^1E\mid [\inn(Z)(\inn(D)\omega_\Lag +
\gamma_\Lag)](\bar x)=0,~\forall Z\in{\cal M}\}\; ,
\]
for any SODE $D$. The constraints that define $S_1$ are of two
different kinds: \bit
 \item
 Lagrangian dynamical constraints:
 $\zeta^{(1)}:=\inn({\cal Z}^{(0)})(\eta-\gamma_\Lag)$,
 ${\cal Z}^{(0)}\in\vf^{\perp}(J^1E)$.
\item
 SODE Lagrangian constraints:
 $\phi^{(1)}:=\inn(Z)\inn(Y)\omega_\Lag$, $Z\in{\cal M}$,
 $Z\not\in\vf^{\perp}(J^1E)$, $Y\in\vf (J^1E)$ with $X^{(0)}_\Lag+Y$
 a SODE.
\eit
 We know that the Lagrangian dynamical constraints can be expressed as
 ${\cal F}\Lag$-projectable functions (Theorem \ref{flprolagcons}).
 But this is not the situation for the SODE Lagrangian constraints, as will
 now be proved. First, the following results are required:

 \begin{lem} For every $V\in\ker{\cal F}\Lag_*$, there exists
 $Z'\in{\cal M}'$ such that ${\cal J}(Z')=V$ and $Z'$ is ${\cal
 F}\Lag$-projectable.
 \label{Zpro}
 \end{lem}
 \proof
 It is evident that if $V\in\ker\pi^1_*$ is ${\cal
 F}\Lag$-projectable, then we can choose an ${\cal
 F}\Lag$-projectable $\bar{\pi}^1$-vertical vector field $Y'$ with
 ${\cal J}(Y')=V$ (in fact, if $Y\in\vf(J^1E)$ is such that ${\cal
 J}(Y)=V$ and $D\in\vf(J^1E)$ is a SODE. Then
 $Y'=Y-(\inn(Y)\eta)D$ and the only problem is to take the
 vertical parts of $Y$ and $D$, which are arbitrary). In
 particular, for every $V\in\ker{\cal F}\Lag_*$, we have ${\cal
 F}\Lag_*(V)=0$, which implies they are ${\cal F}\Lag$-projectable,
 hence there exists some $Z'\in{\cal M}'$ which is ${\cal
 F}\Lag$-projectable and ${\cal J}(Z')=V$.
 \qed

 \begin{lem}
 If $D\in \vf(J^1E)$ is a SODE, then ${\cal J}([V,D])=V$, for every
 $\pi^1$-vertical vector field $V$.
 \label{jvdd}
 \end{lem}
 \proof
 This follows by using the local expressions of the vertical
 endomorphism ${\cal J}$, a SODE and a $\pi^1$-vertical vector
 field.
 \qed

 \begin{lem}
 If $D\in\vf(J^1E)$ is a SODE, then $[V,D]\in{\cal M}$, for every
 $V\in{\cal F}\Lag_*$. Moreover,
 \[
 [\ker{\cal F}\Lag_*,D] + \ker\pi^1_* = {\cal M}'\; .
 \]
 \label{dvm}
 \end{lem}
 \proof
 The first part is a straighforward consequence of the above
 lemma. For the second part, notice that, for every $Z'\in{\cal M}'$,
 ${\cal J}(Z')$ is $\pi^1$-vertical, hence ${\cal J}(Z') = {\cal
 J}([{\cal J}(Z'),D])\in\ker{\cal F}\Lag_*$, i.e., $Z'-[{\cal
 J}(Z'),D]$ is a vertical vector field.
 \qed

 Now, we prove the non-projectability of the
 SODE Lagrangian constraints.

 \begin{prop}
 If $\phi^{(1)}$ is a $1st$-generation SODE Lagrangian constraint,
 then it cannot be expressed as a ${\cal F}\Lag$-projectable
 function.
 \label{noprojecSODE}
 \end{prop}
 \proof
 If $\phi^{(1)}$ is a $1st$-generation SODE Lagrangian constraint,
 then according to Theorem \ref{s1enj1} it can be expressed as
 \[
 \phi^{(1)}=\inn(Z)(\inn(D)\omega_\Lag+\gamma_\Lag)\; ,
 \]
 where $Z\in{\cal M}$, $Z\not\in\vf^{\perp}(J^1E)$ and
 $D\in\vf(J^1E)$ is a SODE.

 First, notice that $Z\not\in\vf^{\perp}(J^1E) +
 \ker\pi^1_*$. If $V$ is a $\pi^1$ vertical vector field
 then $\exists Y\in\vf(J^1E)$ such that ${\cal J}(Y)=Z$ and
 \[
 \inn(Z)(\inn(D)\omega_\Lag+\gamma_\Lag) =
 \inn({\cal J}(Y))(\inn(D)\omega_\Lag+\inn({\cal
 Y})\Omega_\Lag) = - \inn(Y)\inn({\cal J}(D))\omega_\Lag -
 \inn(Y))\inn({\cal J}({\cal Y}))\Omega_\Lag = 0
 \]
 since $D$ and ${\cal Y}$ are SODEs. Therefore, if
 $Z\in\vf^{\perp}(J^1E) + \ker\pi^1_*$, then we obtain a
 Lagrangian dynamical constraint. As a corollary,
 $Z\not\in\ker\pi^1_*$.

 Moreover, $Z$ can be chosen as an element of ${\cal M}'$
 (i.e., $Z\in{\cal M}$ and $Z\in\ker\bar{\pi}^1_*$)
 ${\cal F}\Lag$-projectable. If $Z$ is a
 SODE, ${\cal J}(Z)=0$ and $Z\in{\cal M}$. But
 \[
 \inn(Z)(\inn(D)\omega_\Lag+\gamma_\Lag) =
 \inn(Z)(\inn(D)\Omega_\Lag-\inn(D)(\eta\wedge\gamma_\Lag) +
 \gamma_\Lag) =
 \inn(Z)\inn(D)\Omega_\Lag - \inn(Z)\gamma_\Lag +
 \inn(D)\gamma_\Lag + \inn(Z)\gamma_\Lag)
  \]
 \[
 = \inn(Z)\inn(D)\Omega_\Lag + \inn(D)\inn({\cal Y})\Omega_\Lag =
 \inn(D)\inn({\cal Y}-Z)\Omega_\Lag = 0
 \]
 since ${\cal Y}-Z$ is $\pi^1$-vertical. If
 $Z\not\in\ker\bar{\pi}^1_*$ and it is not a SODE, then the SODE
 Lagrangian constraint $\phi^{(1)}$ can be expressed in an
 equivalent way as
 \[
 \phi^{(1)}=\inn(Z')(\inn(D)\omega_\Lag+\gamma_\Lag)\; ,
 \]
 where $Z'=Z-(\inn(Z)\eta)D$ is $\bar{\pi}^1$-vertical; that is,
 $Z'\in{\cal M}'$. Finally, from Lemma \ref{Zpro} we know that
 vector fields in ${\cal M}'$ can be chosen as ${\cal
 F}\Lag$-projectable (notice that ${\cal F}\Lag$-projectable and
 ${\cal F}\Lag$-nonprojectable vector fields in ${\cal M}'$ give
 rise to the same constraints, since they differ on a vertical
 vector field).

 Now, suppose that $\phi^{(1)}$ is ${\cal F}\Lag$-projectable.
 Then for every $V\in{\cal F}\Lag_*$, we have
 \[
 0 = V(\phi^{(1)}) = V(\inn(Z')(\inn(D)\omega_\Lag+\gamma_\Lag))
 = V[\inn(Z')(\inn(D)\Omega_\Lag - \inn(D)(\eta\wedge\gamma_\Lag) +
 \gamma_\Lag)] = V(\inn(Z')\inn(D)\Omega_\Lag)\; ,
 \]
 since $V[\inn(Z')(\gamma_\Lag -
 \inn(D)(\eta\wedge\gamma_\Lag))] = V[(\inn(Z')(\gamma_\Lag -
 \gamma_\Lag))]=0$ ($Z'$ is $\bar{\pi}^1$ vertical).

 Therefore,
 \[
 0=V(\phi^{(1)}) = V(\inn(Z')\inn(D)\Omega_\Lag) =
 -V(\inn(D)\inn(Z')\Omega_\Lag) = -\inn([V,D])\inn(Z')\Omega_\Lag -
 \inn(D)\Lie(V)(\inn(Z')\Omega_\Lag)
 \]
 but since $Z'$ has been chosen to be ${\cal F}\Lag$-projectable
 and $\Omega_\Lag$ is also ${\cal F}\Lag$-projectable, then \\
 $\Lie(V)(\inn(Z')\Omega_\Lag)=0$. Therefore,
 \[
 0=V(\phi^{(1)}) = -\inn([V,D])\inn(Z')\Omega_\Lag\; .
 \]

 Again using Lemma \ref{dvm}, we know that for every $X'$ in ${\cal M}'$,
 $X'=[V,D]+W$, for some $V\in\ker{\cal F}\Lag_*$ and
 $W\in\ker\pi^1_*$, so we have
 \[
 \inn(X')\inn(Z')\omega_\Lag = \inn(X')\inn(Z')\Omega_\Lag -
 \inn(X')\inn(Z')(\eta\wedge\gamma_\Lag) =
 \inn(X')\inn(Z')\Omega_\Lag =
 \]
 \[
 = \inn([V,D])\inn(Z')\Omega_\Lag +
 \inn(W)\inn(Z')\Omega_\Lag = \inn([V,D])\inn(Z')\Omega_\Lag = 0
 \]
 since $\inn(X')\eta =\inn(Z')\eta=0$ (they are
 $\bar{\pi}^1$-vertical vector fields) and
 $\inn(W)\inn(Z')\Omega_\Lag= \inn({\cal J}(Y))\inn(Z')\Omega_\Lag
 = -\inn(Y)\inn({\cal J}(Z'))\Omega_\Lag=0$ for some
 $Y\in\vf(J^1E)$ (because ${\cal J}(Z')\in\ker{\cal
 F}\Lag_*\subset\ker\Omega_\Lag$). Hence we conclude that
 $\inn(X')\inn(Z')\omega_\Lag=0$, for every $X'\in{\cal M}'$, and
 therefore, from Lemma \ref{con2or3}, there exists a $\pi^1$-vertical
 vector field $V$ such that $Z'- V\in\ker\omega_\Lag$. That is,
 $Z'\in \ker\omega_\Lag+\ker\pi^1_*$. But, for
 $\bar{\pi}^1$-vertical vector fields, $\ker\omega_\Lag =
 \vf^{\perp}(J^1E)$, so $Z'\in\vf^{\perp}(J^1E) + \ker\pi^1_*$, which
 contradicts the hypothesis.
 \qed

\subsection{Stability condition: new generations of constraints}
\protect\label{seccion5.3}

 In general, none of the fields of the family (\ref{family})
 of solutions on $S_1$ is tangent to this submanifold.
 So we must search for the points of $S_1$ where a vector field
 $V^{(0)}\in\ker\,{\cal F}\Lag_*$ exists such that
 ${\mit\Gamma}_\Lag=X_\Lag^{(0)}+\tilde{Y}_\Lag^{(0)}+V^{(0)}$
 is tangent to $S_1$. Then we define:
 \beann
 S_2 &=& \{ \bar x \in S_1\ \mid\ \exists V^{(0)}\in\ker\,{\cal F}\Lag_*\ ,
 \ (X_\Lag^{(0)}+\tilde{Y}_\Lag^{(0)}+V^{(0)})_{\bar x}\in\Tan_{\bar x}S_1\}
 \\ &=&
 \{ \bar x \in S_1\,\mid\,\exists V^{(0)}\in\ker\,{\cal F}\Lag_*\
 \hbox{such that}
\\ & &
 [\Lie(X_\Lag^{(0)}+\tilde{Y}_\Lag^{(0)}+V^{(0)})\zeta^{(1)}](\bar x)=0,\
 \forall\zeta^{(1)},\ \hbox{and}\
 [\Lie(X_\Lag^{(0)}+\tilde{Y}_\Lag^{(0)}+V^{(0)})\phi^{(1)}](\bar x)=0, \
 \forall\phi^{(1)}\}
 \eeann
 where we recall that the $1$st-generation Euler-Lagrange constraints
 are given by
 \beann
 \zeta^{(1)}:=\inn({\cal Z}^{(0)})(\eta-\gamma_\Lag) &\ , \ &
 \forall {\cal Z}^{(0)}\in\vf^{\perp}(J^1E)\ , ~~\hbox{and}
 \\
 \phi^{(1)}:=\inn(Z)\inn(Y)\omega_\Lag  &\ , \ &
 \forall Z\in{\cal M},\ Z\not\in\vf^{\perp}(J^1E),\ Y\in\vf (J^1E),\
 X^{(0)}_\Lag+Y\ \hbox{SODE}\; .
 \eeann
 But, as the Lagrangian dynamical constraints can be expressed as
 ${\cal F}\Lag$-projectable functions (Theorem \ref{flprolagcons}),
 and $V^{(0)}\in\ker\, {\cal F}\Lag_*$,
 using this in the stability condition for these constraints we have that
 $\Lie(V)\zeta^{(1)}=0$, and hence
 this condition reduces to
 $$
 [\Lie(X_\Lag^{(0)}+\tilde{Y}_\Lag^{(0)})\zeta^{(1)}](\bar x)=0 \quad , \quad
 \hbox{for every}\ \zeta^{(1)}\; .
 $$
 Then we have two options:
 \ben
 \item
 $[\Lie(\tilde{Y}_\Lag^{(0)})\zeta^{(1)}](\bar x)=0$,
 but $[\Lie(X_\Lag^{(0)})\zeta^{(1)}](\bar x)\not=0$,
 for every $\bar x \in S_1$, and for some $\zeta^{(1)}$.
 Then we obtain new constraints of the form
 $$
 \zeta^{(2)}:=\Lie(X_\Lag^{(0)})\zeta^{(1)}\in\Cinfty(J^1E)\; .
 $$
 \item
 $[\Lie(\tilde{Y}_\Lag^{(0)})\zeta^{(1)}](\bar x)\not=0$,
 for every $\bar x\in S_1$, and for some $\zeta^{(1)}$.
 Then we obtain new constraints of the form
 $$
 \phi^{(2)}:=
\Lie(X_\Lag^{(0)}+\tilde{Y}_\Lag^{(0)})\zeta^{(1)}\in\Cinfty(J^1E)\;
 .
 $$
 \een
 All these constraints are called the
 {\sl $2nd$-generation Euler-Lagrange constraints}, and define the
 so-called {\sl submanifold of $2nd$-generation Euler-Lagrange constraints}
 $S_2\hookrightarrow S_1$. Recall that we are assuming
 the hypothesis of Assumption \ref{arank}. Then we have:

 \begin{prop}
 \ben
 \item
 The following subsets of $S_1$ are the same:
 \ben
 \item
 ${\cal A}:=\{ \bar x \in S_1\ \mid\
[\Lie(X_\Lag^{(0)})\zeta^{(1)}](\bar x)=0,
 \ \forall\zeta^{(1)}\ \mid\ [\Lie(\tilde
Y_\Lag^{(0)})\zeta^{(1)}](\bar x)=0 \}$.
 \item
 ${\cal B}:=\{ \bar x \in S_1\ \mid\
 [\inn({\cal Z}^{(1)})(\eta-\gamma_\Lag)](\bar x)=0,
 \ \forall {\cal Z}^{(1)}\in\vf^\perp(S_1)\}$.
\item
 ${\cal C}$: the submanifold of $S_1$ of zeros of
 $2$nd-generation ${\cal F}\Lag$-projectable Lagrangian constraints.
 \een
 (This means that the functions $\{\zeta^{(2)}\}$ defining this submanifold
 are the {\rm $2$nd-generation dynamical Lagrangian constraints}).
 \item
 The functions
 $\phi^{(2)}$ cannot be expressed as ${\cal F}\Lag$-projectable functions,
 and they are called {\rm $2$nd-generation SODE Lagrangian constraints}.
 \een
 \end{prop}
 \proof
 \ben
 \item
 ${\cal B}={\cal C}$:

 In order to prove this we need the following:

 \begin{lem}
 If $\jmath_M\colon M\hookrightarrow J^1E$ is a submanifold defined by
 ${\cal F}\Lag$-projectable
 constraints, and $\jmath_S\colon S\hookrightarrow M$
 is a submanifold defined in $M$
 by non ${\cal F}\Lag$-projectable constraints, then
 $\vf^\perp(M)=\vf^\perp (S)$.
 \end{lem}
 \proof
 Recall that
 \beann
 \vf^\perp(M)&:=&\{ Z\in\vf (J^1E)\ \mid\
 \jmath_M^*[\inn(Z)\omega_\Lag-(\inn(Z)\eta)\eta]=0\}\; ,
 \\
 \vf^\perp(S)&:=&\{ Z\in\vf (J^1E)\ \mid\
 \jmath_S^*\jmath_M^*[\inn(Z)\omega_\Lag-(\inn(Z)\eta)\eta]=0\}\; .
 \eeann
 Then it is obvious that $\vf^\perp(M)\subset\vf^\perp (S)$.

 Conversely, if $Z\in\vf^\perp(S)$, then by definition
 $\jmath_M^*[\inn(Z)\omega_\Lag-(\inn(Z)\eta)\eta]\in\df^1(M)$ is a constraint
 $1$-form for $S$, which is non ${\cal F}\Lag_M$-projectable.
 Therefore, for every $X\in\vf (M,S)$ being tangent to $S$ we have
 $$
 \jmath_S^*\inn(X)[\jmath_M^*(\inn(Z)\omega_\Lag-(\inn(Z)\eta)\eta)]=0\; .
 $$
 Now take a local basis of $\vf^\perp (M)$ consisting of vector fields
 $\{ X_i,Y_j\}$, where $\{ X_i\}$ are tangent to $S$, but
 $\{ Y_j\}$ are not. As $M$ is defined in $J^1E$
 by ${\cal F}\Lag$-projectable constraints, then all the vector fields
 tangent to the fibres of ${\cal F}\Lag$ are tangent to $M$ too.
 On the other hand, as $S$ is defined in $M$
 by non ${\cal F}\Lag$-projectable constraints, these constraints
 remove degrees of freedom in the fibres of ${\cal F}\Lag$,
 and hence $\{ Y_j\}$ are vector fields tangent to those fibres
 transverse to $S$. As a consequence
 $$
 \jmath_S^*\inn(Y_j)[\jmath_M^*(\inn(Z)\omega_\Lag-(\inn(Z)\eta)\eta)]=
 \jmath_S^*\jmath_M^*\inn(\jmath_{M*}Y_j)
 [\inn(Z)\omega_\Lag-(\inn(Z)\eta)\eta]=0
 $$
 since $\inn(\jmath_{M*}Y_j)\inn(Z)\omega_\Lag=0$, because
 $\jmath_{M*}Y_j\in\ker\,{\cal F}\Lag_*\subset\ker\,\omega_\Lag$,
 and $\inn(\jmath_{M*}Y_j)\eta=0$, as $\eta$ is a $\bar\pi^1$-semibasic form
 and $\jmath_{M*}Y_j$ are $\bar\pi^1$-vertical vector fields.
 So, we have obtained that, for every $X\in\vf (M)$
 $$
 \jmath_S^*\inn(X)[\jmath_M^*(\inn(Z)\omega_\Lag-(\inn(Z)\eta)\eta)]=0\; ,
 $$
 therefore $\jmath_M^*[\inn(Z)\omega_\Lag-(\inn(Z)\eta)\eta]=0$,
 and thus $Z\in\vf^\perp (M)$. Hence
 $\vf^\perp(S)\subset\vf^\perp (M)$.
 \qed

 Taking this into account, we have that ${\cal B}$ is just
 the submanifold of $2$nd-generation dynamical Lagrangian constraints
 (see item 2.a of Theorem \ref{lagdyncons}),
 and then the result follows from item 4 of Theorem
 \ref{flprolagcons}.

 \quad ${\cal C}\subset{\cal A}$:

 As a consequence of Theorem \ref{eqsol}, we can take a particular
 ${\cal F}\Lag$-projectable solution $X_\Lag^{(0)}\in\vf (J^1E)$.
 Then, as $\zeta^{(1)}$ are $1$st-generation dynamical Lagrangian
 constraints, they can also be expressed as ${\cal F}\Lag$-projectable
 functions, and hence the result follows.

 \quad ${\cal A}\subset{\cal C}$:

 As a consequence of item 2.b of Theorem \ref{lagdyncons},
 and bearing in mind that
 $\tilde Y_\Lag^{(0)}\in\ker\,\Omega_\Lag\cap\ker\,\eta$,
 we have that every constraint defining ${\cal B}$ is also a constraint
 for ${\cal A}$, and then ${\cal A}\subset{\cal B}={\cal C}$.
 \item
 It is immediate from the above item.
 \qed
 \een

\begin{remark}
{\rm  Observe that the expression of the constraints $\zeta^{(2)}$
 depends only on a particular solution $X_\Lag^{(0)}$ of the
 Lagrangian equation, but the expression of $\phi^{(2)}$
 involves the vector field  $\tilde Y_\Lag^{(0)}$, which
 arises from the SODE condition. This fact justifies the above
 terminology.}
\end{remark}

 Furthermore, the stability condition for the
 $1$st-generation SODE constraints gives (on $S_1$)
 \beq
 0=[\Lie(X_\Lag^{(0)}+\tilde{Y}_\Lag^{(0)}+V^{(0)})\phi^{(1)}](\bar x)=
[\Lie(X_\Lag^{(0)}+\tilde{Y}_\Lag^{(0)})\phi^{(1)}+
\Lie(V^{(0)})\phi^{(1)}](\bar
x)
 \label{linsys}
 \eeq
 which is a system of linear equations for $V^{(0)}$, and we have:

 \begin{lem}
 The system (\ref{linsys}) is compatible at all the points of $S_1$.
 \end{lem}
 \label{lemlinsys}
 \proof
 Locally we can take a finite set of independent
 $1$st-generation SODE constraints, $\moment{\phi}{1}{h}$.
 As we have said, these constraints remove $h$
 degrees of freedom on the fibres of ${\cal F}\Lag$.
 Then the matrix of this linear system for $V^{(0)}$ has maximal rank $h$,
 and hence the system is locally compatible.
 However, for every collection of local solutions, a
 global solution can be constructed on $S_1$
 using a partition of unity on this manifold.
Hence the system is compatible at all the points of $S_1$.
 \qed

 Thus from this system we can determine (total or partially)
 the vector field $V^{(0)}$.
 The stability of SODE constraints does not give new constraints but
 removes degrees of freedom in the vector fields solution.
 The solutions can be written in the form
 $$
 {\mit\Gamma}_\Lag:=X_\Lag^{(0)}+\tilde{Y}_\Lag^{(0)}+\tilde V^{(1)}+V^{(1)}
 $$
 where $\tilde V^{(1)}$ is a solution of (\ref{linsys}), and
 $V^{(1)}\in\ker\,{\cal F}\Lag_*$ is any solution of the system
 $\Lie(V^{(0)})(\phi^{(1)})=0$ (on $S_1$),
 and this contains all the gauge freedom.
 The solution ${\mit\Gamma}_\Lag$ is a SODE and is tangent to
 $S_1$ at the points of $S_2$,
 but in general, it is not tangent to $S_2$.

 This discussion has been carried out for the submanifold $S_1$,
 but it can be extended recursively until no new constraints appear.
 In every step, we have a submanifold $S_i$ ($i>1$), the so-called
 {\sl submanifold of $i$th-generation Euler-Lagrange constraints},
 which is defined by constraints of two kinds:
 \bit
 \item
 $\{\zeta^{(i)}\}$, which are the {\sl $i$th-generation dynamical Lagrangian
 constraints}, and can be expressed as the (only)
 ${\cal F}\Lag$-projectable constraints.
 \item
 $\{\phi^{(i)}\}$,  which are the {\sl $i$th-generation SODE
 Lagrangian constraints}, and are not ${\cal F}\Lag$-projectable functions.
 \eit
 Now, we must take the corresponding vector field solutions (on $S_i$)
 $$
 {\mit\Gamma}_\Lag=X_\Lag^{(0)}+\tilde{Y}_\Lag^{(0)}+\tilde V^{(i-1)}+V^{(i-1)}
 $$
 where $\tilde V^{(i-1)}\in\ker\,{\cal F}\Lag_*$ is a fixed vector field,
 and $V^{(i-1)}\in\ker\,{\cal F}\Lag_*$ is undetermined.
 Therefore, the tangency condition for these
 {\sl $i$th-generation Euler-Lagrange constraints}
 leads to similar results as those obtained for $S_1$.

 In this way, we obtain a sequence of
 constrained submanifolds
 \beq
 \cdots \stackrel{\imath_{i+1}^i}{\hookrightarrow} S_i
 \stackrel{\imath_i^{i-1}}{\hookrightarrow} \cdots
 \stackrel{\imath_2^1}{\hookrightarrow} S_1
 \stackrel{\imath_1^0}{\hookrightarrow} M_1
 \stackrel{\jmath_1}{\hookrightarrow} J^1E
 \label{seqsubsode}
 \eeq
 and this procedure will be called the
 {\sl Euler-Lagrange constraint algorithm}.
 As will be seen in the next subsection, if the final
 dynamical constraint submanifold $P_f$ exists, this algorithm ends
 by giving a submanifold $S_f\hookrightarrow P_f$ which is called the
 {\sl final Euler-Lagrange constraint submanifold}.
 In such a case, there exists a vector field
 $X_\Lag\in\vf(J^1E,S_f)$ tangent to $S_f$ such that
 $$
 [\inn({\mit\Gamma}_\Lag)\Omega_\Lag]|_{S_f}=0 \; , \;
 [\inn({\mit\Gamma}_\Lag)\eta]|_{S_f}=1\; ,\;
 [{\cal J}({\mit\Gamma}_\Lag)]|_{S_f}=0\; .
 $$

 As a summary of all the results given in this section,
 we have proved the following:

 \begin{teor}
 Let $\ls$ be an almost regular Lagrangian system. Consider
 the sequence of submanifolds (\ref{seqsubsode}) and assume that,
 for every $i\geq 1$, the distributions
 $\Tan^\perp S_i$ and $\Tan_{S_{i+1}}^\perp S_i\cap\Tan S_{i+1}$
 have constant rank.
 \ben
 \item
 The submanifold $S_1$, where there exist Euler-Lagrange vector
 fields (solutions  of the dynamical Lagrangian equations
 satisfying the SODE condition)
 can be defined (on $J^1E$) as the zero set of the
 {\rm $1st$-generation Euler-Lagrange constraints},
 which are characterized as
 $$
 \inn(Z)(\inn(D)\omega_\Lag +\gamma_\Lag) \ ,\
 \hbox{for every}\ Z\in{\cal M},\ \hbox{and $D\in\vf (J^1E)$ a
 SODE}\; .
 $$
 These constraints are of two kinds:
 \ben
 \item
 The {\rm $1st$-generation dynamical Lagrangian constraints}
 $\{\zeta^{(1)}\}\subset\Cinfty (J^1E)$, which are characterized as
 $$
 \zeta^{(1)}=\inn({\cal Z}^{(0)})(\eta-\gamma_\Lag)
 \ , \
 \hbox{for every}\ {\cal Z}^{(0)}\in\vf^\perp(J^1E).
 $$
 All of them can be expressed as ${\cal F}\Lag$-projectable functions.
 \item
 The {\sl $1st$-generation SODE Lagrangian constraints}
 $\{\phi^{(1)}\}\subset\Cinfty (J^1E)$, which are characterized as
 $$
 \phi^{(1)}:=\inn(Z)\inn(Y)\omega_\Lag\in\Cinfty(J^1E) \ ,\
 \hbox{for every}\ Z\in{\cal M}
 $$
 or equivalently as
 $$
 \phi^{(1)}=\inn(Z)\inn(X_\Lag^{(0)}-D)\omega_\Lag  \ ,\
 \hbox{for every}\ Z\in{\cal M}
 $$
 where $Y\in\vf (J^1E)$, such that $X_\Lag^{(0)}+Y$ is a SODE,
 and $D\in\vf (J^1E)$ is a SODE.

 None of them can be expressed as a ${\cal F}\Lag$-projectable function.
 \een
 \item
 For every $i\geq 1$,
 the Euler-Lagrange vector fields solution on the submanifold $S_i$
 can be written as
 $$
 {\mit\Gamma}_\Lag:=X_\Lag^{(0)}+\tilde{Y}_\Lag^{(0)}+
 \tilde V^{(i-1)}+V^{(i-1)}
 $$
 where $X_\Lag^{(0)}$ is a particular solution of the dynamical
 Lagrangian equations,
 $\tilde{Y}_\Lag^{(0)}\in\ker\,\omega_\Lag\cap\ker\,\eta$ is
 a fixed vector field such that
 ${\cal J}(X_\Lag^{(0)}+\tilde{Y}_\Lag^{(0)})\vert_{S_i}=0$,
 $\tilde V^{(i)}\in\ker\,{\cal F}\Lag_*$ is also a fixed vector field
 (with $\tilde V^{(0)}=0$), and $V^{(i)}\in\ker\,{\cal F}\Lag_*$
 denotes the undetermined part of the solution.
 \item
 Every submanifold $S_{i+1}$ ($i\geq 1$) in this sequence
 can be defined (in $S_i$) as the zero set of the
 {\rm $(i+1)$th-generation Euler-Lagrange constraints},
 which are of two kinds:
 \ben
 \item
 The {\rm $(i+1)$th-generation dynamical Lagrangian constraints},
 $\{\zeta^{(i+1)}\}\subset\Cinfty (J^1E)$, which are obtained
 by making
 $$
 \zeta^{(i+1)}:=\Lie(X_\Lag^{(0)})\zeta^{(i)}
 $$
 for every $\zeta^{(i)}$ such that
 $[\Lie(\tilde{Y}_\Lag^{(0)})\zeta^{(i)}]\vert_{S_i}=0$.

 All of them can be expressed as ${\cal F}\Lag$-projectable functions.
 \item
 The {\rm $(i+1)$th-generation SODE Lagrangian constraints},
 $\{\phi^{(i+1)}\}\subset\Cinfty (J^1E)$, which are obtained
 by making
 $$
 \phi^{(i+1)}:=\Lie(X_\Lag^{(0)}+Y_\Lag^{(0)})\zeta^{(i)}
 $$
 for every $\zeta^{(i)}$ such that
 $[\Lie(\tilde{Y}_\Lag^{(0)})\zeta^{(i-1)}]\vert_{S_i}=0$.

 None of them can be expressed as a ${\cal F}\Lag$-projectable function.
 \een
 \item
 For every $i$th-generation SODE Lagrangian constraint
 $\phi^{(i)}$ ($i\geq 1$), the stability condition
 $$
 \Lie(X_\Lag^{(0)}+\tilde{Y}_\Lag^{(0)}+\tilde V^{(i-1)}+V^{(i-1)})\phi^{(i)}
 \vert_{S_i}=0
 $$
 determines (partially or totally) the undetermined vector field
 $V^{(i-1)}$ (on $S_i$).
 \een
 \label{lagsodecons}
 \end{teor}

 As an evident consequence of this, we have:

 \begin{corol}
 For every $i$ ($1\leq i\leq f$),
 $S_i$ is a submanifold of $M_i$, and
 ${\cal F}\Lag(S_i)={\cal F}\Lag(M_i)=P_i$.
 \end{corol}

 \begin{remark}
{\rm  For autonomous almost regular mechanical systems,
 there is another relation between the Euler-Lagrange constraints and
 the Hamiltonian constraints, which is established using
 the so-called {\sl time-evolution operator}
 \cite{BGPR-86}, \cite{CL-87}, \cite{GP-89}, \cite{GP-92}.
 The generalization of this relation to the present case is in
 progress.}
 \end{remark}

 \subsection{Properties of Euler-Lagrange vector fields}
\protect\label{pelvf}

Summarizing, in Sections \ref{seccion4.5} and \ref{seccion5.3} we
have given two different procedures for obtaining
submanifolds of $M_f\hookrightarrow J^1E$ where
 Euler-Lagrange vector fields exist (satisfying the tangency condition).
Concerning these procedures, the following must be
pointed out:

 \bit
 \item
 The submanifold $S$ in Section \ref{seccion4.5} is constructed
from a previously chosen (${\cal F}\Lag$-projectable)
 vector field $X^f_\Lag\in\vf (M_{f})$.
Hence, in general, there is not only one, but a family of submanifolds
 $\{ S\}$ which are diffeomorphic to $P_f\hookrightarrow J^{1*}E$ (and
${\cal F}\Lag(S)=P_f$). Moreover, for every submanifold $S$ of the
family, there exists a unique Euler-Lagrange vector field on
$S$ (tangent to $S$) which, in addition, is ${\cal F}\Lag_f$-projectable
on the points of $S$.
 \item
The SODE constraints defining the submanifold $S_f$ as
 a submanifold of $M_f$ are not ${\cal F}\Lag$-projectable,
 and hence they remove degrees of freedom on the fibers
 of the submersion ${\cal F}\Lag_f$.
 As a consequence we have that
 ${\cal F}\Lag(S_f)={\cal F}\Lag(M_f)=P_f$.
 Therefore, the submanifolds of the family $\{ S\}$
 are embedded submanifolds of $S_f$.
Furthermore, the Euler-Lagrange vector fields on $S_f$ are not unique.

Of course, in the particular case where $S_f$ is diffeomorphic to $P_f$, then
the family $\{ S\}$ is made up of a unique submanifold which is just $S=S_f$,
and the corresponding Euler-Lagrange vector field is unique
and ${\cal F}\Lag_f$-projectable on the points of $S_f$.
 \item
 Observe also that if $\dim\, S_f>\dim\, S$, then
 the Euler-Lagrange vector field solution on $S_f$ is no longer
 ${\cal F}\Lag_f$-projectable on the points of $S_f$.

 In fact, let $\bar x_1\equiv (t_1,q_1,v_1)$, and
 $\bar x_2\equiv (t_2,q_2,v_2)$ be two different points in $S_f$,
 but in the same fibre of ${\cal F}\Lag_f$. Then,
 $t_1=t_2$, and $q_1=q_2$, but $v_1\not= v_2$.
 Now if $D\in\vf (J^1E)$ is an Euler-Lagrange vector field on $S_f$
 we have that
 $$
 D_{\bar x_1}=\derpar{}{t}\Big\vert_{\bar
x_1}+v^\rho_1\derpar{}{q^\rho}\Big\vert_{\bar x_1}+
 f_1^\rho\derpar{}{v^\rho}\Big\vert_{\bar x_1} \quad , \quad
 D_{\bar x_2}=\derpar{}{t}\Big\vert_{\bar
x_2}+v^\rho_2\derpar{}{q^\rho}\Big\vert_{\bar x_2}+
 f_2^\rho\derpar{}{v^\rho}\Big\vert_{\bar x_2}
 $$
 and ${\cal F}\Lag(\bar x_1)={\cal F}\Lag(\bar x_2)$, but
 ${\cal F}\Lag_*(D_{\bar x_1})\not={\cal F}\Lag_*(D_{\bar x_2})$,
 and the result follows.
 \eit

\section{Dirac brackets and time-dependent constrained
Hamiltonian systems} 

In this Section, we introduce the Dirac
bracket as a local Poisson bracket on the restricted momentum dual
bundle $J^{1*}E$ associated with an almost regular Lagrangian
system. Then we prove that if $g$ is an observable (that is,
$g$ is a  $C^\infty(M,\Real)$-differentiable real function on
$P_f$), the time evolution of $g$ consists essentially of two
terms: the restriction to $P_f$ of the Dirac bracket of $G$ and a
suitable Hamiltonian and the restriction to $P_f$ of the
derivative of $G$ with respect to a fixed  vector field. Here $G$
is an arbitrary extension of $g$ to $J^{1*}E$.

In order to introduce the Dirac bracket, we use a suitable
cosymplectic structure on $J^{1*}E$ and consider a
special basis of constraints for the submanifold $P_f$. In fact,
we classify the constraints into first and second class and
show that the second class constraints are Casimir functions
for the Dirac bracket.

\subsection{Cosymplectic structures on the restricted momentum
dual bundle associated with an almost regular Lagrangian
system}\label{seccion5.1}

Assume that $(J^1E,\Omega_\Lag)$ is an almost regular Lagrangian
system.

We will use the notation of Sections \ref{seccion4.1} and
\ref{seccion4.2}. The following commutative diagram illustrates
the situation $$
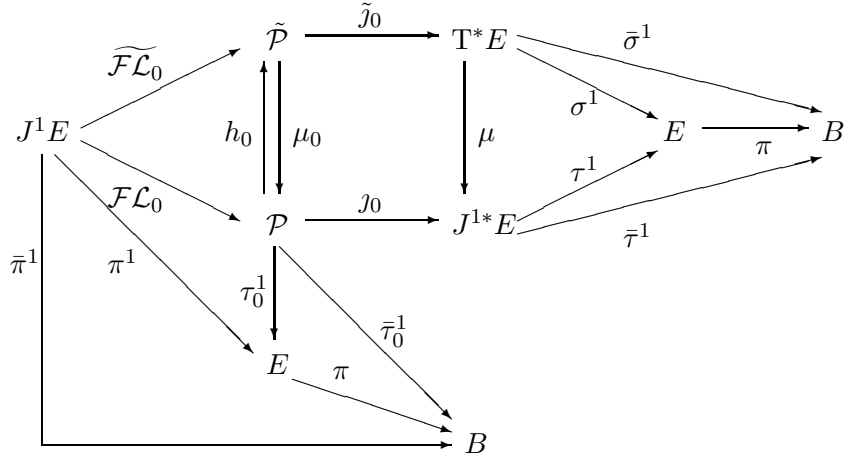
\begin{picture}(305,160)(0,0)
\put(0,100){\mbox{$J^1E$}}

\put(25, 105){\vector(2,1){60}}
\put(35,125){\mbox{$\widetilde{{\cal F}\Lag_0}$}}
\put(25,100){\vector(2,-1){60}} \put(35,75){\mbox{${\cal
F}\Lag_0$}}

\put(95,135){\mbox{$\tilde{\cal P}$}} \put(95,65){\mbox{${\cal
P}$}}
\put(100,130){\vector(0,-1){50}}\put(105,100){\mbox{$\mu_0$}}
\put(94,80){\vector(0,1){50}}\put(79,100){\mbox{$h_0$}}
\put(110,140){\vector(1,0){50}}\put(130,145){\mbox{$\tilde\jmath_0$}}
\put(165,135){\mbox{$\Tan^*E$}}
\put(110,70){\vector(1,0){50}}\put(130,75){\mbox{$\jmath_0$}}
\put(165,65){\mbox{$J^{1*}E$}}
\put(170,130){\vector(0,-1){50}}\put(175,100){\mbox{$\mu$}}
\put(190,140){\vector(4,-1){115}}\put(230,135){\mbox{$\bar{\sigma}^1$}}
\put(190,135){\vector(2,-1){53}}\put(210,110){\mbox{${\sigma}^1$}}
\put(245,100){\mbox{$E$}}
\put(260,105){\vector(1,0){40}}\put(280,95){\mbox{$\pi$}}
 \put(305,100){\mbox{$B$}}
\put(190,65){\vector(4,1){115}}\put(230,60){\mbox{$\bar{\tau}^1$}}
\put(190,70){\vector(2,1){53}}\put(210,85){\mbox{$\tau^1$}}
\put(10,-15){\line(0,0){110}}\put(-2,50){\mbox{$\bar\pi^1$}}
\put(15,95){\vector(1,-1){75}}\put(35,50){\mbox{$\pi^1$}}
\put(10,-15){\vector(1,0){155}}
\put(170,-18){\mbox{$B$}}
\put(98,60){\vector(0,-1){35}}\put(85,40){\mbox{$\tau^1_0$}}
\put(100,60){\vector(1,-1){65}}\put(138,25){\mbox{$\bar{\tau}_0^1$}}
 \put(105,10){\vector(3,-1){60}}\put(120,10){\mbox{$\pi$}}
\put(95,12){\mbox{$E$}}
\end{picture}
$$

Now, let $\nabla$ be a connection in $\pi:E\rightarrow B$ or,
equivalently, a vector field ${\cal Y}_E$ on $E$ such that
$\pi^*(\varpi)({\cal Y}_E)=1.$

For every point $x\in E$, we have a splitting of the tangent
space $\Tan_xE$
\[
\Tan_xE=H_x(\nabla)\oplus V_x(\pi),
\]
where $V(\pi)$ is the $\pi$-vertical subbundle and $H(\nabla)$ is
the horizontal subbundle associated with $\nabla$. Denote by
$Hor_x^\nabla:\Tan_xE\rightarrow H_x(\nabla)$ and by
$Ver_x^\nabla:\Tan_xE\rightarrow V_x(\pi)$ the horizontal and
vertical projectors. Then,
\begin{equation}\label{5.1}
Hor_x^\nabla(X)=(\pi^*(\varpi)_x(X)){\cal Y}_E(x), \;\;\;\;
Ver_x^\nabla(X)=X-Hor_x^\nabla(X).
\end{equation}

Thus, if $\;^t(Ver^\nabla):\Tan^*E\to \Tan^*E$ is the adjoint
homomorphism of the homomorphism of vector bundles
$Ver^\nabla:\Tan E\to V(\pi)\subseteq \Tan E$, then it follows that
\begin{equation}\label{5.2'}
\;^t(Ver^\nabla)_x(\alpha)=\alpha-(\inn {\cal
Y}_E)(\alpha)\pi^*(\varpi)_x, \end{equation} for $x\in E$ and
$\alpha\in \Tan^*_xE,$ where $\inn {\cal Y}_E:\Tan^*E\to \Real$ is the
function defined by
\begin{equation}\label{5.2''}
(\inn {\cal Y}_E)(\beta)=\beta({\cal Y}_E(x)),\makebox[1cm]{}
\mbox{for }\beta\in \Tan_x^*E.
\end{equation}
 From (\ref{5.2'}), (\ref{5.2''}) and the definition of the
Liouville $1$-form $\Theta$ of $\Tan^*E$, we deduce that
\begin{equation}\label{5.2'''}
\begin{array}{rcl}
\;^t(Ver^\nabla)^*(\Theta)&=&\Theta-(\inn {\cal
Y}_E)(\bar{\sigma}^1)^*(\varpi)\\
\;^t(Ver^\nabla)^*(\Omega)&=&\Omega+d(\inn{\cal Y}_E)\wedge
(\bar{\sigma}^1)^*(\varpi).
\end{array}
\end{equation}
Using (\ref{5.2'}) it is also easy to prove that
$\;^t(Ver^\nabla)$ induces a smooth map $\widetilde{\,^t(Ver^
\nabla)}:J^{1*}E=\displaystyle\frac{\Tan^*E}{\Lambda_0\Tan^*E}\to \Tan^*E$
in such a way that
\begin{equation}\label{5.2 0I}
\widetilde{\,^t(Ver^ \nabla)}\circ \mu=\,^t(Ver^ \nabla),\;\;\;
\mu\circ \widetilde{\,^t(Ver^\nabla)}=Id.
\end{equation}
As a consequence, $\widetilde{\,^t(Ver^\nabla)}$ is a global section
of the submersion $\mu:\Tan E\to J^{1*}E$ and, in particular,
$S_E^\nabla=\widetilde{\,^t(Ver^\nabla)}(J^{1*}E)$ is an embedded
submanifold of $\Tan^*E$ of codimension $1.$ Moreover, the map
$\widetilde{\,^t(Ver^\nabla)}:J^{1*}E\to S_E^\nabla$ is a
diffeomorphism. Note that
\begin{equation}\label{5.3}
S_E^\nabla=\{\alpha\in \Tan^*E/ (\inn {\cal Y}_E)(\alpha)=0\}.
\end{equation}
Now, we introduce the $1$-form $\tilde{\Theta}^\nabla$ and the
$2$-form $\tilde{\omega}^\nabla$ on $J^{1*}E$ given by
\begin{equation}\label{5.4}
\tilde{\Theta}^\nabla=\widetilde{\,^t(Ver^\nabla)^*}(\Theta),\;\;\;\;
\tilde{\omega}^\nabla=\widetilde{\,^t(Ver^\nabla)^*}(\Omega)=
-d\tilde{\Theta}^\nabla.
\end{equation}

The $1$-form $\tilde\Theta^\nabla$ (resp. the $2$-form
$\tilde{\omega}^\nabla)$ is called the {\sl Liouville $1$-form }
(resp. {\sl $2$-form }) of $J^{1*}E$ associated with the
connection $\nabla$.

Denote by $\tilde\eta$ the $1$-form on $J^{1*}E$ given by
$\tilde\eta=(\bar{\tau}^1)^*(\varpi)$ (see Section
\ref{seccion4.1}).

\begin{teor}\label{teorema9}
\begin{enumerate}
\item
The couple $(\tilde{\omega}^\nabla, \tilde\eta)$ is a cosymplectic
structure on $J^{1*}E.$
\item
If $\tilde{{\cal R}}^\nabla$ is the Reeb vector field of
$(\tilde{\omega}^\nabla, \tilde\eta)$ we have that
\[
\widetilde{\,^t(Ver^\nabla)}_*(\tilde{\cal R}^\nabla)=X_{\inn
{\cal Y}_E}\vert_{S_E^\nabla},\]
 where $X_{\inn {\cal Y}_E}$ is the
Hamiltonian vector field on the symplectic manifold
$(\Tan^*E,\Omega)$ associated with the function $\inn {\cal
Y}_E:\Tan^*E\to \Real.$
\end{enumerate}
\end{teor}
\proof If $\inn_E^\nabla:S_E^\nabla\to \Tan^*E$ is the canonical
inclusion, then since
$\widetilde{\,^t(Ver^\nabla)}^*((\bar\sigma^1)^*(\varpi))=\tilde\eta,$
we must prove that the couple $((\inn_E^\nabla)^*(\Omega),
(\inn_E^\nabla)^*((\bar{\sigma}^1)^*(\varpi)))$ is a cosymplectic
structure on $S_E^\nabla$ with Reeb vector field $X_{\inn {\cal
Y}_E}\vert_{S_E^\nabla}.$

It is clear that $(\inn_E^\nabla)^*(\Omega)$ (respectively,
$(\inn_E^\nabla)^*((\bar{\sigma}^1)^*(\varpi))$) is a closed
$2$-form (respectively, $1$-form) on $S_E^\nabla$ and that the
restriction of $X_{\inn {\cal Y}_E}$ to $S_E^\nabla$ is tangent to
$S_E^\nabla$ (see (\ref{5.3})). Moreover, from (\ref{5.2''}) and
since $\pi^*(\varpi)({\cal Y}_E)=1,$ it follows that
$((\bar{\sigma}^{1})^*(\varpi))(X_{\inn {\cal Y}_E})=1$, which
implies that
$((\inn_E^\nabla)^*((\bar{\sigma}^1)^*(\varpi)))(X_{\inn {\cal
Y}_E}\vert_{S_E^\nabla})=1.$

Furthermore, using (\ref{5.3}), we obtain that
$\ker((\inn_E^\nabla)^*(\Omega))=\langle X_{\inn {\cal
Y}_E}\vert_{S_E^\nabla}\rangle$.

This ends the proof of the result. \qed

\begin{remark}
{\rm From  (\ref{5.2'''}), (\ref{5.2 0I}) and (\ref{5.4}), we
deduce that
\begin{equation}\label{r5.4}
\begin{array}{rcl}
\mu^*(\tilde{\Theta}^\nabla)&=&\Theta-(\inn{\cal
Y}_E)((\bar{\sigma}^1)^*(\varpi)),\\
\mu^*({\tilde\omega^\nabla})&=&\Omega+d(\inn {\cal
Y}_E)\wedge(\bar{\sigma}^1)^*(\varpi).
\end{array}
\end{equation}}
\end{remark}
Next we study the relation between the cosymplectic
structure $(\tilde{\omega}^\nabla, \tilde\eta)$ and the
cosymplectic structure on $J^{1*}E$ defined by another connection
$\nabla'$ on $\pi:E\to B.$

Suppose that ${\cal Y}_E'$ is the vector field on $E$ associated
with $\nabla'$. Then, $\pi^*(\varpi)({\cal Y}'_E)=1$ and
\[
V={\cal Y}'_E-{\cal Y}_E
\]
is a vertical vector field with respect to $\pi:E\to B$. This
implies that the function $\inn V:\Tan^*E\to \Real$ given by
$$
(\inn V)(\alpha)=\alpha(V(x)),\makebox[1cm]{} \mbox{ for
}\alpha\in \Tan_x^*E
$$
induces a smooth function $\widetilde{\inn V}:J^{1*}E\to \Real$ in
such a way that
\begin{equation}\label{5.4'''}
\mu^*(\widetilde{\inn V})=\inn V.
\end{equation}
Moreover, we obtain that
\begin{teor}\label{t10}
\begin{enumerate}
\item[$(i)$]
If $(\tilde{\omega}^{\nabla'}, \tilde\eta)$ is the cosymplectic
structure on $J^{1*}E$ defined by the connection $\nabla'$ then
\[
\tilde{\omega}^{\nabla'}=\tilde{\omega}^\nabla+ d(\widetilde{\inn
V})\wedge\tilde\eta.
\]
\item[$(ii)$]
If $f:J^{1*}E\to \Real $ is a real $C^\infty$-function and
$X_f^{\tilde{\;}\nabla}$ (respectively, $X_f^{\tilde{\;}\nabla'}$)
is the Hamiltonian vector field of $f$ with respect to the
cosymplectic structure $(\tilde{\omega}^\nabla, \tilde\eta )$
(respectively, $(\tilde{\omega}^{\nabla'},\tilde\eta)$) then
$X_f^{\tilde{\;}\nabla}=X_f^{\tilde{\;}\nabla'}$.
\item[$(iii)$]
If $\{\;,\;\}^{\tilde{\;}\nabla}$ (respectively,
$\{\;,\;\}^{\tilde{\;}\nabla'}$) is the Poisson bracket associated
with the cosymplectic structure $(\widetilde{w}^\nabla,
\tilde\eta)$ (respectively, $(\widetilde{w}^{\nabla'},
\tilde\eta)$) then
\[
\{\;,\;\}^{\tilde{\;}\nabla'}=\{\;,\;\}^{\tilde{\;}\nabla}.
\]
\end{enumerate}
\end{teor}
\proof Using (\ref{r5.4}) and (\ref{5.4'''}), it follows that
$$
\mu^*(\widetilde{\omega}^{\nabla'})=\mu^*(\tilde{\omega}^\nabla)+
d(\inn V)\wedge \tilde\eta\\ = \mu^*(\tilde{\omega}^\nabla+
d(\widetilde{\inn V}))\wedge\tilde\eta.
$$
Therefore, since $\mu^*$ is injective, we conclude that
\[
\widetilde{\omega}^{\nabla'}=\widetilde{\omega}^\nabla+
d(\widetilde{\inn V})\wedge \tilde\eta.
\]
This proves $(i)$.

$(ii)$ and $(iii)$ follow from $(i)$ and Proposition \ref{14'''}
(see Appendix \ref{C}). \qed

Let us now recall the definition of the {\sl Hamiltonian density
} and of the {\sl Hamiltonian function } associated with the
Lagrangian system, the $1$-form $\varpi$ and a connection $\nabla$
on $\pi:E\to B$ (see \cite{CCI-91,EMR-sdtc,JMP}).

 Using the connection $\nabla,$ we can
introduce the {\sl Hamilton-Cartan $1$-form }
$\Theta_{h_0}^\nabla$ on ${\cal P}$ associated with the Lagrangian
system, the $1$-form $\varpi$ and the connection $\nabla.$ This
$1$-form is given by
\begin{equation}\label{5.2}
(\Theta_{h_0}^\nabla)_{\tilde{x}}(\tilde{X})=h_0(\tilde{x})
(Ver_x^\nabla((\tau_0^1)_{*\tilde{x}}(\tilde{X}))),
\end{equation}
for $\tilde{x}\in {\cal P}$ and $\tilde{X}\in \Tan_{\tilde{x}}{\cal
P}.$ The $1$-form $\Theta_{h_0}^\nabla$ allows us to introduce, in a
natural way, the {\sl Hamilton-Cartan $2$-form }
$\Omega_{h_0}^\nabla$ on ${\cal P}$ which is defined by
$\Omega_{h_0}^\nabla=-d\Theta_{h_0}^\nabla.$

If $\tilde{X}$ is vertical with respect to the projection
$\tau_0^1:{\cal P}\to B,$ a direct computation, using (\ref{5.1}),
(\ref{5.2}) and the fact that
$\Theta_{h_0}=(\widetilde{\jmath_0}\circ h_0)^*(\Theta),$ shows
that
\[
(\Theta_{h_0}^\nabla-\Theta_{h_0})_{\tilde{x}}(\tilde{X})=0.
\]
Therefore, there exists $h_0^\nabla\in C^\infty({\cal P})$ such
that
\begin{equation}\label{5.2 0}
\Theta_{h_0}^\nabla-\Theta_{h_0}=h_0^\nabla((\bar{\tau_0}^1)^*(\varpi))=
h_0^\nabla(\eta^0).
\end{equation}
${\frak h}_0^\nabla=h_0^\nabla(\eta^0)$ and $h_0^\nabla$ are
called the Hamiltonian density and the Hamiltonian function
associated with the Lagrangian system, the connection $\nabla$ and
the $1$-form $\varpi$ (see \cite{CCI-91,EMR-sdtc,JMP}).

\begin{remark}\label{r8'}
{\rm If ${\cal Y}'_E$ is the vector field on $E$ associated with
another connection $\nabla'$ on $\pi:E\to B$ and $V={\cal
Y}_E'-{\cal Y}_E$ is the corresponding vertical vector field with
respect to $\pi:E\to B$, then using (\ref{5.2}) and (\ref{5.2 0}),
and the fact that $\Theta_{h_0}=(\widetilde{\jmath_0}\circ
h_0)^*(\Theta)$, we deduce that
\begin{equation}\label{e43'}
h_0^{\nabla'}=h_0^\nabla-\widetilde{iV}\vert_{\cal P},
\end{equation}
where $\widetilde{iV}:J^{1*}E\to \Real$ is the real function on
$J^{1*}E$ induced by the vector field $V.$ This implies that \[
\Theta_{h_0}^{\nabla'}=\Theta_{h_0}^\nabla-(\widetilde{\inn
V})\vert_P\eta^0,\makebox[1cm]{}\Omega_{h_0}^{\nabla'}=\Omega_{h_0}^\nabla
+ d(\widetilde{\inn V})\vert_P\wedge\eta^0.
\]}
\end{remark}

Next, we will introduce an extension $\tilde{h}^\nabla:J^{1*}E\to
\Tan^*E$ of the diffeomorphism $h_0:{\cal P}\to \tilde{\cal P}.$ For
this purpose, we consider an arbitrary extension
$h^\nabla:J^{1*}E\to \Real$ of the Hamiltonian function
$h_0^\nabla:{\cal P}\to \Real.$ Then, we define the map
$H^\nabla:\Tan^*E\to \Tan^*E$ given by
\begin{equation}\label{5.5}
H^\nabla(\alpha)=\alpha - (h^\nabla(\mu(\alpha))+(\inn {\cal
Y}_E)(\alpha))\pi^*(\varpi)(\sigma^1(\alpha)).
\end{equation}
It is clear that $\sigma^1\circ H^\nabla=\sigma^1.$ Furthermore,
we have
\begin{lem}
The following relations hold
\begin{eqnarray}
\mu\circ H^\nabla&=&\mu\label{5.6}\\ (H^\nabla)^*(\Theta)&=&
\Theta - (h^\nabla\circ \mu+\inn {\cal
Y}_E)(\bar{\sigma}^1)^*(\varpi)\label{5.7}\\
(H^\nabla)^*(\Omega)&=& \Omega + (\mu^*(dh^\nabla)+d(\inn {\cal
Y}_E))\wedge (\bar\sigma^1)^*(\varpi)\label{5.8}\\ H^\nabla\circ
\widetilde{{\cal F}\Lag_0}&=&\widetilde{{\cal F}\Lag_0}\label{5.9}
\end{eqnarray}
\end{lem}
\proof From (\ref{5.5}) it follows (\ref{5.6}).

Furthermore, using (\ref{5.5}), the definition of $\Theta$
and the fact that $\sigma^1\circ H^\nabla=\sigma^1,$ we deduce
that (\ref{5.7}) holds. Therefore,
\[
(H^\nabla)^*(\Omega)=-d((H^\nabla)^*(\Theta))=\Omega +
(\mu^*(dh^\nabla)+d(\inn {\cal
Y}_E))\wedge(\bar{\sigma}^1)^*(\varpi).
\]
Finally, we will prove that (\ref{5.9}). Let $\bar{y}$ be a point
of $J^1E$. Assume that $\pi^1(\bar{y})=\sigma^1(\widetilde{{\cal
F}\Lag_0}(\bar{y}))=x$ and that $\bar{\cal Y}_E(\bar{y})$ is a
tangent vector to $J^1E$ at $\bar{y}$ such that $(\pi^1)_{*\bar
y}(\bar{\cal Y}_E(\bar{y}))={\cal Y}_E(x).$

 From (\ref{23'}), (\ref{pback0}), (\ref{5.2}) and (\ref{5.2 0}),
we obtain that
\begin{equation}\label{5.10}
\begin{array}{lcl}
h^\nabla(\mu(\widetilde{{\cal F}\Lag_0}(\bar{y})))+
\widetilde{{\cal F}\Lag_0}(\bar{y})({\cal
Y}_E(x))&\kern-3pt=&\kern-3pt h_0^\nabla({\cal
F}\Lag_0(\bar{y})))+ \Theta_\Lag(\bar y)(\bar{\cal Y}_E(\bar y))\\
&\kern-3pt=&\kern-3pt h_0^\nabla({\cal
F}\Lag_0(\bar{y}))+\Theta_{h_0}({\cal F}\Lag_0(\bar{y}))(({\cal
F}\Lag_0)_{*\bar y}(\bar{\cal
Y}_E(\bar{y})))\\&\kern-3pt=&\kern-3pt\Theta_{h_0}^\nabla({\cal
F}\Lag_0(\bar{y}))(({\cal F}\Lag_0)_{*\bar{y}} (\bar{\cal
Y}_E(\bar{y})))\\ &\kern-3pt=&\kern-3pth_0({\cal
F}\Lag_0(\bar{y}))(Ver_x^\nabla({\cal Y}_E(x)))=0.
\end{array}
\end{equation}
As a consequence, using (\ref{5.5}) and (\ref{5.10}), we conclude that
$H^\nabla(\widetilde{{\cal F}\Lag_0}(\bar{y}))=\widetilde{{\cal
F}\Lag_0}(\bar y).$ \qed

 Suppose that $\mu(\alpha)=\mu(\alpha'),$
for $\alpha,\alpha'\in \Tan^*E.$ Then,
$H^\nabla(\alpha)=H^\nabla(\alpha')$ (see (\ref{5.5})). Thus,
there exists a mapping $\tilde{h}^\nabla: J^{1*}E\to \Tan^*E$ such
that
\begin{equation}\label{5.11}
\tilde{h}^\nabla\circ \mu=H^\nabla.
\end{equation}
Moreover, from (\ref{5.6}), we have that $(\mu\circ
\widetilde{h}^\nabla)\circ \mu=\mu$, which implies that
\begin{equation}\label{5.12}
\mu\circ \widetilde{h}^\nabla=Id.
\end{equation}
Since $\mu$ is submersion, we deduce that $\tilde{h}^\nabla$ is a
smooth mapping, and therefore $\tilde{h}^\nabla$ is a global
section of $\mu:\Tan^*E\to J^{1*}E.$

Furthermore, using (\ref{5.6}), (\ref{5.9}) and the fact
that $\mu(\tilde{\cal P})={\cal P},$ it follows that
$\tilde{h}^\nabla({\cal P})=\tilde{\cal P}.$ Furthermore, using
(\ref{5.9}), (\ref{5.11}) and (\ref{5.12}), we obtain that
\[
\mu_0\circ \tilde{h}^\nabla\vert_{\cal P}=Id,\;\;\;\;
\tilde{h}^\nabla\vert_{\cal P}\circ \mu_0=Id,
\]
that is, $\tilde{h}^\nabla\vert_{\cal P}=\mu_0^{-1}=h_0.$ We now
introduce the $1$-form $\Theta_{h^\nabla}$ and the $2$-form
$\Omega_{h^\nabla}$ on $J^{1*}E$ given by
\[
\Theta_{h^\nabla}=(\tilde{h}^\nabla)^*(\Theta),\;\;\;\;
\Omega_{h^\nabla}=(\tilde{h}^\nabla)^*(\Omega)=-d\Theta_{h^\nabla}.
\]
It is clear that
\begin{equation}\label{51'}
\jmath_0^*(\Theta_{h^\nabla})=\Theta_{h_0},\;\;\;
\jmath_0^*(\Omega_{h^\nabla})=\Omega_{h_0}. \end{equation}
  From
(\ref{5.8}) and (\ref{5.11}), we deduce that
\[
\mu^*(\Omega_{h^\nabla})=\Omega + (\mu^*(dh^\nabla)+d(\inn {\cal
Y}_E))\wedge(\bar{\sigma}^1)^*(\varpi)
\]
and using (\ref{r5.4}), it follows that
\[
\mu^*(\Omega_{h^\nabla})=\mu^*(\tilde{\omega}^\nabla)+
\mu^*(dh^\nabla)\wedge(\bar{\sigma}^1)^*(\varpi)\\ =
\mu^*(\tilde{\omega}^\nabla + dh^\nabla \wedge\tilde\eta).
\]
Since $\mu^*$ is injective, this implies that
\begin{equation}\label{51''}
\Omega_{h^\nabla}=\tilde{\omega}^\nabla + dh^\nabla\wedge
\tilde\eta.
\end{equation}
Thus (see Proposition \ref{14'''} in Appendix \ref{C}), we have
the following

\begin{teor}\label{t11}
\begin{enumerate}
\item The couple $(\Omega_{h^\nabla}, \tilde\eta)$ is a
cosymplectic structure on $J^{1*}E$ with Reeb vector field ${\cal
R}_{h^\nabla}$ given by
\[
{\cal R}_{h^\nabla}=\tilde{\cal R}^\nabla +
X_{h^\nabla}^{\tilde{\;}\nabla} =E_{h^\nabla}^{\tilde{\;}\nabla},
\]
where $X_{h^\nabla}^{\tilde{\;}\nabla}$ (respectively,
$E_{h^\nabla}^{\tilde{\;}\nabla})$is the Hamiltonian vector field
(respectively, the evolution vector field) of $h^\nabla$ with
regard to the cosymplectic structure $(\tilde{\omega}^\nabla,
\tilde\eta).$
\item
If $f:J^{1*}E\to \Real$ is a real $C^\infty$-differentiable
function and $X_f^{h^\nabla}$ (respectively,
$X_{f}^{\tilde{\;}\nabla}$) is the Hamiltonian vector field of $f$
with regard to the cosymplectic structure $(\Omega_{h^\nabla},
\tilde\eta)$ (respectively, $(\tilde{\omega}^\nabla,\tilde\eta)$
then $X_f^{h^\nabla}=X_f^{\tilde{\;}\nabla}.$
\item
If $\{\;,\;\}^{\tilde{\;}\nabla}$ (respectively,
$\{\;,\;\}^{h^\nabla}$) is the Poisson bracket associated with the
cosymplectic structure $(\tilde{\omega}^\nabla, \tilde\eta)$
(respectively, $(\Omega_{h^\nabla}, \tilde\eta)$), then
\[
\{\;,\;\}^{\tilde{\;}\nabla}=\{\;,\;\}^{h^\nabla}.
\]
\end{enumerate}
\end{teor}
\begin{remark}
{\rm If $(J^1E,\Omega_\Lag)$ is a hyperregular Lagrangian system
then ${\cal P}=J^{1*}E,$ $\jmath_0=Id$ and
$\Omega_{h_0}=\Omega_{h^\nabla}.$ Thus, $(\Omega_{h_0}, \eta^0)$
is a cosymplectic structure on $J^{1*}E$ and ${\cal
R}_{h^\nabla}=E_{h^\nabla}^{\tilde{\;}\nabla}$ is the unique
solution of the Hamilton equations
\[
\inn({\cal R}_{h^\nabla})\Omega_{h_0}=0,\;\;\;\; \inn({\cal
R}_{h^\nabla})\eta^0=1.
\]}
\end{remark}

Next we will write the local
expressions of some of the geometric structures introduced above.
Assume that $(t,q^j,p_j)$ and $(t,q^j,p,p_j)$ are natural
coordinates on $J^{1*}E$ and $\Tan^*E$, respectively. If the local
expression of ${\cal Y}_E$ is
\[
{\cal Y}_E=\frac{\partial}{\partial t}+ {\cal
Y}_E^i(t,q^j)\frac{\partial}{\partial q^i}
\]
then
\begin{equation}\label{54'}
\begin{array}{rcl}
\;^t(Ver^\nabla)(t,q^i,p,p_i)&=&
\widetilde{\;^t(Ver^\nabla)}(t,q^i,p_i)=(t,q^i,-{\cal
Y}^j_E(t,q^i)p_j,p_i)\\

\tilde{\Theta}^\nabla&=&p_idq^i-({\cal Y}_E^i(t,q^j)p_i)dt\\

\tilde{\omega}^\nabla&=&dq^i\wedge dp_i-dt\wedge \left({\cal
Y}^i_E(t,q^j)dp_i + p_i\displaystyle\frac{\partial {\cal
Y}_E^i}{\partial q^j}dq^j\right)\\

\tilde{\cal R}^\nabla&=&\displaystyle\frac{\partial }{\partial t}
+ {\cal Y}_E^i(t,q^j)\displaystyle\frac{\partial}{\partial
q^i}-p_i\displaystyle\frac{\partial {\cal Y}_E^i}{\partial
q^j}\displaystyle\frac{\partial }{\partial p_j}\\

H^\nabla(t,q^i,p,p_i)&=&(t, q^i,-h^\nabla(t,q^j,p_j)-{\cal
Y}^i_E(t,q^j)p_i, p_i)\\

\tilde{h}^\nabla(t,q^i,p_i)&=& (t,q^i, -h^\nabla (t,q^j,p_j)-{\cal
Y}^i_E(t,q^j)p_i,p_i)\\

\Theta_{h^\nabla}&=&p_idq^i + (h^\nabla(t,q^j,p_j)- {\cal
Y}_E^i(t,q^j)p_i)dt\\

\Omega_{h^\nabla}&=&dq^i\wedge dp_i - dt\wedge \left({\cal
Y}^i_E(t,q^j)dp_i + p_i\displaystyle\frac{\partial {\cal
Y}_E^i}{\partial q^j}dq^j\right) + dh^\nabla\wedge dt\\

X_{h^\nabla}^{\tilde{\;}\nabla}&=&\displaystyle\frac{\partial
h^\nabla}{\partial p_i}\displaystyle\frac{\partial}{\partial q^i}
+ \displaystyle\frac{\partial h^\nabla}{\partial q^i}
\displaystyle\frac{\partial}{\partial p_i}\\

{\cal R}_{h^\nabla}&=&\displaystyle\frac{\partial }{\partial t} +
\left({\cal Y}_E^j(t,q^i) +\displaystyle\frac{\partial
h^\nabla}{\partial p_j}\right)\displaystyle\frac{\partial }{\partial
q^j}-\left(\displaystyle\frac{\partial h^\nabla}{\partial q^j} +
p_i\displaystyle\frac{\partial {\cal Y}_{E}^i}{\partial
q^j}\right)\displaystyle\frac{\partial }{\partial p_j}
\end{array}
\end{equation}

\subsection{First and second class constraints and the solutions of
the Hamiltonian dynamics}\label{seccion5.2}

Assume that $(J^1E,\Omega_{\cal L})$ is an almost regular
Lagrangian system as in Section \ref{seccion5.1}.

If $\nabla$ is a connection on $\pi:E\to B$, we will denote by
$(\tilde{\omega}^\nabla,\tilde\eta)$ the cosymplectic structure on
$J^{1*}E$ introduced in Section \ref{seccion5.1} (see
Theorem \ref{teorema9}), by $\tilde\Lambda^\nabla$ the Poisson
$2$-vector associated with such a structure, and by
$\{\;,\;\}^{\tilde{\;}\nabla}$ the corresponding Poisson bracket
(see Appendix \ref{B}).

Let $P_0$ (resp. $P_f$) be the first (resp. final) Hamiltonian
dynamical constraint submanifold. If $p_0$ (resp. $p_f$) is the
dimension of $P_0$ (resp. $P_f$), then since the pull-back of the
$1$-form $\tilde\eta$ to $P_0$ (resp. $P_f$) is not zero at every
point of $P_f$, it follows that the distribution
$\#_{\tilde\Lambda^\nabla}((\Tan P_0)^0)$ (resp.
$\#_{\tilde\Lambda^\nabla}((\Tan P_f)^0)$) has constant rank
$2n+1-p_0$ (resp. $2n+1-p_f$) along $P_f$. We will suppose that
the distributions $\Tan P_0\cap \#_{\tilde\Lambda^\nabla}((\Tan P_0)^0)$,
$\Tan P_f\cap \#_{\tilde\Lambda^\nabla}((\Tan_{P_f}P_0)^0)$ and $\Tan P_f\cap
\#_{\tilde\Lambda^\nabla}((\Tan P_f)^0)$ also have constant rank
$k_0, k_{0f}$ and $k_f,$ respectively, along $P_f.$
\begin{remark}\label{r11-0}
{\rm If $\nabla'$ is another connection on $\pi:E\to B$, then from
Theorem \ref{t10} we deduce that
$\tilde\Lambda^\nabla=\tilde\Lambda^{\nabla'}.$ Thus,
\[
\begin{array}{lll}
\Tan P_0\cap \#_{\tilde\Lambda^{\nabla'}}((\Tan P_0)^0)&=&\Tan P_0\cap
\#_{\tilde\Lambda^\nabla}((\Tan P_0)^0)\\
\Tan P_f\cap\#_{\tilde\Lambda^{\nabla'}}((\Tan_{P_f}P_0)^0)&=&\Tan P_f\cap
\#_{\tilde\Lambda^\nabla}((\Tan_{P_f}P_0)^0)\\
\Tan P_f\cap\#_{\tilde\Lambda^{\nabla'}}((\Tan P_f)^0)&=&\Tan P_f\cap
\#_{\tilde\Lambda^\nabla}((\Tan P_f)^0).\\
\end{array}
\]}
\end{remark}

With the above hypotheses, we will show that it is possible to
choose a suitable basis of constraints for the submanifold $P_f$.
We will proceed in three steps.

{\bf First step}: Assume that $\{\xi_i^{(0)}\}_{i=1,\dots,
2n+1-p_0}$ is a set of local independent constraint functions
defining  $P_0$ as a submanifold of $J^{1*}E$. Then the rank of
the matrix
$$(\{\xi_i^{(0)},\xi_{i'}^{(0)}\}^{\tilde{\;}\nabla}\vert_{P_f})_{1\leq
i,i'\leq 2n+1-p_0}$$
 is $l_0=2n+1-p_0-k_0.$ We can suppose,
without loss of generality, that the $l_0$ first rows of this
matrix are independent. With this hypothesis, it is easy to prove
that for every $j, 1\leq j\leq k_0,$ there exist local real
functions $\{f_j^i\}_{1\leq i\leq l_0}$ on $J^{1*}E$ such that the
matrix
 $$(\{\xi_i^{(0)},\xi_{i'}^{(0)}\}
^{\tilde{\;}\nabla}\vert_{P_f})_{1\leq i,i'\leq l_0}$$
 is regular. In addition,
$\{X^{\tilde{\;}\nabla}_{\tilde{\xi}_{j}^{(0)}|P_f}\}_{1\leq j\leq
k_0}$ is a local basis of the distribution $\Tan P_0\cap
\#_{\tilde\Lambda^\nabla}((\Tan P_0)^0)$ along $P_f$, where
\[
\tilde\xi_j^{(0)}=
\xi_{l_0+j}^{(0)}-f_j^i\xi_i^{(0)},\makebox[1cm]{}
\mbox{ for } 1\leq j\leq k_0.
\]
Thus, if $\bar\xi_i^{(0)}=\xi_i^{(0)},$  for $1\leq i\leq l_0,$ it
is clear that
$\{\bar\xi_i^{(0)},\tilde{\xi}_j^{(0)}\}_{\tiny\begin{array}{l}
1\leq i\leq l_0\\1\leq j\leq k_0\end{array}}$ is a set of local
independent constraint functions defining $P_0$ as a submanifold
of $J^{1*}E.$

\medskip

{\bf Second step}: Assume that
$\{\bar\xi_i^{(0)},\tilde\xi_j^{(0)},\xi_r^{(f)}\}_{\tiny\begin{array}{l}1\leq
i\leq l_0\\1\leq j\leq k_0\\1\leq  r\leq p_0-p_f\end{array}}$ is a
set of local independent constraint functions defining $P_f$ as a
submanifold of $J^{1*}E$. Then the rank of the matrix
$$(\{\tilde\xi_j^{(0)},\xi_r^{(f)}\}^{\tilde{\;}\nabla}\vert_{P_f})
_{\tiny\begin{array}{l} 1\leq j\leq k_0\\1\leq r \leq
p_0-p_f\end{array}}$$ is $k_0-k_{0f}.$ We can suppose, without
loss of generality, that the $k_0-k_{0f}$ first rows of this
matrix are independent. With this hypothesis, it is easy to prove
that for every $t$, $1\leq t\leq k_{0f}$,  there exist local real
functions $\{g_t^j\}_{1\leq j\leq k_0-k_{0f}},$ such that
\[
\{X_{\hat\xi_{t}^{(0)}}^{\tilde{\;}\nabla}\vert_{P_f}\}_{1\leq t\leq
k_{0f}}
\]
is a local basis of the distribution $\Tan P_f\cap
\#_{\tilde\Lambda^\nabla}((\Tan_{P_f}P_0)^0)$, where
\[
\hat\xi_t^{(0)}=\tilde\xi_{k_0-k_{0f}+t}^{(0)}-
g_t^j\tilde\xi_j^{(0)}, \makebox[1cm]{}
\mbox{ for } 1\leq t\leq k_{0f}.
\]
It is clear that $\{\bar\xi_i^{(0)}, \tilde\xi_j^{(0)},
\hat\xi_t^{(0)}, \xi_r^{(f)}\},$ with $1\leq i\leq l_0,$ $1\leq
j\leq k_0-k_{0f},$ $1\leq t\leq k_{0f}$ and $1\leq r\leq p_0-p_f,$
is a set of local independent constraint functions defining $P_f$
as a submanifold of $J^{1*}E$.

\medskip

{\bf Third step}: We consider the $(l_0 +
(k_0-k_{0f})+(p_0-p_f))\times (l_0+(k_0-k_{0f})+(p_0-p_f))$ matrix
\[
\left(
\begin{array}{ccc}
\{\bar\xi_i^{(0)},\bar{\xi}_{i'}^{(0)}\}^{\tilde{\;}\nabla}\vert_{P_f}&0&
\{\bar\xi_i^{(0)},\xi_{r}^{(f)}\}^{\tilde{\;}\nabla}\vert_{P_f}\\ 0&0
&\{\tilde\xi_j^{(0)},\xi_{r}^{(f)}\}^{\tilde{\;}\nabla}\vert_{P_f}\\
-\{\bar\xi_i^{(0)},\xi_{r}^{(f)}\}^{\tilde{\;}\nabla}\vert_{P_f}&
-\{\tilde\xi_j^{(0)},\xi_{r}^{(f)}\}^{\tilde{\;}\nabla}\vert_{P_f}&
\{\xi_r^{(f)},\xi_{r'}^{(f)}\}^{\tilde{\;}\nabla}\vert_{P_f}\end{array}
\right)\] A direct computation shows that the $l_0+(k_0-k_{0f})$
first rows are independent and that the rank of this matrix is
$l_0+(k_0-k_{0f})+s_f$, with $s_f=(p_0-p_f)-(k_f-k_{0f}).$ We can
suppose, without loss of generality, that the
$l_0+(k_0-k_{0f})+s_f$ first rows are independent. Then, it is
easy to prove that for every $u$, $1\leq u\leq k_f-k_{0f},$ there
exist local real functions $\{h_u^r\}_{1\leq r\leq s_f}$ such that
\[
\{X_{\hat\xi_t^{(0)}}^{\tilde{\;}\nabla}\vert_{P_f},
X_{\tilde\xi_u^{(f)}}^{\tilde{\;}\nabla}\vert_{P_f}
\}_{\tiny\begin{array}{l} 1\leq t\leq k_{0f}\\1\leq u\leq
k_f-k_{0f}\end{array}}
\]
is a local basis of the distribution $\Tan P_f\cap
\#_{\tilde\Lambda^\nabla}((\Tan P_f)^0),$ where
\[
\tilde\xi_u^{(f)}=\xi_{s_f+u}^{(f)}-h_u^r\xi_r^{(f)}.
\]
In conclusion, if $\bar\xi_r^{(f)}=\xi_r^{(f)},$ for $1\leq r\leq
s_f,$ we have that
$\{\bar\xi_i^{(0)},\tilde\xi_j^{(0)},\hat\xi_t^{(0)},
\bar\xi_r^{(f)},\tilde\xi_u^{(f)}\}$ with $1\leq i\leq l_0,$
$1\leq j\leq k_0-k_{0f}, 1\leq t\leq k_{0f},$ $1\leq r\leq s_f$
and $1\leq u\leq k_f-k_{0f},$ is a set of local independent
constraint functions defining $P_f$ as a submanifold of
$J^{1*}E.$ Moreover, along $P_f$
\begin{equation}\label{+1}
\begin{array}{lcl}
\vspace{0,2cm}
\Tan P_0\cap \#_{\tilde\Lambda^\nabla}((\Tan P_0)^0)&=&
\langle X_{\tilde\xi_{j}^{(0)}}^{\tilde{\;}\nabla
}\vert_{P_f},X_{\hat\xi_{t}^{(0)}}^{\tilde{\;}\nabla
}\vert_{P_f}\rangle_{\tiny\begin{array}{l}1\leq j\leq k_0-k_{0f}\\1\leq t\leq
k_{0f}\end{array}} \\
\vspace{0,2cm}
 \Tan P_f\cap\#_{\tilde\Lambda^\nabla}((\Tan_{P_f}P_0)^0)
&=&
\langle X_{\hat\xi_t^{(0)}}^{\tilde{\;}\nabla }\vert_{P_f}\rangle_{1\leq t\leq
k_{0f}} \\
 \Tan P_f\cap \#_{\tilde\Lambda^\nabla}((\Tan P_f)^0)&=&
\langle X_{\hat\xi_t^{(0)}}^{\tilde{\;}\nabla
}\vert_{P_f},X_{\tilde\xi_u^{(f)}}^{\tilde{\;}\nabla
}\vert_{P_f}\rangle_{\tiny\begin{array}{l}1\leq t\leq k_{0f}\\1\leq u\leq
k_f-k_{0f}\end{array}}
\end{array}
\end{equation}
In addition, if we denote by
\begin{equation}\label{+2}
\{\bar {\cal X}_\alpha\}_{1\leq \alpha\leq
l_0+(k_0-k_{0f})+s_f}=\{\bar\xi_i^{(0)},\tilde
\xi_j^{(0)},\bar\xi_r^{(f)}\}_{\tiny\begin{array}{l}1\leq j\leq
l_0\\1\leq j\leq k_0-k_{0f}\\1\leq r\leq s_f\end{array}}
\end{equation}
then we deduce that the matrix
\[
(\bar {\cal C}_{\alpha\beta}=\{\bar {\cal X}_\alpha,\bar {\cal
X}_\beta\}^{\tilde{\;}\nabla}\vert_{P_f})_{1\leq \alpha,\beta\leq
l_0+(k_0-k_{0f})+s_f}
\]
is regular.

Following Dirac's terminology,
$\{\bar {\cal X}_\alpha\}_{1\leq \alpha\leq l_0+(k_0-k_{0f})+s_f}$
and
$\{\hat\xi_t^{(0)},\tilde\xi_u^{(f)}\}_{\tiny\begin{array}{l}1\leq
t\leq k_{0f}\\ 1\leq u\leq k_f-k_{0f} \end{array}}$ are
said to be {\sl second-class } and {\sl first-class constraints } for
the submanifold $P_f$, respectively.

We now consider the completely integrable distribution $D$ whose
annihilator $D^0$ is generated by $$\{d\bar {\cal
X}_\alpha\}_{1\leq \alpha\leq l_0+(k_0-k_{0f}) +s_f}.$$ Since the
matrix $(\bar {\cal C}_{\alpha\beta})$ is regular, it is follows
that there exists an open neighbourhood $U$ of $P_f$ in $J^{1*}E$
such that $\Tan_UJ^{1*}E=D\oplus \#_{\tilde \Lambda^\nabla}(D^0)$.

Thus, we have the corresponding projectors
\[
{\Bbb P}:\Tan_UJ^{1*}E\to D,\makebox[1cm]{} {\Bbb Q}:\Tan_UJ^{1*}E\to
\#_{\tilde\Lambda^\nabla}(D^0).
\]
A direct computation shows that
\begin{equation}\label{+2'}
{\Bbb Q}=\bar{\cal C}^{\alpha\beta}X_{\bar
{\cal X}_\alpha}^{\tilde{\;}\nabla}\otimes d\bar {\cal
X}_\beta,\makebox[1cm]{} {\Bbb P} =
Id-\bar{\cal C}^{\alpha\beta}X_{\bar {\cal
X}_\alpha}^{\tilde{\;}\nabla}\otimes d\bar {\cal X}_\beta,
\end{equation}
where $(\bar{\cal C}^{\alpha\beta})$ is the inverse of matrix
$(\bar{\cal C}_{\alpha\beta}).$
\begin{remark}\label{r11-0'}
{\rm If $X\in {\frak X}(U)$, we have that
\[
\begin{array}{lcl}
\tilde{\omega}^\nabla({\Bbb P}(\tilde{\cal R}^\nabla), {\Bbb
P}(X))&=&-\tilde\omega^\nabla({\Bbb Q}(\tilde{\cal
R}^\nabla),{\Bbb P}(X))=-\bar{\cal
C}^{\alpha\beta}\tilde{\cal R}^\nabla({\cal X}_\beta)
(i(X_{\bar{\cal
X}_\alpha}^{\tilde{\;}\nabla})\tilde\omega^\nabla)({\Bbb P}(X))\\
&=&(\bar{\cal C}^{\alpha\beta}\tilde{\cal
R}^\nabla(\bar{\cal X}_\beta) \tilde{\cal R}^\nabla(\bar{\cal
X}_{\alpha}))\tilde\eta({\Bbb P}(X)).
\end{array}
\]
Thus, using that $\bar{\cal C}^{\alpha\beta}=-\bar{\cal
C}^{\beta\alpha},$ it follows that
\[
\tilde\omega^\nabla({\Bbb P}(\tilde{\cal R}^\nabla),{\Bbb
P}(X))=0,\makebox[1cm]{}\tilde\eta({\Bbb P}(\tilde{\cal
R}^\nabla))=1.
\]
Therefore, if $\Tan P_f\cap \#_{\tilde\Lambda^\nabla}((\Tan P_f)^0)=\{0\}$
and $\jmath_f:P_f\to J^{1*}E$ is the canonical inclusion then
${\Bbb P}(\tilde{\cal R}^\nabla)\vert_{P_f}$ is tangent to $P_f$ and
\[
i({\Bbb
P}(\tilde{R}^\nabla)\vert_{P_f})\jmath_f^*(\tilde\omega^\nabla)=0,
\makebox[1cm]{}
i({\Bbb P}(\tilde{\cal
R}^\nabla)\vert_{P_f})\jmath_f^*(\tilde\eta)=1.\] As a consequence, from
Lemmas \ref{10''} and \ref{14'} (see Appendices \ref{A} and
\ref{B}), we deduce that the couple
$(\jmath_f^*(\tilde\omega^\nabla),\jmath_f^*(\tilde\eta))$ is a
cosymplectic structure on $P_f$, and that ${\Bbb P}(\tilde{\cal
R}^\nabla)\vert_{P_f}$ is the Reeb vector field of this structure.}
\end{remark}

\begin{remark}\label{r11-0''}
{\rm If $\nabla'$ is another connection on $\pi:E\to B$, and we
consider the same initial set of constraints
$\{\xi_i^{(0)}\}_{i=1,\dots, 2n+1-p_0}$ defining $P_0$ (as a
submanifold of $J^{1*}E$), then following the above process we
obtain the same final set of constraints
$\{\bar\xi_i^{(0)},\tilde\xi_j^{(0)},\hat\xi_t^{(0)},
\bar{\xi}_r^{(f)},\tilde{\xi}_u^{(f)}\}$
defining the submanifold $P_f$ (see Theorem \ref{t10}). Therefore,
we also obtain the same splitting of $\Tan_UJ^{1*}E$  and the
same projectors
\[
{\Bbb P}:\Tan_UJ^{1*}E\to D,\makebox[1cm]{}{\Bbb Q}:\Tan_UJ^{1*}E\to
\#_{\tilde\Lambda^{\nabla'}}(D^0)=\#_{\tilde{\Lambda}^\nabla}(D^0).
\]}
\end{remark}

 Next, suppose that $h^\nabla:J^{1*}E\to \Real$ is an
arbitrary extension of the Hamiltonian function $h_0^\nabla:P_0\to
\Real$ \vspace{6pt} and that
$E_{h^\nabla}^{\tilde{\;}\nabla}=\tilde{\cal R}^\nabla +
X_{h^\nabla}^{\tilde{\;}\nabla}$ is the evolution vector field of
$h^\nabla$ with respect to the cosymplectic structure
$(\tilde\omega^\nabla, \tilde\eta)$ (see Section
\ref{seccion5.1}).

We will prove that the restriction to $P_f$ of ${\Bbb
P}(E_{h^\nabla}^{\tilde{\;}\nabla})$ is tangent to $P_f$, and that
such a restriction is a solution of the constrained Hamilton
equations. For this purpose, we will use the following result.

\begin{lem}\label{l11-1}
If $x_f$ is a point of $P_f$, then
\[
\{X\in \Tan_{x_f}P_0/ i(X)(\Omega_{h_0}(x_f))\vert_{\Tan_{x_f}P_0}=0,\;
i(X)(\eta^0(x_f))=0\}=\Tan_{x_f}P_0\cap
\#_{{\tilde\Lambda}^\nabla}(x_f)((\Tan_{x_f}P_0)^0).
\]
\end{lem}
\proof Suppose that $X_{h_0}^f$ is a solution of the constrained
Hamilton equations on the submanifold $P_f$; that is,
$X_{h_0}^f\in{\frak X}(P_f)$ and
\[
(i(X_{h_0}^f)\Omega_{h_0})\vert_{P_f}=0,\makebox[1cm]{}
(i(X_{h_0}^f)\eta^0)\vert_{P_f}=1.
\]
We consider the cosymplectic
structure $(\bar\omega^\nabla(x_f),\eta^0(x_f))$ on the vector space
 $\Tan_{x_f}J^{1*}E$,  where
\[
\bar\omega^\nabla(x_f)=\Omega_{h_0}(x_f)-(i(Y_{h_0}^f)\Omega_{h_0})(x_f)\wedge
\eta^0(x_f).
\]
The Reeb vector of this structure is $Y_{h_0}^f(x_f).$ Since
$Y_{h_0}^f(x_f)\in \Tan_{x_f}P_0$ and
$$\bar{\omega}^\nabla(x_f)(Y,Z)=\Omega_{h_0}(x_f)(Y,Z),$$ for
$Y,Z\in \Tan_{x_f}P_0,$ it follows that
\[
\begin{array}{l}
\{X\in \Tan_{x_f}P_0/ (i(X)(\Omega_{h_0}(x_f)))\vert_{\Tan_{x_f}P_0}=0,\;
i(X)(\eta^0(x_f))=0\}\\=\{X\in \Tan_{x_f}P_0/
i(X)(\bar\omega^\nabla(x_f))\vert_{\Tan_{x_f}P_0}=0, \;
i(X)(\tilde\eta(x_f))=0\}\\ = \Tan_{x_f}P_0\cap
\#_{\bar\Lambda^\nabla(x_f)}((\Tan_{x_f}P_0)^0),
\end{array}
\]
where $\bar\Lambda^\nabla(x_f)$ is the $2$-vector on
$\Tan_{x_f}J^{1*}E$ associated with the cosymplectic structure
$(\bar\omega^\nabla(x_f),$ $\tilde\eta(x_f))$ (see Appendix
\ref{B}). Therefore, using Proposition \ref{14''} (see Appendix
\ref{B} ) and the fact that
\[
\bar\omega^\nabla(x_f)=\tilde\omega^\nabla(x_f)+ (dh^\nabla(x_f)-
(i(Y_{h_0}^f)\Omega_{h_0})(x_f))\wedge \tilde\eta(x_f),
\]
we deduce the result. \qed

\begin{teor}\label{t11-2}
The restriction to the submanifold $P_f$ of the vector field
${\Bbb P}(E_{h^\nabla}^{\tilde{\;}\nabla})$ is tangent to $P_f$
and, in addition, this restriction is a solution of the
constrained Hamilton equations; that is,
\begin{equation}
\label{+3} [i({\Bbb
P}(E_{h^\nabla}^{\tilde{\;}\nabla}))\Omega_{h_0}]\vert_{P_f}=0,\;\;
[i({\Bbb P}(E_{h^\nabla}^{{\tilde{\;}}\nabla}))\eta^0]\vert_{P_f}=1.
\end{equation}
\end{teor}
\proof The definition of the distribution $D$ implies that
\begin{equation}\label{x1}
{\Bbb P}(E_{h^\nabla}^{\tilde{\;}\nabla})(\bar {\cal X}_\alpha)=0,
\makebox[1cm]{}\mbox{ for every } \alpha.
\end{equation}
Furthermore, we have that
\begin{equation}\label{*1}
{\Bbb
P}(E_{h^\nabla}^{\tilde{\;}\nabla})=E_{h^\nabla}^{\tilde{\;}\nabla}-
\lambda^iX_{\bar\xi_i^{(0)}}^{\tilde{\;}\nabla}
-\mu^jX_{\tilde{\xi}_j^{(0)}}^{\tilde{\;}\nabla}-
\nu^rX_{\bar\xi_r^{(f)}}^{\tilde{\;}\nabla}.
\end{equation}

As a consequence, from (\ref{+1}) and (\ref{*1}), we obtain
\begin{equation}\label{*2}
\begin{array}{ll}
{\Bbb
P}(E_{h^\nabla}^{\tilde{\;}\nabla})\vert_{P_f}(\hat{\xi}_t^{(0)})=
E_{h^\nabla}^{\tilde{\;}\nabla}\vert_{P_f}(\hat{\xi}_t^{(0)}),& 1\leq
t\leq k_{0f}\\ {\Bbb
P}(E_{h^\nabla}^{\tilde{\;}\nabla})\vert_{P_f}(\tilde{\xi}_u^{(f)})=
E_{h^\nabla}^{\tilde{\;}\nabla}\vert_{P_f}(\tilde{\xi}_u^{(f)}),& 1\leq
u\leq k_f-k_{0f}.
\end{array}
\end{equation}
Now, let $X_{h_0}^f$ be a solution of the constrained Hamilton
equations on the submanifold $P_f.$

Using (\ref{51'}), (\ref{51''})  and the fact that
\[
\Omega_{h_0}(X_{h_0}^f,X_{\tilde{\xi}_j^{(0)}}^{\tilde{\;}\nabla})\vert_{P_f}=
\Omega_{h_0}(X_{h_0}^f,X_{\hat{\xi}_t^{(0)}}^{\tilde{\;}\nabla})\vert_{P_f}=
\Omega_{h_0}(X_{h_0}^f,X_{\tilde{\xi}_u^{(f)}}^{\tilde{\;}\nabla})\vert_{P_f}=0
\]
we deduce that
\begin{equation}\label{*3}
E_{h^\nabla}^{\tilde{\;}\nabla}\vert_{P_f}(\tilde\xi_j^{(0)})=
E_{h^\nabla}^{\tilde{\;}\nabla}\vert_{P_f}(\hat\xi^{(0)}_t)=
E^{\tilde{\;}\nabla}_{h^\nabla}\vert_{P_f}(\tilde\xi_{u}^{(f)})=0
\end{equation}
for $1\leq j\leq k_0-k_{0f},$ $1\leq t\leq k_{0f}$ and $1\leq
u\leq k_f-k_{0f}.$

Therefore, from (\ref{x1}), (\ref{*2}) and (\ref{*3}), we conclude
that the restriction of ${\Bbb
P}(E_{h^\nabla}^{\tilde{\;}\nabla})$ to $P_f$ is tangent to $P_f.$

Next, we will prove that (\ref{+3}) holds.
Using (\ref{51'}), (\ref{51''}) and (\ref{*3}), we have that
\begin{equation}
\label{*4}
(i(X^{\tilde{\;}\nabla}_{\tilde{\xi}_j^{(0)}})\Omega_{h_0})\vert_{P_f}=0,
\makebox[1cm]{}
\mbox{for every $j$, $1\leq j\leq k_0-k_{0f}.$}
\end{equation}
Furthermore, since the matrix
$(\{\bar\xi_i^{(0)},
\bar{\xi}_{i'}^{(0)}\}\vert_{P_f}^{\tilde{\;}\nabla})_{1\leq
i,i'\leq l_0}$ is regular, one can find local real functions
$\{\bar\lambda^i\}_{1\leq i\leq l_0}$ on $J^{1*}E$ such that the
restriction to $P_f$ of the vector field
\[
X^0_{h_0}=E_{h^\nabla}^{\tilde{\;}\nabla}-\bar\lambda^i
X_{\bar\xi_i^{(0)}}^{\tilde{\;}\nabla}
\]
is tangent to $P_0.$ Moreover, it follows that (see (\ref{51'})
and Theorem \ref{t11})
\[
(i(X_{h_0}^0)\Omega_{h_0})\vert_{P_f}=(\bar\lambda^j
E_{h^\nabla}^{\tilde{\;}\nabla}(\bar{\xi}_j^{(0)})\eta^0)\vert_{P_f}.
\]
But as $X_{h_0|P_0}^0$ is tangent to $P_0,$ we have
\[
0=(\bar\lambda^jX_{h_0}^0(\bar\xi_j^{(0)}))\vert_{P_f}=
(\bar\lambda^jE_{h^\nabla}^{\tilde{\;}\nabla}(\bar\xi_j^{(0)}))\vert_{P_f}
-(\bar\lambda^i\bar\lambda^j\{\bar\xi_j^{(0)},
\bar\xi_i^{(0)}\}^{\tilde{\;}\nabla})\vert_{P_f}=
(\bar\lambda^jE_{h^\nabla}^{\tilde{\;}\nabla}(\bar\xi^{(0)}_j))\vert_{P_f}.
\]
Thus, we deduce that
\begin{equation}\label{*5}(i(X_{h_0}^0)\Omega_{h_0})\vert_{P_f}=0,
\makebox[1cm]{}
(i(X_{h_0}^0)\eta^0)\vert_{P_f}=1.
\end{equation}

 This implies that
\[
(i(X_{h_0}^f-X_{h_0}^0)\Omega_{h_0})\vert_{P_f}=0,\makebox[1cm]{}
(i(X_{h_0}^f-X_{h_0}^0)\eta^0)\vert_{P_f}=0 \] and consequently (see
(\ref{+1}) and Lemma \ref{l11-1})
\begin{equation}\label{*6}
X_{h_0}^f=(E_{h_\nabla}^{\tilde{\;}\nabla}-\bar\lambda^i
X^{\tilde{\;}\nabla}_{{\bar\xi}_i^{(0)}}-
\bar\mu^jX^{\tilde{\;}\nabla}_{{\tilde\xi}_j^{(0)}}-
\bar\kappa^kX^{\tilde{\;}\nabla}_{{\hat\xi}_k^{(0)}})\vert_{P_f},
\end{equation}
where $\bar\mu^j$ and $\bar\kappa^k$ are local real functions on
$J^{1*}E.$

Now, from (\ref{*1}), (\ref{*6}), and since
$(X^{\tilde{\;}\nabla}_{\hat\xi_t^{(0)}})\vert_{P_f}$ and $({\Bbb
P}(E_{h^\nabla}^{\tilde{\;}\nabla}))\vert_{P_f}-X_{h_0}^f$ are tangent
to $P_f$, we obtain that the restriction to $P_f$ of the vector
field
\[
Z=(\bar\lambda^i-\lambda^i)X^{\tilde{\;}\nabla}_{\bar\xi_i^{(0)}}
+(\bar\mu^j-\mu^j)X^{\tilde{\;}\nabla}_{\tilde\xi_j^{(0)}}-
\nu^rX_{\bar\xi_r^{(f)}}^{\tilde{\;}\nabla}
\]
is also tangent to $P_f.$ Hence, it follows that $Z(x_f)\in
D(x_f)\cap \#_{\tilde\Lambda^\nabla}(x_f)(D^0(x_f))=\{0\},$ for
all $x_f\in P_f,$ and therefore
\[
(\bar\lambda^i-\lambda^i)\vert_{P_f}=0,\makebox[1cm]{}
(\bar\mu^j-\mu^j)\vert_{P_f}=0,\makebox[1cm]{}(\nu^r)\vert_{P_f}=0,
\]
for every $i,j$ and $r$. This implies that (see (\ref{*1}))
\begin{equation}\label{*7}
({\Bbb P}(E_{h^\nabla}^{\tilde{\;}\nabla}))\vert_{P_f}=
(X_{h_0}^0-\bar\mu^jX_{\tilde\xi_j^{(0)}}^{\tilde{\;}\nabla})\vert_{P_f}.
\end{equation}
Finally, using (\ref{*4}), (\ref{*5}) and (\ref{*7}), we conclude
that
\[
[i({\Bbb
P}(E^{\tilde{\;}\nabla}_{h^\nabla}))\Omega_{h_0}]\vert_{P_f}=0,\makebox[1cm]{}
[i({\Bbb P}(E^{\tilde{\;}\nabla}_{h^\nabla}))\eta^0]\vert_{P_f}=1.
\]
\qed

\begin{remark}\label{r11-2'}
{\rm If $\Tan P_f\cap \#_{\tilde\Lambda^\nabla}((\Tan P_f)^0)=\{0\}$ and
$\jmath_f:P_f\to J^{1*}E$ is the canonical inclusion, then the couple
$(\jmath_f^*(\tilde\omega^\nabla),\jmath_f^*(\tilde\eta))$ defines
a cosymplectic structure on $P_f$ (see Remark \ref{r11-0'}), and
$({\Bbb P}(E_{h^\nabla}^{\tilde{\;}\nabla}))\vert_{P_f}$ is just the
evolution vector field of $(h_0^\nabla)\vert_{P_f}$ with respect to
the structure
$(\jmath_f^*(\tilde\omega^\nabla),\jmath_f^*(\tilde\eta)).$}
\end{remark}

\begin{remark}\label{r11-2''}
{\rm Let ${\cal Y}_{E}'$ be the vector field on $E$ associated
with another connection $\nabla'$ on $\pi:E\to B$, and
$h_0^{\nabla'}:P_0\to \Real$ be the corresponding Hamiltonian
function. Using Remark \ref{r8'}, we deduce that
\[
h^{\nabla'}=h^\nabla-\widetilde{iV}
\]
is an extension to $J^{1*}E$ of the Hamiltonian function
$h_0^{\nabla'}$, where $\widetilde{iV}:J^{1*}E\to \Real$ is the
real function on $J^{1*}E$ induced by the $\pi$-vertical vector
field $V={\cal Y}_{E}'-{\cal Y}_E$. Thus, from Theorem \ref{t10}
and Proposition \ref{14'''} (see Appendix \ref{B}), we obtain that
the evolution vector fields $E_{h^\nabla}^{\tilde{\;}\nabla}$ and
$E_{h^{\nabla'}}^{\tilde{\;}\nabla'}$ of $h^{\nabla}$ and
$h^{\nabla'}$ with respect to the cosymplectic structures
$(\tilde\omega^\nabla, \tilde\eta)$ and
$(\tilde\omega^{\nabla'},\tilde\eta)$ coincide. As a consequence, it
follows that
\[
{\Bbb P}(E_{h^\nabla}^{\tilde{\;}\nabla})={\Bbb
P}(E_{h^{\nabla'}}^{\tilde{\;}{\nabla'}}). \]} \end{remark}

\subsection{Dirac brackets and evolution of an observable}

Splitting $\Tan_UJ^{1*}E=D\oplus
\#_{\tilde{\Lambda}^\nabla}(D^0)$ allows us to introduce a {\sl
Dirac bracket } $\{\;,\;\}_D^{\tilde{\;}\nabla}$, which can be
defined as follows (see Appendix \ref{B}). If $F$ and $G$ are
$C^\infty$-differentiable real functions on $U$, then
\[
\{F,G\}_D^{\tilde{\;}\nabla}=\tilde\omega^\nabla({\Bbb
P}(X_{F}^{\tilde{\;}\nabla}), {\Bbb P}(X_G^{\tilde{\;}\nabla})),
\]
where ${\Bbb P}: \Tan_UJ^{1*}E\to D$ is the projector considered in
Section \ref{seccion5.2}.

A direct computation, using (\ref{+2'}), proves that
\begin{equation}\label{*8}
\{F,G\}_D^{\tilde{\;}\nabla}=\{F,G\}^{\tilde{\;}\nabla} +
\bar{\cal C}^{\alpha\beta}\{F,\bar {\cal
X}_\beta\}^{\tilde{\;}\nabla}\{\bar{\cal
X}_\alpha,G\}^{\tilde{\;}\nabla}.
\end{equation}
Moreover, we have
\begin{teor}\label{t11-3}
$(i)$ The Dirac bracket $\{\;,\;\}_D^{\tilde{\;}\nabla}$ is a
Poisson bracket and the Hamiltonian vector field of a
$C^\infty$-differentiable real function $F$ on $U$ (with respect
to $\{\;,\;\}_D^{\tilde{\;}\nabla}$) is ${\Bbb
P}(X_F^{\tilde{\;}\nabla}).$

$(ii)$ If $g$ is an observable (that is, $g$ is a
$C^\infty$-differentiable real function on $P_f$), its evolution
is given by the formula
\[
\dot{g}={\Bbb P}(\tilde{\cal R}^\nabla)\vert_{P_f}(G) +
(\{G,h^\nabla\}_D^{\tilde{\;}\nabla})\vert_{P_f},
\]
where $h^\nabla:U\to \Real$ and $G:U\to \Real$ are arbitrary
extensions to $U$ of the Hamiltonian function $h_0^\nabla:P_0\to
\Real$ and of $g:P_f\to \Real$, respectively.

$(iii)$ If $F$ is a $C^\infty$-differentiable real function on $U$
such that $X_{F}^{\tilde{\;}\nabla}(x)\in
\#_{\tilde\Lambda^\nabla}(D^0)(x),$ for every $x\in U$, then $F$ is
a Casimir function for the Dirac bracket, i.e.,
\[
\{F,G\}_{D}^{\tilde{\;}\nabla}=0,\makebox[1cm]{} \mbox{for every
$G\in C^\infty(U).$}
\]
$(iv)$ If $F$ is a $C^\infty$-differentiable real function on $U$
such that the restriction to $P_f$ of $X_F^{\tilde{\;}\nabla}$ is
tangent to $P_f$, then
$(\{F,G\}_{D}^{\tilde{\;}\nabla})\vert_{P_f}=
(\{F,G\}^{\tilde{\;}\nabla})\vert_{P_f},$
for every $G\in C^\infty(U).$
\end{teor}
\proof $(i)$ Using Proposition \ref{p11-B} (see Appendix \ref{B}),
and the fact that $D$ is a completely integrable distribution, we
deduce that $\{\;,\;\}_D^{\tilde{\;}\nabla}$ is a Poisson bracket.
Furthermore, from (\ref{+2'}) and (\ref{*8}), we obtain that
\[
{\Bbb
P}(X_F^{\tilde{\;}\nabla})(G)=\{G,F\}_D^{\tilde{\;}\nabla},
\makebox[1cm]{}\mbox{for
} F,G\in C^\infty(U).
\]
\medskip
$(ii)$ If $E_{h^\nabla}^{\tilde{\;}\nabla}=\tilde{\cal R}^\nabla +
X_{h^\nabla}^{\tilde{\;}\nabla}$ is the evolution vector field of
$h^\nabla$ with respect to the cosymplectic structure
$(\tilde{\omega}^\nabla,\tilde\eta)$, then from Theorem
\ref{t11-2} it follows that
\[
\dot{g}={\Bbb P}(E_{h^\nabla}^{\tilde{\;}\nabla})\vert_{P_f}(g).
\]
Thus, using the first part of this Theorem, we have that
\[
\dot{g}={\Bbb P}(\tilde{\cal R}^\nabla)\vert_{P_f}(G)+ {\Bbb
P}(X_{h^\nabla}^{\tilde{\;}\nabla})\vert_{P_f}(G)= {\Bbb
P}(\tilde{\cal R}^\nabla)\vert_{P_f}(G) +
(\{G,h^\nabla\}_D^{\tilde{\;}\nabla})\vert_{P_f}.
\]
$(iii)$ If $F\in C^\infty(U)$ and $X_F^{\tilde{\;}\nabla}(x)\in
\#_{\tilde \Lambda^\nabla}(D^0)(x),$ for every $x\in U$, then it is
clear that ${\Bbb P}(X_F^{\tilde{\;}\nabla})=0.$ Therefore,
\[
\{F,G\}_D^{\tilde{\;}\nabla}=-\{G,F\}_D^{\tilde{\;}\nabla}=-{\Bbb
P}(X_F^{\tilde{\;}\nabla})(G)=0,\makebox[1cm]{}\mbox{ for every
$G\in C^\infty(U).$}
\]
$(iv)$ The condition $(X_F^{\tilde{\;}\nabla})\vert_{P_f}\in {\frak X}
(P_f)$ implies that $({\Bbb
P}(X_F^{\tilde{\;}\nabla}))\vert_{P_f}=(X_{F}^{\tilde{\;}\nabla})\vert_{P_f}$,
and consequently
\[
(\{F,G\}_D^{\tilde{\;}\nabla})\vert_{P_f}=
-(X_F^{\tilde{\;}\nabla})\vert_{P_f}(G)=
(\{F,G\}^{\tilde{\;}\nabla})\vert_{P_f}.
\]
\qed

 From Theorem \ref{t11-3}, we deduce that
\[
\{\bar{\cal X}_\alpha, F\}_D^{\tilde{\;}\nabla}=0,
\]
for every $\alpha$ and $F\in C^\infty(U)$. In other words, the
second class constraints $\{\bar{\cal X}_\alpha\}$ are Casimir
functions for the Dirac bracket $\{\;,\;\}_D^{\tilde{\;}\nabla}.$
Using Theorem \ref{t11-3}, we also obtain that
\[
(\{\hat\xi_t^{(0)},F\}_D^{\tilde{\;}\nabla})\vert_{P_f}=
(\{\hat\xi_t^{(0)},F\}^{\tilde{\;}\nabla})\vert_{P_f},\makebox[1cm]{}
\{\tilde\xi_u^{(f)},F\}_D^{\tilde{\;}\nabla})\vert_{P_f}=
\{\tilde\xi_u^{(f)},F\}^{\tilde{\;}\nabla})\vert_{P_f},
\]
for every $t$ and $u$, $1\leq t\leq k_{0f}$ and $1\leq u\leq
k_f-k_{0f}.$

\begin{remark}\label{r11-3'}
{\rm If $\Tan P_f\cap \#_{{\tilde{\Lambda}}^\nabla}(\Tan (P_f)^0)=\{0\}$
and $\jmath_f:P_f\to J^{1*}E$ is the canonical inclusion, we can
consider the cosymplectic structure
$(\jmath_f^*(\tilde\omega)^\nabla),\jmath_f^*(\tilde\eta))$ on
$P_f$ whose Reeb vector field is ${\Bbb P}(\tilde{\cal
R}^\nabla)\vert_{P_f}$ (see Remark \ref{r11-0'}). Furthermore,
if $X\in {\frak X}(U)$ and $G\in C^\infty(U),$
\[
(i({\Bbb P}(X_G^{\tilde{\;}\nabla}))\tilde\omega^\nabla)({\Bbb
P}(X))=(dG-\tilde{\cal R}^\nabla(G)\tilde\eta)({\Bbb
P}(X))-\tilde\omega^\nabla({\Bbb Q}(X_G^{\tilde{\;}\nabla}), {\Bbb
P}(X)).
\]
Thus, from (\ref{+2'}), we have that
\[
(i({\Bbb P}(X_G^{\tilde{\;}\nabla}))\tilde\omega^\nabla)({\Bbb
P}(X))=dG({\Bbb P}(X))-{\Bbb P}(\tilde{\cal
R}^\nabla)(G)\tilde\eta({\Bbb P}(X)) \] and, since ${\Bbb
P}(X_G^{\tilde{\;}\nabla})\vert_{P_f}$ is tangent to $P_f$, it follows
that ${\Bbb P}(X_G^{\tilde{\;}\nabla})\vert_{P_f}$ is the Hamiltonian
vector field of $g=G\vert_{P_f}$ with respect to the cosymplectic
structure
$(\jmath_f^*(\tilde\omega)^\nabla),\jmath_f^*(\tilde\eta))$.
Therefore, if on $P_f$ (resp. $J^{1*}E$) we consider the Poisson
structure induced by the cosymplectic structure
$(\jmath_f^*(\tilde\omega)^\nabla),\jmath_f^*(\tilde\eta))$ (resp.
by the Dirac bracket $\{\;,\;\}_D^{\tilde{\;}\nabla}$) then the
canonical inclusion $\jmath_f$ is a Poisson morphism.}
\end{remark}

\begin{remark}\label{r11-3''}
{\rm If $\nabla'$ is another connection on $\pi:E\to B$, then we
can obtain the same splitting of $\Tan_{U}(J^{1*}E)$ and the same
projectors ${\Bbb P}:\Tan_UJ^{1*}E\to D$ and ${\Bbb Q}:\Tan_UJ^{1*}E\to
\#_{\tilde\Lambda^{\nabla'}}(D^0)=\#_{\tilde\Lambda^\nabla}(D^{0}).$
As a consequence, using (\ref{*8}) and Theorem \ref{t10}, we conclude
that the Dirac brackets coincide, i.e.,
\[
\{\;,\;\}_D^{\tilde{\;}\nabla}=\{\;,\;\}_D^{\tilde{\;}\nabla'}.
\]}
\end{remark}

\section{Examples}

In this Section, we illustrate some aspects of the theory
with two examples. In the first, we treat the particular
case where the base manifold $B$ is $\Real$ and the submersion
$\pi$ is a trivial fibration of $E$ on $\Real$ and, as a
consequence, we recover some results obtained in
\cite{CLM-94,LMM-96}. In the second example, we consider the
special case of time-dependent Lagrangians which are affine on the
velocities.

\subsection{Example 1}
Suppose that $B=\Real$ and that $\pi:E\to \Real$ is a trivial
fibration; that is, $Q$ is a differentiable manifold of dimension
$n$, $E=\Real\times Q$ and $\pi:E=\Real\times Q\to \Real$ is the
canonical projection over the first factor. Then, the evolution
phase space $J^1E$ of the system can be identified with the
product manifold $\Real\times \Tan Q$ in such a way that the fibration
$\pi^1:J^1E\to \Real$ is the map $\bar\tau_Q=Id\times \tau_Q:\Real
\times \Tan Q\to \Real\times Q,$ where $\tau_Q:\Tan Q\to Q$ is the
canonical projection (see \cite{Sa-89}).

Now, let $t$ be the usual coordinate on $\Real$ and $\varpi$ the
volume form given by $\varpi=dt.$ If $\Lag$ is a Lagrangian
density on $J^1E\cong \Real \times \Tan Q,$ we have that ${\cal
L}=\lag\eta,$ with $\eta=(\bar\pi^1)^*(dt)$ and $\lag:\Real \times
\Tan Q\to \Real$ the Lagrangian function.

The Lagrangian function and the {\sl Liouville vector field} $C$
of $\Tan Q$ allow us to introduce the {\sl Lagrangian energy}
$E_\Lag:\Real \times \Tan Q\to \Real$ as the real function on
$\Real\times \Tan Q$ defined by
\[
E_\Lag=C(\lag)-\lag.
\]
If $(t,q^\nu,v^\nu)$ are fibered coordinates on $\Real\times \Tan Q$,
we have that
\[
E_\Lag=v^\nu\frac{\partial \lag}{\partial v^\nu}-\lag.
\]
Furthermore, the vector field ${\cal Y}_E=\frac{\partial
}{\partial t}$ (resp. ${\cal Y}=\frac{\partial}{\partial t})$ on
$\Real \times Q$ (resp. $\Real\times \Tan Q$) induces a distinguished
connection $\nabla$ (resp. $\bar\nabla$) on the fibration
$\pi:\Real \times Q\to \Real$ (resp. $\bar\tau_Q:\Real\times \Tan Q\to
\Real).$) Thus, if $M_i$ is the $i$th-Lagrangian dynamical
constraint submanifold, one can give several descriptions of $M_i$
as the zero set of the $i$th-generation dynamical Lagrangian
constraints $\{\zeta^{(i)}\}$ (see Theorem \ref{lagdyncons}). In
particular, using Theorem \ref{lagdyncons}, we deduce that
\[
\zeta^{(i)}=\inn({\cal
Z}^{(i)})(\eta-\gamma_\Lag),\makebox[1cm]{}\mbox{ for } {\cal
Z}^{(i)}\in {\frak X}^\perp(M_i),
\]
where $\gamma_\Lag$ is the $1$-form on $\Real \times \Tan Q$ given by
$\gamma_{\Lag}=\inn(\frac{\partial}{\partial t}){\Omega_{\cal L}}$,
and $\Omega_{\Lag}$ is the Poincare-Cartan $2$-form. This
description of the constraints $\zeta^{(i)}$ was obtained in
\cite{CLM-94}.
On its turn, the description of the $i$th-generation of
 SODE Lagrangian constraints is given in Theorem \ref{lagsodecons}.

Next, we will assume that the Lagrangian system
$(J^1E,\Omega_{\Lag})$ is almost regular.

Note that in the particular case where the fibration $\pi$ is
trivial, the restricted momentum dual bundle associated with $\pi$
can be identified with the product manifold $\Real\times \Tan^*Q$.
In addition, the map
\[
\begin{array}{rcl}
J^{1*}E\times \Real\cong (\Real\times \Tan^*Q)\times
\Real&\longrightarrow &\Tan^*E=\Tan^*(\Real\times Q)\\
((\lambda,\alpha),\mu)&\longrightarrow & \alpha + \mu dt(\lambda)
\end{array}
\]
is a diffeomorphism. Moreover, in \cite{CLM-94} it was proved that
the Lagrangian energy $E_\Lag$ is ${\cal FL}_0$-projectable onto a
real function $\tilde{h}_0^\nabla:{\cal P}\to \Real.$ Using this
function, a diffeomorphism $\tilde{h}_0:{\cal
P}\subseteq \Real\times \Tan^*Q\to \tilde{\cal P}\subseteq \Tan^*E\cong
(\Real\times \Tan^*Q)\times \Real$ can be defined as follows
\[
\tilde{h}_0(\tilde{x})=(\tilde{x},-\tilde{h}_0^\nabla(\tilde{x})),
\makebox[1cm]{}\mbox{ for every }\tilde{x}\in {\cal P}.
\]
A direct computation shows that $\tilde{h}_0=h_0=\mu_0^{-1},$
where $\mu_0:\tilde{\cal P}\to {\cal P}$ is the restriction to
$\tilde{\cal P}$ of the canonical projection
 $\mu:\Tan^*(\Real\times Q)\to \Real\times \Tan^*Q$
 (see Section \ref{seccion4.1}).
Therefore, the Hamilton-Cartan forms $\Theta_{h_0}$  and
$\Omega_{h_0}$ are given by $\Theta_{h_0}=(\tilde\jmath_0\circ
\tilde h_0)^*\Theta$ and
 $\Omega_{h_0}=(\tilde\jmath_0\circ\tilde h_0)^*\Omega,$
 $\Theta$ and $\Omega$ being the Liouville
$1$-form and the Liouville $2$-form, respectively, of
 $\Tan^*(\Real\times Q).$

Now, if $\tilde{\cal Y}$ is a vector field on ${\cal P}$ which is
$\tau_0^1$-projectable onto the vector field
 ${\cal Y}_E=\frac{\partial}{\partial t}$, then $\tilde{\cal Y}$
induces a connection $\tilde\nabla$ in the fibration
 $\bar\tau_0^1:{\cal
P}\to \Real.$ Using this connection, one can give several
descriptions of the $i$th-Hamiltonian dynamical constraint
submanifold $P_i$ as the zero set of the $i$th-generation
dynamical Hamiltonian constraints $\{\xi^{(i)}\}$ (see Theorem
\ref{hamcons}). In particular, from Theorem \ref{hamcons}, it
follows that
\[
\xi^{(i)}=\inn (\tilde{\cal
Z}^{(i)})(\eta^0-\gamma_{h_0}),\makebox[1cm]{} \mbox{ for }
\tilde{\cal Z}^{(i)}\in {\frak X}^\perp(P_i),
\]
where $\eta^0=(\bar\tau_{0}^1)^*dt$ and $\gamma_{h_0}$ is the
$1$-form on ${\cal P}$ given by  $\gamma_{h_0}=\inn(\tilde{\cal
Y})\Omega_{h_0}.$ This description of the constraints $\xi^{(i)}$
was obtained in \cite{CLM-94}.

Moreover, it is clear that $\tilde{h}_0^\nabla$ is just
the Hamiltonian function $h_0^\nabla$ on ${\cal P}$ associated
with the Lagrangian system, the connection $\nabla$ on $\pi:E\to
\Real$ and the $1$-form $\varpi=dt.$ In addition, if
$h^\nabla:\Real \times \Tan^*Q\to \Real$
 is an extension to
 $\Real\times \Tan^*Q$ of the Hamiltonian function $h_0^\nabla$, then we
deduce that $\tilde\Theta^\nabla$ and $\tilde\omega^\nabla$ are
the Liouville $1$-form and the Liouville $2$-form of $\Tan^*Q$, respectively,
that $\tilde{\cal R}^\nabla=\frac{\partial }{\partial t}$,
and that
\[
\Theta_{h^\nabla}=\tilde\Theta^\nabla + h^\nabla
dt,\makebox[1cm]{} \Omega_{h^\nabla}=\tilde\omega^\nabla +
dh^\nabla\wedge dt.
\]
As a consequence, the bracket $\{\;,\;\}^{\tilde{\;}\nabla}$ on $\Real
\times \Tan^*Q$ is the usual Poisson bracket on $\Real\times \Tan^*Q$
(see Section \ref{seccion5.1}).

Finally, suppose that $\{\bar{\cal X}_\alpha\}$ is a maximal set
of local independent second class constraints for the final
Hamiltonian dynamical constraint submanifold $P_f$, and denote by
${\cal C}_{\alpha\beta}$ the real function defined by
\[
{\cal C}_{\alpha\beta}=\bar{\cal C}_{\alpha\beta} + \frac{\partial
 \bar{\cal X}_\alpha}{\partial t}\frac{\partial
 \bar{\cal X}_\beta}{\partial t}=\{\bar{\cal X}_\alpha,\bar{\cal
 X}_\beta\}^{\tilde{\;}\nabla} + \frac{\partial
 \bar{\cal X}_\alpha}{\partial t}\frac{\partial
 \bar{\cal X}_\beta}{\partial t},\makebox[1cm]{} \mbox{for every
 }\alpha \mbox{ and } \beta.
 \]
 Then, since the matrix $(\bar{\cal C}_{\alpha\beta})$ is regular,
 we deduce that the matrix $({\cal C}_{\alpha\beta})$ also is
 regular. In fact, if $(\bar{\cal C}^{\alpha\beta})$ is the inverse
 matrix of $(\bar{\cal C}_{\alpha\beta})$ and
 \[
 {\cal C}^{\alpha\beta}=
\bar{\cal C}^{\alpha\beta}
 +\bar{\cal C}^{\alpha\gamma}\bar{\cal
 C}^{\beta\nu}
 \frac{\partial
 \bar{\cal X}_\gamma}{\partial t}\frac{\partial
 \bar{\cal X}_\nu}{\partial t}
\]
then $({\cal C}^{\alpha\beta})$ is the inverse matrix of $({\cal
C}_{\alpha\beta})$ (note that the skew-symmetric character of the
matrix $(\bar{\cal C}^{\alpha\beta})$ implies that
$\displaystyle\bar{\cal C}^{\gamma'\nu'}\frac{\partial
 \bar{\cal X}_{\gamma'}}{\partial t}\frac{\partial
 \bar{\cal X}_{\nu'}}{\partial t}=0).$ Thus, if
 $\{\;,\;\}^{\tilde{\;}\nabla}_D$ is the corresponding Dirac
 bracket on $\Real \times \Tan^*Q,$ using (\ref{*8}), we conclude that
 \[
 \begin{array}{lcl}
 \{F,G\}_D^{\tilde{\;}\nabla}&=&\{F,G\}^{\tilde{\;}\nabla}-\displaystyle
 {\cal C}_{\alpha\beta}\{\bar{\cal
 X}_\beta,G\}^{\tilde{\;}\nabla}\{\bar{\cal X}_\alpha,F\}^{\tilde{\;}\nabla}
\\&&
 -\displaystyle{\cal C}_{\alpha\beta}{\cal
 C}_{\alpha'\beta'}\{\bar{\cal
 X}_\beta,G\}^{\tilde{\;}\nabla}\{\bar{\cal X}_{\beta'},F\}^{\tilde{\;}\nabla}
\frac{\partial
 \bar{\cal X}_{\alpha}}{\partial t}\frac{\partial
 \bar{\cal X}_{\alpha'}}{\partial t},
 \end{array}\]
 for $F,G\in C^\infty(\Real \times \Tan^*Q).$ This expression for the
 Dirac bracket $\{\;,\;\}_D^{\tilde{\;}\nabla}$ was obtained in
 \cite{LMM-96}.

\subsection{Example 2}
Let $\gamma$ be a $1$-form on the configuration space $E$ and
$\lag:J^1E\to \Real$ be the Lagrangian function defined by
\[
\lag(j_t^1\phi)=\hat\gamma(j_t^1\phi)=\gamma(\dot{\phi}(t)),\makebox[1cm]{}
\mbox{
for } j_t^1\phi\in J^1E.
\]
If we take fibered coordinates $(t,q^i,v_i)$ on $J^{1}E$ such that
$\gamma=\gamma_i(t,q)dq^i + \gamma_0(t,q)dt$ then
\[
\hat{\gamma}(t,q^i,v^i)=\gamma_i(t,q)v^i + \gamma_0(t,q).
\]
Thus $\lag=\hat\gamma$ is an {\sl affine Lagrangian on the
velocities}.

A direct computation shows that $\Theta_\Lag=(\pi^1)^*\gamma$ and
hence $\Omega_\Lag=-d\Theta_\Lag=(\pi^1)^*(-d\gamma).$ We also
obtain that $\widetilde{\cal FL}=\gamma\circ \pi^1$. Therefore,
$\tilde{\cal P}$ is an embedded submanifold of $\Tan^*E$ which is
diffeomorphic to $E$. The mapping $\widetilde{\cal FL}_0:J^1 E\to
\tilde{\cal P}=\gamma(E)$ may be viewed as the composition
\[
J^1 E\stackrel{\pi^1}\longrightarrow
E\stackrel{\gamma}\longrightarrow\gamma(E)=\tilde{\cal P}.
\]
Since $\gamma:E\to \gamma(E)$ is a diffeomorphism and $\pi^1$ is a
surjective submersion with connected fibers, we deduce that
$\widetilde{\cal FL}_0:J^1E \to \tilde{\cal P}$ is also  a
surjective submersion with connected fibers. Moreover, it
is easy to prove that $\widetilde{\cal FL}_0^{-1}(\widetilde{\cal
FL}_0(x))={\cal FL}_0^{-1}({\cal
FL}_0(x))=(\pi^1)^{-1}(\pi^1(x)),$ for every $x\in J^1E.$
As a consequence, the Lagrangian is almost-regular. Note that the
diagram \[
\begin{picture}(200,150)(0,0)
\put(0,100){\mbox{$E$}}

\put(10,105){\vector(2,1){60}} \put(30,125){\mbox{$\gamma$}}
\put(75,135){\mbox{$\tilde{\cal P}$}}

\put(70,65){\mbox{$J^1E$}} \put(70,70){\vector(-2,1){60}}
\put(30,70){\mbox{$\pi^1$}}

\put(80,80){\vector(0,1){50}} \put(85,105){\mbox{$\widetilde{\cal
FL}_0$}}

\put(90,140){\vector(1,0){75}} \put(120,145){\mbox{$\mu_0$}}
\put(170,135){\mbox{${\cal P}={\cal FL}_0(J^1 E)$}}
\put(85,75){\vector(3,2){85}} \put(130,90){\mbox{${\cal FL}_0$}}
\end{picture}\]
\vspace{-80pt}

\noindent is commutative, that the mappings $\gamma:E\to
\tilde{\cal P}$ and $\mu_0:\tilde{\cal P}\to {\cal P}$ are
diffeomorphisms and that $(\mu_0\circ
\gamma)^*(\Omega_{h_0})=-d\gamma$, and  $(\mu_0\circ
\gamma)^*(\eta^0)=\pi^*(\varpi).$ Thus, the submanifold ${\cal P}$
can be identified with $E$ and,with this identification, the
mapping ${\cal FL}_0:J^1E\to {\cal P}$ is the submersion
$\pi^1:J^1E\to E$ and the $2$-form $\Omega_{h_0}$ (resp. the
$1$-form $\eta^0$) on ${\cal P}$ is the $2$-form $-d\gamma$ (resp.
the $1-$form  $\pi^*(dt))$. Taking these identifications into account,
 the constrained Hamilton equations on $E$ are
\begin{equation}\label{.1}
i(Y)(d\gamma)=0,\makebox[1cm]{} i(Y)(\pi^*(dt))=1.
\end{equation}
Next, we will  assume that the couple $(d\gamma,\pi^*(dt))$ is a
cosymplectic structure on $E$ with Reeb vector field ${\cal
R}_\gamma$. Then, a direct computation shows that the matrix
$(\gamma_{ij}=\displaystyle\frac{\partial \gamma_j}{\partial
q^i}-\displaystyle\frac{\partial\gamma_i}{\partial q^j})_{1\leq
i,j\leq n}$ is regular and that
\begin{equation}\label{.2}
{\cal R}_\gamma=\frac{\partial}{\partial t} +
\frac{1}{2}\gamma^{ij}\left(\frac{\partial \gamma_0}{\partial
q^i}-\frac{\partial\gamma_i}{\partial t}\right)\frac{\partial }{\partial
q^j},
\end{equation}
where $(\gamma^{ij})_{1\leq i,j\leq n}$ is the inverse matrix of
$(\gamma_{ij})_{1\leq i,j\leq n}.$

Furthermore, it is clear that the final constraint
submanifold for the Hamiltonian problem is $P_0={\cal P}$ and that
there exists a unique solution $Y_{h_0}$ of the constrained
Hamiltonian equations (\ref{.1}); namely, the Reeb vector field
${\cal R}_\gamma$.

 Moreover, if $X_{\Lag}$ is a vector field on
$J^{1}E$ which projects via ${\cal FL}_0$ onto ${\cal R}_\gamma=Y_{h_0}$,
 then $X_{\Lag}$ is a  solution of the
dynamical Lagrangian equations on $J^1E$. In addition, the
corresponding submanifold $S$ of $J^1E$ (see Section
\ref{seccion4.5}) is ${\cal R}_{\gamma}(E)$ and the unique vector
field $X_\Lag^S\equiv\Gamma_\Lag$ on $S$ satisfying
\[
\inn(\Gamma_\Lag)\Omega_\Lag\vert_S=0\; ,\;
 \inn(\Gamma_\Lag)\eta\vert_S=1,\makebox[1cm]{}
 {\cal J}(\Gamma_\Lag)\vert_S=0
\]
is $\Gamma_\Lag=({\cal R}_\gamma)^c\vert_S$, where $({\cal R}_\gamma)^c$
 denotes the complete lift of ${\cal R}_\gamma$ to $\Tan E$ (see
\cite{LMM-96b}).
 From (\ref{.2}) and the results of Section \ref{seccion4.5}, we
deduce that $S$ can be locally described as follows
\[
S=\{(t,q^j,v^j)\in
J^1E\ \mid\ v^j=\frac{1}{2}\gamma^{ij}\left(\frac{\partial
\gamma_0}{\partial q^i}-\frac{\partial \gamma_i}{\partial t}\right),\;\;
\forall j\}.
\]

Another way of  finding Euler-Lagrange vector fields is by applying
the procedure described in Section \ref{sodeprob}. As there is solution
of the dynamical Lagrangian equations (\ref{eq01})
everywhere in $J^1E$, we look for the submanifold of $J^1E$
where solutions satisfying the SODE condition exist.
The corresponding compatibility conditions ($1$st-generation
SODE Lagrangian constraints) are given by Prop. \ref{submS1}.
In this case, these constraints are
$$
\phi^{(1)_j}:=v^j-\frac{1}{2}\gamma^{ij}\left(\frac{\partial
\gamma_0}{\partial q^i}-\frac{\partial \gamma_i}{\partial t}\right)
$$
Hence $S_1=S$. The corresponding Euler-Lagrange vector fields are
$$
\Gamma_\Lag\vert_{S_1}=\frac{\partial}{\partial t} +
\frac{1}{2}\gamma^{ij}\left(\frac{\partial \gamma_0}{\partial
q^i}-\frac{\partial\gamma_i}{\partial t}\right)\frac{\partial }{\partial
q^j}+F^j\derpar{}{v^j}.
$$
The tangency conditions \dst\Gamma_\Lag(\phi^{(1)_j})\vert_{S_1}=0\) determine
all the coefficients $F^j$, then we obtain the unique solution
$\Gamma_\Lag\vert_{S_1}=({\cal R}_\gamma)^c\vert_{S_1}$. Thus, $S_F=S_1=S$
(see the comments made in Section \ref{pelvf}).

Now, let $\nabla$ be a connection on $\pi:E\to B$, and ${\cal Y}_E$
be the associated vector field on $E$ such that
$\pi^*(\varpi)({\cal Y}_E)=1.$

Using (\ref{5.2}) and (\ref{5.2 0}), we obtain that
\[
h_0^\nabla=-\gamma({\cal Y}_E),
\]
$h_0^\nabla:P_0\cong E\to \Real$ being the Hamiltonian function.

Moreover, a set of local independent constraint functions
defining $P_0={\cal P}$ as a submanifold of $J^{1*}E$ is given by
\begin{equation}\label{.3}
\xi_i^{(0)}=p_i-\gamma_i(t,q),\makebox[1cm]{} 1\leq i\leq n.
\end{equation}
 From (\ref{54'}) and (\ref{.3}) it follows that
\begin{equation}\label{.4}
\{\xi_i^{(0)},\xi_j^{(0)}\}^{\tilde{\;}\nabla}\vert_{P_0(\cong E)}=
\frac{\partial \gamma_j}{\partial
q^i}-\frac{\partial \gamma_i}{\partial q^j}=\gamma_{ij}
\end{equation}
and, hence we conclude that the matrix
$(\{\xi_i^{(0)},\xi_j^{(0)}\}^{\tilde{\;}\nabla}\vert_{P_0})_{1\leq
i,j\leq n}$ is regular. Thus, all the constraints
$\{\xi_i^{(0)}\}_{1\leq i\leq n}$ are second class and $\Tan P_0\cap
\#_{\tilde{\Lambda}^\nabla}((\Tan P_0)^0)=\{0\}.$ Furthermore, with
the identification $P_0={\cal P}\cong E$, the
cosymplectic structure on $E$ induced by the couple
$(\tilde\omega^\nabla,\tilde\eta)$ (see Remark
\ref{r11-0'}) is
\[
(-d\gamma + d(\gamma({\cal Y}_E))\wedge \pi^*(dt),\pi^*(dt)).
\]
Finally, if $\{\;,\;\}_D^{\tilde{\;}\nabla}$ is the Dirac  bracket
on $J^{1*}E$, then using (\ref{54'}), (\ref{*8}) and (\ref{.4}),
we deduce that
$$
\{F,G\}_D^{\tilde{\;}\nabla}=\left(\displaystyle\frac{\partial
F}{\partial q^i}\displaystyle\frac{\partial G}{\partial
p_i}-\displaystyle\frac{\partial F}{\partial
p_i}\displaystyle\frac{\partial G}{\partial q^i}\right)
-\gamma^{ij} \left(\displaystyle\frac{\partial F}{\partial
q^i} + \displaystyle\frac{\partial \gamma_i}{\partial
q^k}\frac{\partial F}{\partial p_k}\right)\left(\displaystyle\frac{\partial
G}{\partial q^j} + \displaystyle\frac{\partial
\gamma_j}{\partial q^l}\frac{\partial F}{\partial p_l}\right)
$$
for $F,G\in C^\infty(J^{1*}E).$

 \subsection*{Acknowledgments}

We are grateful for the financial support of the CICYT PB98-0821, PB97-1257,
PB98-0920, BFM2000-0808 and PGC2000-2191-E.
We wish to thank Mr. Jeff Palmer for his
assistance in preparing the English version of the manuscript.

\section*{Appendices}

\appendix
\section{Precosymplectic vector spaces}\label{A}

Let $V$ be a real vector space of finite dimension, and suppose
that $\omega$ is a $2$-form on $V$ and $\eta\in V^*$. We can
consider the linear map $\flat_{(\eta,\omega)}:V\to V^*$ given by
\[
\flat_{(\eta,\omega)}(v)=\inn (v)\omega + (\inn (v)\eta)\eta,
\makebox[1cm]{}\mbox{ for every } v\in V.
\]
If $S$ is a subspace of $V$, we will denote by $S^0$ the
annihilator of $S$; that is,
\[
S^0=\{\alpha\in V^*/\inn (u)\alpha=0,\;\;\forall u\in S\},
\]
and, in the same way, for a subspace $W$ of $V^*$,
\[
W^0=\{u\in V/i(u)\alpha=0,\;\; \forall \alpha\in W\}.
\]
Then, if $K$ is a subspace of $V$ we define the {\sl orthogonal
subspace } of $K$ with respect to $\eta$ and $\omega$ as
\begin{equation}\label{1}
K^\perp=(\flat_{(\eta,\omega)}(K))^0=\{v\in V/
(\inn(v)\omega-(\inn (v)\eta)\eta)\vert_K=0\}.
\end{equation}

A {\sl precosymplectic structure } on a real vector space on $V$
is a triad $(\eta,\omega, {\cal R})$, where $\eta\in V^*,$
$\omega$ is a $2$-form on $V$, ${\cal R}\in V$ and
\[
\inn ({\cal R})\omega=0,\;\;\;\; i({\cal R})\eta=1.
\]
Note that, in this case, the rank of $\flat_{(\eta,\omega)}$ is
odd. Moreover, we have
\begin{lem}\label{10'}
Let $(\eta,\omega,{\cal R})$ be a precosymplectic structure on a
real vector space. Then:
\begin{enumerate}
\item[$(i)$]
$V^\perp=\ker \omega\cap \ker\eta.$
\item[$(ii)$]
If $K$ is a subspace of $V$, then
$\dim K^\perp=\dim V-\dim K+ \dim(V^\perp\cap K)$.
\item[$(iii)$]
If $K$ and $K'$ are subspaces of $V$, $K'\subseteq K$ then
$K^\perp\subseteq (K')^\perp.$
\end{enumerate}
\end{lem}
\proof $(i)$ It is clear that $\ker\omega\cap \ker\eta\subseteq
V^\perp$. Conversely, if $v\in V^\perp$ then
\[
0=(\inn(v)\omega-(\inn (v)\eta)\eta)({\cal R})=-\inn (v)\eta ,
\]
which implies that $\inn (v)\omega=0,$ that is, $v\in
\ker\omega\cap \ker\eta.$

$(ii)$ Consider the linear map
\[
\flat_{(\eta,\omega)|K}:K\to V^*,\;\;\;\; u\in K\to
\flat_{(\eta,\omega)}(u)=\inn(u)\omega + (\inn(u)\eta)\eta\in V^*.
\]
A direct computation shows that
$\ker(\flat_{(\eta,\omega)|K})=V^\perp\cap K.$ Moreover,
we have
\[
\flat_{(\eta,\omega)|K}(K)=\flat_{(\eta,\omega)}(K)=
((\flat_{(\eta,\omega)}(K))^0)^0=(K^\perp)^0.
\]
Therefore,
\[
\dim K-\dim(V^\perp\cap K)=\dim V-\dim K^\perp.
\]

$(iii)$ It follows from (\ref{1}). \qed

 Now, let $\Omega$ a $2$-form on a real vector space $V$ of
finite dimension and $\eta\in V^*.$ If ${\cal R}$ is a vector of
$V$ such that $\eta({\cal R})=1,$ we can consider the $2$-form
$\omega_{\cal R}$ defined by
\begin{equation}\label{a2}
\omega_{\cal R}=\Omega-\eta\wedge \gamma_{\cal R}
\end{equation}
where $\gamma_{\cal R} \in V^*$ is given by
\begin{equation}\label{a3}
\gamma_{\cal R}=\inn({\cal R})(\Omega).
\end{equation}
It is clear that the triad $(\eta,\omega_{\cal R}, {\cal R})$ is a
precosymplectic structure on $V.$

If $K$ is a subspace of $V$, we will denote by $K_{(\omega_{\cal
R},\eta)}^\perp$ the orthogonal subspace of $K$ with respect to
$\omega_{\cal R}$ and $\eta.$

\begin{lem}\label{10'''}
Suppose that ${\cal R}$ and ${\cal R}'$ are vectors of $V$ such
that $\eta({\cal R})=\eta({\cal R}')=1.$ Then:
\begin{enumerate}
\item[$(i)$] $V_{(\omega_{\cal
R},\eta)}^\perp=V_{(\omega_{{\cal R}',\eta)}}^\perp.$
\item[$(ii)$] If $K$ is a subspace of $V,$ we have
\[
\dim K_{(\omega_{\cal R},\eta)}^\perp=\dim K_{(\omega_{{\cal
R}'},\eta)}^\perp.
\]
\end{enumerate}
\end{lem}
\proof $(i)$ We must prove that $\ker\omega_{\cal R} \cap
\ker\eta=\ker \omega_{{\cal R}'}\cap \ker\eta$ (see the first part
of Lemma \ref{10'}).

 From (\ref{a2}) and (\ref{a3}), we deduce that
\[
\omega_{{\cal R}'}=\omega_{\cal R} + \eta\wedge (\gamma_{\cal
R}-\gamma_{{\cal R}'}).
\]
Thus, if $v\in \ker\omega_{\cal R}\cap \ker \eta$ then
$\inn(v)\omega_{{\cal R}'}=-\inn(v)(\gamma_{\cal R}-\gamma_{\cal
R'})\eta$. In particular,
\[
0=\inn({{\cal R}'})\inn(v)\omega_{{\cal R}'}=-\inn(v)(\gamma_{\cal
R}-\gamma_{{\cal R}'})
\]
which implies that $\inn(v)\omega_{{\cal R}'}=0.$

This shows that $\ker \omega_{\cal R}\cap \ker\eta\subseteq
\ker\omega_{{\cal R}'}\cap \ker\eta.$ The converse is proved in a
similar way.

$(ii)$ It  follows from $(i)$ and the second part of Lemma
\ref{10'}. \qed

Next, we will exhibit the definition and some properties of
cosymplectic structures on a real vector space (for more
information, see \cite{A-89,CLL-92,LMM-96}).

Let $V$ be a real vector space of dimension $2n+1$.

A {\sl cosymplectic structure } on $V$ is a couple $(\eta,\omega)$,
where $\eta\in V^*$, $\omega$ is a $2$-form on $V$ and the linear
map $\flat_{(\eta,\omega)}:V\to V^*$ is a linear isomorphism or,
equivalently,
$\eta\wedge\omega^n=\eta\wedge\omega\wedge\dots^{(n}\dots \wedge
\omega\not=0.$ If $(V,\eta,\omega)$ is a cosymplectic vector space
then ${\cal R}=\flat_{(\eta,\omega)}^{-1}(\eta)$ is called the
{\sl Reeb vector } of $V$. ${\cal R}$ is the unique vector of $V$
which satisfies the conditions
\[
\inn({\cal R})\eta=1,\;\;\;\; \inn ({\cal R})\omega=0.
\]
In particular the triad $(\eta,\omega,{\cal R})$ is a
precosymplectic structure on $V$. Furthermore, using Lemma
\ref{10'}, it follows that
\begin{lem}\label{10''}\cite{LMM-96} Let $(\eta,\omega)$ be a
cosymplectic structure on a real vector space $V$. Then:
\begin{enumerate}
\item $V^\perp=\{0\}.$
\item If $K$ is a subspace of $V$, we have that $\dim K^\perp=\dim V-\dim K.$
\item If $K$ is a subspace of $V$ and $K\cap K^\perp=\{0\}$ then
$V=K\oplus K^\perp.$
\end{enumerate}
\end{lem}
We also get
\begin{lem}\label{13'}
Let $(\eta,\omega)$ be a cosymplectic structure on a real vector
space $V$. Suppose that $K$ is a subspace of $V$ such that $K\cap
K^\perp=\{0\}$ and that there exists ${\cal R}_K\in K$ satisfying
\begin{equation}\label{72'}
i({\cal R}_K)\omega_K=0,\;\; i({\cal R}_K)\eta_K=1,
\end{equation}
where $\omega_K=\omega\vert_{K\times K}$ and $\eta_K=\eta\vert_K.$ Then
the couple $(\eta_K,\omega_K)$ is a cosymplectic structure on $K$
and ${\cal R}_K$ is the Reeb vector of $K.$
\end{lem}
\proof Let $\flat_{(\eta_K,\omega_K)}:K\to K^*$ be the linear map
induced by the $2$-form $\omega_K$ and the $1$-form $\eta_K.$ If
$X\in \ker \flat_{(\eta_K,\omega_K)}$ then we deduce that
$i(X)\omega_K=0,$ $i(X)\eta_K=0.$ Now, since $K\cap
K^{\perp}=\{0\}$, it is easy to prove that
\begin{equation}\label{72''}
\ker\omega_K\cap \ker\eta_K=\{0\}
\end{equation}
and thus $X=0.$ Therefore, the map $\flat_{(\eta_K,\omega_K)}$ is
a linear isomorphism.

Furthermore, if $X\in \ker\omega_K$ then, from (\ref{72'}),
it follows $X-\eta_K(X){\cal R}_K\in \ker\omega_K\cap
\ker\eta_K$ and consequently, from (\ref{72''}), we obtain that
$X=\eta_K(X){\cal R}_K.$ Thus, $\ker\omega_K=\langle{\cal R}_K\rangle$, which
implies that the dimension of $K$ is odd. \qed

\section{Poisson and cosymplectic structures and Dirac
brackets}\label{B}

 A {\sl Poisson structure } on a differentiable
manifold $M$ is a bracket of functions
$\{\;,\;\}:C^\infty(M)\times C^\infty(M)\to C^\infty(M)$ which
satisfies the following properties:
\begin{itemize}
\item
{\sl Skew-symmetry: }  $\{F,G\}=-\{G,F\}.$
\item
{\sl Leibniz rule: } $\{FF',G\} =F\{F',G\} + F'\{F,G\}.$
\item
{\sl Jacobi identity: } $\{F,\{G,H\}\} + \{G,\{H,F\}\} +
\{H\{F,G\}\}=0.$
\end{itemize}
If $\{\;,\}$ is a Poisson bracket and $F\in C^\infty (M)$ one can
define the {\sl Hamiltonian vector field } $X_F$ associated with
$F$ as follows
\[
X_F(G)=\{G,F\},\;\; \mbox{ for every } G\in C^\infty(M).
\]
In addition, one also can define a $2$-vector $\Lambda$
characterized  by the condition
\begin{equation}\label{B1}
\Lambda(dF,dG)=\{F,G\}, \makebox{ for } F,G\in C^\infty(M),
\end{equation}
and it follows that $[\Lambda,\Lambda]=0,$ where $[\;,\;]$ is the
{\sl Schouten-Nijenhuis bracket}.

Conversely, if $\Lambda$ is a $2$-vector on $M$ such that
$[\Lambda,\Lambda]=0$ and $\{\;,\;\}$ is the bracket of functions
given by (\ref{B1}), then $\{\;,\;\}$ is a Poisson bracket (for
more details, see \cite{Li-77,V-94}).

Next, we show that a cosymplectic structure on manifold $M$
induces a Poisson bracket on $C^\infty(M)$ (see
\cite{A-89,CLL-92}). We need some results about cosymplectic
vector spaces.

Let $(V,\eta,\omega)$ be a cosymplectic vector space with Reeb
vector ${\cal R}.$ If $\alpha\in V^*$, we can consider the vectors
$X_\alpha$ and $E_\alpha$ defined by
\[
X_\alpha=\flat_{(\eta,\omega)}^{-1} (\alpha-\alpha({\cal R})\eta),
\;\;\; E_\alpha=\flat^{-1}_{(\eta,\omega)}(\alpha +
(1-\alpha({\cal R}))\eta).\]
These vectors are characterized by
the following conditions
\[
\begin{array}{lcl}
i(X_\alpha)\omega=\alpha-\alpha({\cal R})\eta,&&
i(X_\alpha)\eta=0,\\ i(E_\alpha)\omega=\alpha-\alpha({\cal
R})\eta,&&i(E_\alpha)\eta=1. \end{array}
\]
Note that $E_\alpha={\cal R} +X_\alpha.$

The cosymplectic structure $(\eta,\omega)$ also allows us to
introduce a $2$-vector $\Lambda:V^*\times V^*\to \Real$ on $V$
given by
\[
\Lambda(\alpha,
\beta)=\omega(X_\alpha,X_\beta)=\omega(E_\alpha,E_\beta),
\makebox[1cm]{} \mbox{for }\alpha,\beta\in V^*.
\]
The $2$-vector $\Lambda$ induces a linear map $\#_\Lambda:V^*\to
V$ defined by $\beta(\#_\Lambda(\alpha))=\Lambda(\alpha,\beta),$
for $\alpha,\beta\in V^*$. We have that
\[
\flat_{(\eta,\omega)}(\#_\Lambda(\alpha))=\alpha-\alpha({\cal
R})\eta.
\]
In particular, $\#_\Lambda(V^*)=\ker\eta$, $\ker\#_\Lambda=\langle\eta\rangle$
and the map $\#_\Lambda:\langle{\cal R}\rangle^0\to \ker\eta$ is a linear
isomorphism.

If $K$ is a subspace of $V$ which satisfies certain conditions,
then the map $\#_\Lambda$ can be used in order to obtain the
orthogonal subspace of $K$ with respect to $(\eta,\omega).$

\begin{lem}\label{14'}
If $K$ is a subspace of $V$ and there exists ${\cal R}_K\in K$
such that
\begin{equation}\label{B2}
(i({\cal R}_K)\omega)\vert_K=0,\;\;\; i({\cal R}_K)\eta=1
\end{equation}
then $K^\perp =\#_\Lambda(K^0).$ \end{lem}
 \proof
 Suppose that
$\alpha\in K^0$. Since $\eta(\#_\Lambda (\alpha))=0,$ it follows
that
\[
\begin{array}{lcl}
(i(\#_\Lambda(\alpha))\omega-\eta(\#_\Lambda(\alpha))\eta)\vert_K&=&
(i(\#_\Lambda(\alpha))\omega)\vert_K=
(\flat_{(\eta,\omega)}(\#_\Lambda(\alpha))\vert_K\\ &=&
(\alpha-\alpha({\cal R})\eta)\vert_{K}=(-\alpha({\cal R})\eta)\vert_K.
\end{array}
\]
 Therefore, from (\ref{B2}), we deduce that
 \[
 0=(-\alpha({\cal R})\eta)({\cal R}_K)=-\alpha({\cal R})
 \] and
 $(i(\#_\Lambda(\alpha))\omega-\eta(\#_\Lambda(\alpha))\eta)\vert_K=0,$
 i.e, $\#_\Lambda(\alpha)\in K^\perp.$ We have proved that
 $\#_\Lambda(K^0)\subseteq K^\perp.$

 Now, using Lemma \ref{10''} and the fact that $K^0\cap
 \langle\eta\rangle=\{0\},$ we obtain that
 \[
 \dim \#_\Lambda(K^0)=\dim K^0=\dim K^\perp.
 \]
 \qed
 Moreover, a direct computation shows the following.
 \begin{prop}\label{14''}
 Let $(\eta,\omega)$ be a cosymplectic structure on a real vector
space $V$  and ${\cal R}$ be the Reeb vector of $V$. Suppose that
$\alpha$ is  a $1$-form on $V$ and  that $\tilde{\omega}$ is the
$2$-form on $V$ defined by $\tilde{\omega}=\omega + \alpha\wedge
\eta.$ Then:
\begin{enumerate}
\item
The couple $(\eta,\tilde\omega)$ is a cosymplectic structure on $V$
with the Reeb vector $\tilde{\cal R}$ given by $\tilde{\cal
R}=E_\alpha,$ where $E_\alpha=\flat^{-1}_{(\eta,\omega)} (\alpha +
(1-\alpha({\cal R}))\eta).$
\item
If $\beta$ is a $1$-form on $V$ and $X_\beta$ (resp.
$\tilde{X}_\beta$) is the vector of $V$ defined by
$X_\beta=\flat_{(\eta,\omega)}^{-1}(\beta-\beta({\cal R})\eta)$
(resp.
$\tilde{X}_\beta=\flat_{(\eta,\tilde{\omega})}^{-1}(\beta-\beta({\cal
R})\eta))$ then $X_\beta=\tilde{X}_\beta.$
\item If $\Lambda$ (resp. $\tilde\Lambda$) is the $2$-vector on $V$
induced by the structure $(\eta,\omega)$ (resp.
$(\eta,\tilde\omega))$ then $\Lambda= \tilde\Lambda.$
\end{enumerate}
\end{prop}
 Now, let $M$
be an odd-dimensional differentiable manifold, and $(\eta,\omega)$
a {\sl cosymplectic structure } on $M$; that is, $\eta$ is a
closed $1$-form, $\omega$ is a closed $2$-form and the couple
$(\eta_x,\omega_x)$ is a cosymplectic structure on the vector
space $\Tan_xM$, for every $x\in M$. Then there exists a unique vector
field ${\cal R}$ on $M$, the {\sl Reeb vector field}, satisfying
the conditions $i({\cal R})\omega=0$ and $i({\cal R})\eta=1.$
Moreover, using the above constructions on cosymplectic vector
spaces, one can take the $2$-vector $\Lambda$ on $M$ induced
by the couple $(\eta,\omega)$, and it follows that $\Lambda$ defines
a Poisson structure on $M$ (see \cite{A-89,CLL-92}). In addition,
if $F\in C^\infty(M)$ we also can consider the vector fields $X_F$
and $E_F$ on $M$ characterized by the conditions
\[
\begin{array}{lcllcl}
i(X_F)\omega&=&dF-{\cal R}(F)\eta, & i(X_F)\eta&=&0,\\
i(E_F)\omega&=&dF-{\cal R}(F)\eta,&i(E_F)\eta&=&1.
\end{array}
\]
$X_F$ is just the Hamiltonian vector field of $F$ with respect to
the Poisson structure $\Lambda.$ $E_F$ is called the {\sl
evolution vector field } of $F$ with respect to the cosymplectic
structure $(\eta,\omega)$. We have that $E_F={\cal R} + X_F.$

Furthermore, from Proposition \ref{14''}, we deduce the following
result.
\begin{prop}\label{14'''}\cite{CLMM-00}
Let $(\eta,\omega)$ be a cosymplectic structure on a
differentiable manifold $M$ and ${\cal R}$ be the Reeb vector
field of $M$. Suppose that $F$ is $C^\infty$-differentiable real
function on $M$ and that $\tilde\omega$ is the $2$-form on $M$
defined by $\tilde\omega=\omega + dF\wedge \eta$. Then:
\begin{enumerate}
\item
The couple $(\eta,\tilde\omega)$ is a cosymplectic structure on $M$
with Reeb vector field $\tilde{\cal R}$ given by
\[
\tilde{\cal R}=E_F,
\]
where $E_F$ is the evolution vector field of $F$ with respect to
the cosymplectic structure $(\eta,\omega).$
\item
If $G$ is a $C^\infty$-differentiable real function on $M$ and
$X_G$ (resp. $\tilde{X}_G$) is the Hamiltonian vector field of $G$
with respect to the structure $(\eta,\omega)$ (resp.
$(\eta,\tilde\omega))$ then $X_G=\tilde{X}_G.$
\item
If $\Lambda$ (resp. $\tilde\Lambda$) is the Poisson $2$-vector on
$M$ associated with the cosymplectic structure $(\eta, \omega)$
(resp. $(\eta,\tilde\omega)$) then $\Lambda=\tilde\Lambda.$
\end{enumerate}
\end{prop}

Next, assume that $(\eta,\omega)$ is a cosymplectic structure on a
manifold $M$ and that $D$ is a distribution on $M$ of dimension
$d$ satisfying
\begin{equation}\label{B3}
\Tan M=D\oplus \#_\Lambda(D^0),
\end{equation}
$\Lambda$ being the Poisson $2$-vector associated with the
structure $(\eta,\omega).$ The splitting (\ref{B3}) allows us
to introduce two projectors
\[
{\Bbb P}:\Tan M\to D,\;\;\; {\Bbb Q}:\Tan M\to \#_\Lambda(D^0)
\]
in such a way that $X={\Bbb P}X + {\Bbb Q}X,$ for every $X\in \Tan M.$

Moreover, one can define a {\sl Dirac  bracket } of functions
$\{\;,\;\}_D$ as follows
\[
\{F,G\}_D=\omega({\Bbb P}(X_F),{\Bbb P}(X_G)),
\]
for $F,G\in C^\infty(M)$, where $X_F$ and $X_G$ are the
Hamiltonian vector fields of $F$ and $G$ with respect to the
cosymplectic structure $(\eta,\omega)$. It is clear that the
bracket $\{\;,\;\}_D$ is skew-symmetric and that it satisfies the
Leibniz rule.

Moreover, since $(\eta(\#_\Lambda(\alpha))=0$ for every
$\alpha$, we deduce that $\eta({\Bbb P}({\cal R}))=1.$ Thus, it is
easy to prove that $D(x)=\bar D(x)\oplus\langle{\Bbb P}({\cal R})(x)\rangle$,
for every $x\in M,$ where $\bar{D}(x)=D(x)\cap \ker\eta_x.$
Therefore, the assignment
\[
x\in M\longrightarrow \bar D(x)\subseteq \Tan_xM
\]
defines a distribution  $\bar D$ on $M$ of dimension $d-1$. In
addition, we have
\begin{prop}\cite{CLMM-00}\label{p11-B}
The distribution $\bar D$ is completely integrable if and only if
$\{\;,\;\}_D$ is a Poisson bracket.
\end{prop}
\begin{remark}
{\rm If the distribution $D$ is completely integrable then it is
clear that the distribution $\bar{D}$ is also completely
integrable.}
\end{remark}

\section{Auxiliary results}\label{C}

 Taking into account the characterization of the submanifolds
 $C_i$ given in Proposition \ref{cons},
 we can prove the following results
 (remember that $C_0\equiv F$):

\begin{lem}
For every $x\in C_0$,
$\ker\,\Omega_x\cap\ker\,\eta_x\subseteq
\ker\,\omega_x\cap\ker\,\eta_x=\Tan_x^\perp C_0$.
\label{a}
\end{lem}
\proof The last equality follows from Lemma \ref{10'}.

For the first inclusion,  if $X\in\ker\,\Omega_x\cap\ker\,\eta_x$
then $$
\inn(X)\omega_x=\inn(X)\Omega_x-\inn(X)(\eta_x\wedge\gamma_x)=
\inn(X)\gamma_x\wedge\eta_x = \inn(X)\inn({\cal
Y})\Omega_x\wedge\eta_x = - \inn({\cal
Y})\inn(X)\Omega_x\wedge\eta_x = 0 $$ and thus $X\in\Tan_x^\perp
C_0$. \qed

\begin{lem}
For every $x\in C_1$,
$\ker\,\Omega_x\cap\ker\,\eta_x=
\ker\,\omega_x\cap\ker\,\eta_x=\Tan_x^\perp C_0$.
\label{b}
\end{lem}
\proof If $X\in\Tan_x^\perp C_0=\ker\,\omega_x\cap\ker\,\eta_x$
then $$
\inn(X)\Omega_x=\inn(X)\omega_x+\inn(X)(\eta_x\wedge\gamma_x)=
-[\inn(X)\gamma_x]\eta_x $$ but as $x\in C_1$, there exists
$Y\in\Tan_xC_0$ such that $\inn(Y)\Omega_x=0$ and
$\inn(Y)\eta_x=1$; therefore the contraction of the above equality
with this $Y$ leads to $\inn(X)\gamma_x=0$, and hence
$\inn(X)\Omega_x=0$. \qed

\begin{lem}
For every $Z,Y\in\ker\,\omega\cap\ker\,\eta$, we get
\ben
\item
$\inn(Y)\inn(Z)\d\gamma\vert_{C_1}=0$.
\item
$[Z,Y]\vert_{C_1}\in(\ker\,\omega\cap\ker\,\eta)\vert_{C_1}$.
\een
\label{c}
\end{lem}
\proof
\ben
\item
As $Z,Y\in\ker\,\omega\cap\ker\,\eta$, we obtain that
$\inn(Y)\inn(Z)\Omega=\inn(Y)\inn(Z)\omega+\inn(Y)\inn(Z)(\eta\wedge\gamma)=0$.
 Therefore
\beann
\inn(Y)\inn(Z)\d\gamma&=&
\inn(Y)\inn(Z)\d\inn({\cal Y})\Omega=
\inn(Y)\inn(Z)[\Lie({\cal Y})\Omega]
\\ &= &
\Lie({\cal Y})\inn(Y)\inn(Z)\Omega-\inn(Y)\inn([{\cal Y},Z])\Omega
-\inn([{\cal Y},Y])\inn(Z)\Omega
\\ &=&
\inn([{\cal Y},Z])\inn(Y)\Omega-\inn([{\cal Y},Y])\inn(Z)\Omega
\eeann
 However, as a consequence of the above Lemma, since
$Z\vert_{C_1},Y\vert_{C_1}\in
(\ker\,\Omega\cap\ker\,\eta)\vert_{C_1}$, hence
$\inn(Y)\inn(Z)\d\gamma\vert_{C_1}=0$.
\item
Observe that as $Z,Y\in\ker\,\omega\cap\ker\,\eta$, and $\eta$ is
a closed form, $$
\inn([Z,Y])\eta=\Lie(Z)\inn(Y)\eta-\inn(Y)\Lie(Z)\eta=
-\inn(Y)\inn(Z)\d\eta-\inn(Y)\d\inn(Z)\eta=0\; , $$ then
$[Z,Y]\in\ker\,\eta$. Moreover, $$
\inn([Z,Y])\omega=\Lie(Z)\inn(Y)\omega-\inn(Y)\Lie(Z)\omega=
-\inn(Y)\inn(Z)\d\omega
-\inn(Y)\d\inn(Z)\omega=-\inn(Y)\inn(Z)\d\omega. $$ Now, for every
$X\in\vf(C_0)$, as
 $\d\omega=\d\Omega-\d(\eta\wedge\gamma)=\eta\wedge\d\gamma$, we have
$$ \inn(X)\inn(Y)\inn(Z)\d\omega=
\inn(X)\inn(Y)\inn(Z)(\eta\wedge\d\gamma)=
[\inn(X)\eta]\inn(Y)\inn(Z)\d\gamma\; . $$ Therefore, for every
$x\in C_1$, taking into account the first item we obtain $$
\inn(X_x)\inn(Y_x)\inn(Z_x)(\d\gamma)_x=
[\inn(X_x)\eta_x]\inn(Y_x)\inn(Z_x)(\d\gamma)_x = 0 $$ and the
result follows. \qed \een


 \end{document}